%% file: paper1.tex
\documentclass[twocolumn]{aastex62}
\shorttitle{{\it Spitzer} Galactic Center}
\shortauthors{Simpson}

\usepackage{latexsym}
\usepackage{graphicx}
\usepackage{amssymb}
\usepackage{amsmath}
\usepackage{longtable}
\usepackage{epsf}

\begin{document}

\title{{\it Spitzer} Infrared Spectrograph Observations of the Galactic Center: 
Quantifying the Extreme Ultraviolet/Soft X-ray Fluxes}

\author[0000-0001-8095-4610]{Janet P. Simpson}
\affil{SETI Institute \\
189 Bernardo Ave. \\
Mountain View, CA 94043, USA}
\email{janet.p.simpson@gmail.com}

\begin{abstract}

It has long been shown that the extreme ultraviolet spectrum of the ionizing stars of \ion{H}{2} regions can be estimated by comparing the observed line emission to detailed models. 
In the Galactic Center (GC), however, previous observations have shown that the ionizing spectral energy distribution (SED) of the local photon field is strange, producing both very low excitation ionized gas (indicative of ionization by late O stars) and also widespread diffuse emission from atoms too highly ionized to be found in normal \ion{H}{2} regions. 
This paper describes the analysis of all the GC spectra taken by {\it Spitzer}'s Infrared Spectrograph and downloaded from the Spitzer Heritage Archive. 
In it, \ion{H}{2} region densities and abundances are described, and serendipitously discovered candidate planetary nebulae, compact shocks, and candidate young stellar objects are tabulated. 
Models were computed with Cloudy, using SEDs from Starburst99 plus additional X-rays, and compared to the observed mid-infrared forbidden and recombination lines. 
The ages inferred from the model fits do not agree with recent proposed star formation sequences 
(star formation in the GC occurring along streams of gas with density enhancements caused by close encounters with the black hole, Sgr A*), 
with Sgr~B1, Sgr~C, and the Arches Cluster being all about the same age, around 4.5 Myr old, with similar X-ray requirements. 
The fits for the Quintuplet Cluster appear to give a younger age, but that could be caused by 
%rejuvenated binary O stars or 
higher-energy photons from shocks from stellar winds or from a supernova.

\end{abstract}


\section{Introduction}

Owing to the deep potential well of the Galactic Center (GC), 
all aspects of the environment are found at extreme conditions compared to what is seen in the Galactic spiral arms: 
the dense nuclear star cluster at whose center lies a $4 \times 10^6$ M$_\odot$ black hole 
at a distance of {approximately 8} kpc ({Reid et al. 2014; Boehle et al. 2016,} Bland-Hawthorn \& Gerhard 2016), 
massive dense molecular clouds (residing in the so-called the Central Molecular Zone, CMZ, Morris \& Serabyn 1996),
and a radiation field including both synchrotron radiation (Yusef-Zadeh, Hewitt, \& Cotton 2004) 
and substantial X-rays (e.g., Ponti et al. 2015).
 
Figure 1 shows the prominent features of the GC that will be discussed in this paper:
blue, green, and red indicate the 21 \micron\ emission observed by the {\it MidCourse Space Experiment} 
({\it MSX}, Price et al. 2001) and
the 70 and 500 \micron\ emission from {\it Herschel} (Molinari et al. 2011).
The 21 and 70 \micron\ emission shows the warm dust heated by nearby massive stars; 
it consequently delineates the star-forming regions Sgr B and Sgr C and the dust heated by the massive star clusters
known as the Quintuplet and Arches Clusters as well as the nucleus of the Galaxy itself, Sgr A
(Morris \& Serabyn 1996).
The far-infrared emission from the cold dust shows the locations of the dense molecular clouds that 
have very few indicators of active star formation, even though their overall masses and densities 
seem to be adequately high.
These clouds have the appearance of a lopsided ring of gas and dust 
circling the center (Molinari et al. 2011).

%Figure 1
\begin{figure*}
\includegraphics[width=173mm]{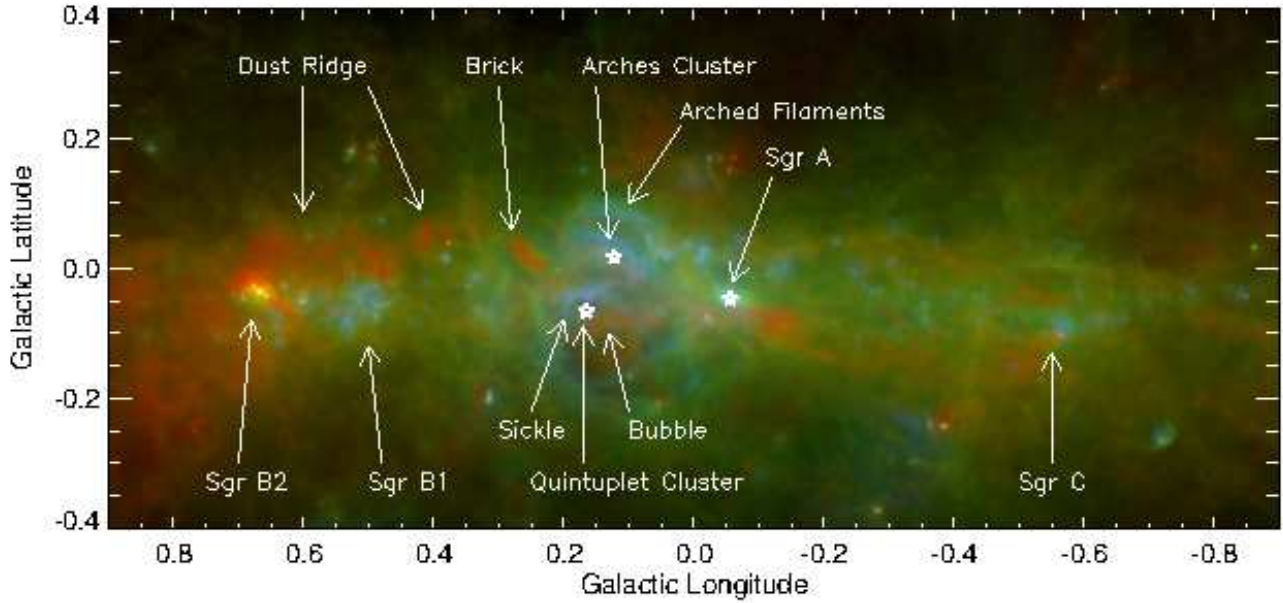}
\caption{Three-color image of the Galactic Center. 
Blue is the 21\micron\ Band E {\it MSX} image from Price et al. (2001), 
green is the 70 \micron\ image from 
Hi-GAL (Molinari et al. 2011) taken with the Photodetector Array Camera and Spectrometer 
(PACS, Poglitsch et al. 2010) on {\it Herschel Space Observatory} (Pilbratt et al. 2010), 
and red is the 500 \micron\ image from Hi-GAL taken with the 
Spectral and Photometric Imaging Receiver (SPIRE, Griffin et al. 2010).
} 
\end{figure*}

Longmore et al. (2013) showed that if one describes the positions of the molecular clouds 
at positive Galactic longitudes as lying on the Earth-side of the cloud orbit, 
the amount of star formation in the clouds increases as a function of time since the clouds 
passed the pericenter of the Galaxy's exact center, Sgr A.
They suggested that passing the pericenter compresses the clouds, thereby triggering stars to form.
Kruijssen, Dale, \& Longmore (2015) then showed from the radial velocities of the clouds 
that the gas orbit could not be closed; instead the cloud motions can be better portrayed 
as streams of gas in orbit about Sgr A.
Modeling the orbits, they estimated times since pericenter passage of 
0.30 Myr for the Brick, 0.74 Myr for Sgr B2, and 3.58 Myr for Sgr C (Figure~1). 
The Arches and Quintuplet Clusters formed from gas clouds that passed the pericenter 
at even earlier times. 

Krumholz \& Kruijssen (2015) and Krumholz, Kruijssen, \& Crocker (2017) 
expanded the discussion of gas streams in the GC to include the effects of the Milky Way's Bar.
Gas flows in along the Bar and is highly turbulent within the inner Lindblad resonance 
at $\sim 1000$~kpc from the center owing to the high shear.
However, the turbulence dissipates at $\sim 100$~pc (the edge of the CMZ) 
where the rotation curve becomes close to solid-body,
thereby allowing the massive molecular clouds to form with resulting star formation.
This processs is episodic because feedback from winds from the stars and supernovae 
blow much of the gas out, greatly reducing the star formation rate.
Krumholz et al. (2017) suggest that the GC is currently in a low star-formation state, 
the last major star formation having occurred $\sim 8$~Myr ago.
This explains the apparent low star-formation rate of the GC (Kruijssen et al. 2014).

This is an exciting model because of the wealth of important observations that it explains. 
However, there are still uncertainties with the details of this model.
The model posits that the positions in the foreground of Sgr A at positive longitudes 
are all very young, $< 1$~Myr, 
with Sgr B1 (already on the back loop of the orbit) somewhat but not greatly older than Sgr B2 
(1.5 Myr vs. 0.7 Myr, Barnes et al. 2017) 
and Sgr C closing in on its second pass by the pericenter (Sgr A).
However, the appearance of Sgr B1 is that of an \ion{H}{2} region in the process of dispersal,
indicating a much older age than that of Sgr B2, 
even though both it and Sgr C contain substantial numbers of young stellar object (YSO) candidates 
(An et al. 2011; Lu et al. 2017).

There are additional suggestions for the causes of the low star-formation rate in the GC:
Federrath et al. (2016) suggest that solenoidal driving of the turbulence owing to the high shear in the GC
compared to the compressive turbulence in spiral arm star forming regions is the cause
of the low star formation rate (see also Kruijssen et al. 2014).
High levels of turbulence in the GC is the usual suggestion for the high gas temperature
(line widths much larger than thermal) 
in molecular clouds compared to the dust temperature (e.g., Immer et al. 2016).
In addition, recent higher resolution observations indicate that 
the temperature and density conditions in the GC gas clouds 
are not as simple as the previous theories assumed: 
Kauffmann et al. (2017a) observe that although the line widths are large when integrated 
over whole clouds, the line widths for individual clumps within a cloud can be of normal width 
(compared to spiral arm molecular clouds).
Moreover, Kauffmann et al. (2017b) find that the densities in the clumps do not increase fast enough 
towards their centers to be able to initiate star formation
and that there is no clear trend of increased and then decreased star formation 
as predicted by the Kruijssen et al. (2015) model. 
They suggest that initial conditions may be as important as position on the orbital path since pericenter 
and that data on more clouds are needed to test the model.

Further insight can be ascertained from mid-infrared (MIR) spectroscopy.
In typical observing programs, sources are first identified with imaging 
on either the {\it Infrared Space Observatory}'s (ISO) ISOGAL survey (Schuller et al. 2006)
or {\it Spitzer Space Telescope}'s (Werner et al. 2004) Infrared Array Camera (IRAC, Fazio et al. 2004); 
then spectra are taken with the Infrared Spectrograph (IRS, Houck et al. 2004).
The IRS has four modules: short-low, SL, with resolution $R \sim 100$ and range 5.2 -- 14.5 \micron, 
long-low, LL, with $R \sim 100$ and spectral range 14.0 -- 37 \micron, 
short-high, SH, with $R \sim 600$ and range 9.9 -- 19.6 \micron, and 
long-high, LH, with $R \sim 600$ and range 19 -- 36 \micron.
In spectra from 5 -- 20 \micron\ one can identify young stellar objects (YSOs) 
from the presence of ice features  (e.g., Simpson et al. 2012) 
and organic molecules evaporated from ice (e.g., An et al. 2009) 
in absorption in their dusty envelopes. 
From the ionic abundances of a number of elements of differing ionization potential ($IP$), 
one can determine the ionization structure of a region, get probable locations of 
the exciting stars, and, if the cluster of stars that produce the energetic photons is large enough to 
include a representative sample of the initial mass function, get an estimate of the age of the cluster.

A number of {\it Spitzer} programs observed targets in the GC with the IRS. 
Immer et al. (2012) found 14 YSOs from IRS spectra of 68 ISOGAL sources, based on the shapes 
of their spectral energy distributions (SEDs) and the presence or lack of 
polycyclic aromatic hydrocarbon (PAH) features or atomic fine structure lines characteristic of \ion{H}{2} regions. 
Out of 107 candidates, An et al. (2011) found 16 YSOs and 19 possible YSOs from the presence of 
methanol mixed with the CO$_2$ ice in their deep 15 \micron\ absorption features. 
Note that these same candidate YSOs also have deep ice features at 5.8 and 6.8 \micron, 
features that are also characteristic of YSOs (e.g., Simpson et al. 2012 and references therein).
An, Ram\'irez, \& Sellgren (2013) used the SH and LH background spectra from the same data set 
to test whether the GC is ionized by flux from an active galactic nucleus (AGN); 
by comparing their flux ratios to the external galaxy fluxes measured by Dale et al. (2009), 
they concluded that there is no significant AGN activity in the GC. 

In Simpson et al. (2007) we analyzed SH and LH spectra of the Arched Filaments, Arches Cluster, 
regions near the Quintuplet Cluster, and the Radio Arc Bubble. 
In it we showed that the Arches Cluster is the source of the ionizing photons of the Arched Filaments, 
and the Quintuplet Cluster is likewise the source of the ionizing photons of the Bubble. 
The Quintuplet Cluster also ionizes the rim of the Bubble at its lowest Galactic latitudes, 
from which we inferred that the G0.10$-$0.08 molecular cloud is totally unrelated along the line of sight. 
From measurements of the Fe/Ne ratio we concluded that 
strong shocks have contributed to destroying grains in the Bubble, 
thereby increasing the abundance of gas-phase iron compared to abundances 
in the Arched Filaments and the Bubble rim. 

Simpson et al. (2007) also found that
that even though the SED of the photons ionizing the observed gas 
could be described as having the effective temperature, $T_{\rm eff}$, of $\sim 36,000$~K 
(late O, Sternberg et al. 2003), 
essentially all the observed positions contained detectable [\ion{O}{4}] 25.9 \micron\ lines. 
This is surprising because 
O$^{3+}$ has an $IP = 54.9$ eV and this ionization stage is only observed in \ion{H}{2} regions 
ionized by the most massive O stars, and then only by stars with low metallicity, 
which the GC is not 
(O$^{3+}$ is sometimes detected in the \ion{H}{2} regions of low metallicity dwarf galaxies, Lutz et al. 1998).
In fact, An et al. (2013) showed that the [\ion{O}{4}] line is observed at many positions in the GC; 
nhey used its presence compared to its immediate (and partially blended) neighbor [\ion{Fe}{2}] 26.0 \micron\ 
to distinguish starbursts from LINERS (low-ionization nuclear emission-line regions) or Seyferts 
by their relative excitation.
In this paper I address why there is such a dichotomy between 
the generally observed, very low excitation gas interspersed with quite high excitation gas.

The {\it Spitzer} Heritage Archive contains a great deal more data of the same high quality.
Although both the SH and LH modules have compact apertures, measuring barely more 
than one spatial resolution element in both directions, 
the SL and LL modules are long slit, with slit lengths $57''$ and $168''$, respectively. 
Moreover, both SL and LL have two grating orders each, displaced by $22''$ or $24''$ on the sky, 
such that when one is observing some source in one order, the other order is taking data 
a slit length away. 
The result is that although the papers referenced above analyze the point source targets of each program, 
a substantial amount of the GC was actually covered when one includes 
the full slit lengths of the SL and LL modules and both orders, the order 
planned for the astronomical observing request (AOR) and the `other' order, 
observed but not normally analyzed. 
No one to this time has analyzed any of that data!

This is the first of several papers in which we analyze the IRS spectra of all four modules from programs 
0018 (PI: J. Houck), 3121 (PI: K. Kraemer), 3189 (PI: F. Schuller),  
3295 (PI: J. Simpson), 3616 (PI: J. Chiar), and 40230 (PI: S. Ram\'irez).
%Program 3397 (PI: W.-R. Hamann) has SH spectra only of WR stars in the Sickle and ?? Steinke et al. (2016) think this star is one of the ionizing stars of the Sickle but it is not part of the Quintuplet Cluster.
Subsequent papers will discuss individual objects such as Sgr A and Sgr B1 (J. Simpson et al., in preparation),
and exceptionally-highly excited sources that could be shocks (e.g., Allen et al. 2008) 
or candidate planetary nebulae (D. An et al., in preparation).
Section 2 describes the data and its analysis, 
Section 3 discusses the results, including the observation that [\ion{O}{4}] line emission is 
prevalent over the whole GC, 
Section 4 presents \ion{H}{2} region models computed with the code Cloudy {(Ferland et al. 2017)} 
and compares them to the observed line ratios to estimate the parameters of the 
exciting star clusters, 
and Section 5 presents the summary and conclusions. 

\section{Observations}

\input tab1.tex

The observations were taken with the {\it Spitzer} IRS in Cycles 1 and 4 and were 
downloaded from the {\it Spitzer} Heritage Archive\footnote{sha.ipac.caltech.edu/applications/Spitzer/SHA/}.
The spectra from programs 3295 (line fluxes published in Simpson et al. 2007) and 0018 
(mentioned in Simpson et al. but published in this paper) 
were reduced and calibrated with the {\it Spitzer} S13.2 pipeline; 
all the other data were reduced and calibrated with the final version of the pipeline, S18.18.
SH and LH spectra are found in programs 0018, 3295, and 40230, and 
SL and LL spectra are found in programs 3121, 3189, 3616, and 40230.

The data reduction subsequent to the pipeline for the SH and LH spectra from program 40230 
follows that described by Simpson et al. (2007) for program 3295 (and unpublished 0018): 
the basic calibrated data (bcds) for each telescope pointing 
(all program 40230 spectra were taken in staring mode)  
were median combined and cleaned of rogue and exceptionally noisy pixels 
(rejecting all spectra where the bcds show `jailbars').
Background subtraction was not performed because no background spectra at large enough distances 
from the GC were ever taken for any of these programs and the GC line-of-sight itself 
has at least some emission at all locations, as will be discussed later.
{Sample SH and LH spectra are shown in figure 2 of Simpson et al. (2007).}

CUBISM (Smith et al. 2007a) was used to extract the spectra of {both} the low resolution SL and LL 
{bcds and the high resolution SH and LH bcds}.
CUBISM was written to analyze the spectral maps of galaxies in the 
SIRTF Nearby Galaxies Survey (SINGS) program (Kennicutt et al. 2003); 
the software takes the bcds of the map and 
produces a three-dimensional cube of x, y, and wavelength,
where x and y refer to coordinates parallel and perpendicular to the 
{various slits.}
%SL or LL slit. 
The GC programs observed in both mapping (multiple adjacent slit positions on the sky) 
and staring (single slit placed on a requested target) modes.
{ For the low resolution modules,}
after cleaning the bad pixels, 
CUBISM was used to extract spectra from both modes at multiple positions along the slit, 
both for the order that had the requested target and also for the `other' order a slit length 
(plus $22''$ or $24''$) distant.
For staring mode (or for mapping mode where the multiple slit positions did not touch), 
the x-y slit in the CUBISM GUI appears as a line of about 32 pixels long and 2 pixels wide, 
where the pixel size is $1.8''$ for SL and $5.1''$ for LL; the physical slit widths 
and spatial resolution are approximately 2 pixels for all modules.
The extraction boxes were 3 pixels by 2 pixels for the single slits 
(or sometimes 3 by 3 pixels for the small maps of program 3189)
spaced by 2 pixels along the slit, thus producing typically 15 spectra per slit per order. 
Care was taken to have the extraction boxes of the two SL or LL orders overlap on the sky 
so that the spectra could be joined into single 5.3 -- 14.5 \micron\ or 14 -- 37 \micron\ spectra.
However, since the SL and LL slits are close to orthogonal on the sky, 
spectra from the two modules could not be joined except at the target positions. 
See figure 2 of An et al. (2011) for the layout of the slits as used by program 40230.
Not all SL pointings could produce usable spectra because the IRS peak-up cameras 
are on the same module and saturation in the peak-up arrays produced uncalibratable spectra.
{For the high resolution modules, the CUBISM extraction boxes were 
5 pixels by 2 pixels for SH and 3 pixels by 2 pixels for LH.}

Line and continuum intensities were obtained from all spectra by fitting Gaussian profiles 
and a sloping continuum. 
The lines measured are given in Table 1. 
The uncertainties for the line fluxes were estimated from the rms deviation of the data 
from the fit. 
Because there were hundreds of SH and LH spectra and thousands of SL and LL spectra, 
this fitting had to be done with an automated line fitting program without hand checking,
using a fixed set of wavelengths for each line such that there would be no interference 
with another line or PAH feature.
{A tool in the Spectroscopic Modeling Analysis and Reduction Tool 
(SMART, Higdon et al. 2004) can also be used to fit Gaussian line profiles to spectra, 
both single lines and partially blended lines with multiple Gaussians.}
Because the weak H 7--6 lines are crucial to the estimation of abundances with respect to hydrogen
and because the baseline for this line 
can be affected by the close, usually stronger H$_2$ S(2) line,
they were measured by hand with SMART in the SH spectra.
Additionally, those lines that are known to be blends were measured by hand with SMART:   
[\ion{O}{4}] 25.9 and [\ion{Fe}{2}] 26.0 \micron, [\ion{P}{3}] 17.89 \micron\ and [\ion{Fe}{2}] 17.94 \micron,    
and, rarely, [\ion{Ne}{5}] 14.32 and [\ion{Cl}{2}] 14.37 \micron, all in the high resolution spectra.
The first two blends for program 3295 are plotted in figure 10 of Simpson et al. (2007).
Tests of measuring the same line with both the automated line fitting program and with SMART 
show a difference in intensities of only a few percent for lines with signal/noise $> 3$ 
and less than $2\%$ for bright lines,
the difference being due to the flexible choice of baseline wavelengths with SMART 
versus the uniform baseline wavelengths with the automatic measuring program. 
{The weak H 7-6 recombination line exhibits a larger difference in intensities with a standard deviation
(SD) of $\sim 6\%$.}
However, whereas hand measurement with SMART could always identify those cases where 
no line could be fit, that was not the case for the automated Gaussian fitting program, 
which always found some sort of fit, good or bad. 
Consequently, its bad fits were identified as having  
negative fluxes, too big or too small line widths, or extreme radial velocities
and were removed from the data set. 

{The SH and LH spectra from program 40230 were also extracted with SMART.
This presents a good test of the reliability of spectral extraction with the various extraction software programs.
The spectra are dissimilar enough that the estimated line intensities differ by 
typically 4 or 5\% (SD) for the brighter lines and up to 14\% (SD) for the weaker lines, such as the H 7-6 
recombination line at 12.37 \micron. 
These systematic uncertainties are included in quadrature with the measured rms uncertainties
from the Gaussian fits for the computation of abundances in the next section. 
For the sake of consistency, almost all line intensities in this paper 
are from spectra extracted with CUBISM.
The exceptions are those from the previously described SH and LH spectra from programs 0018 and 3295, 
the blended line pair [P III] 17.89 and [Fe II] 17.94 \micron\ (e.g., Simpson et al. 2007),
and the quite faint [S I] 25.25 \micron\ line, which line is often obscured by the bad fringing in the LH spectra
that can be removed by SMART but not by CUBISM (the absolute intensities of this line are equal to the 
intensities measured from the SMART spectra times a scale factor computed from the 
other lines measured from CUBISM-extracted spectra).
}

\section{Results}

\subsection{Line Intensities}

%Figure 2 abc
\begin{figure*}
\includegraphics[width=184mm]{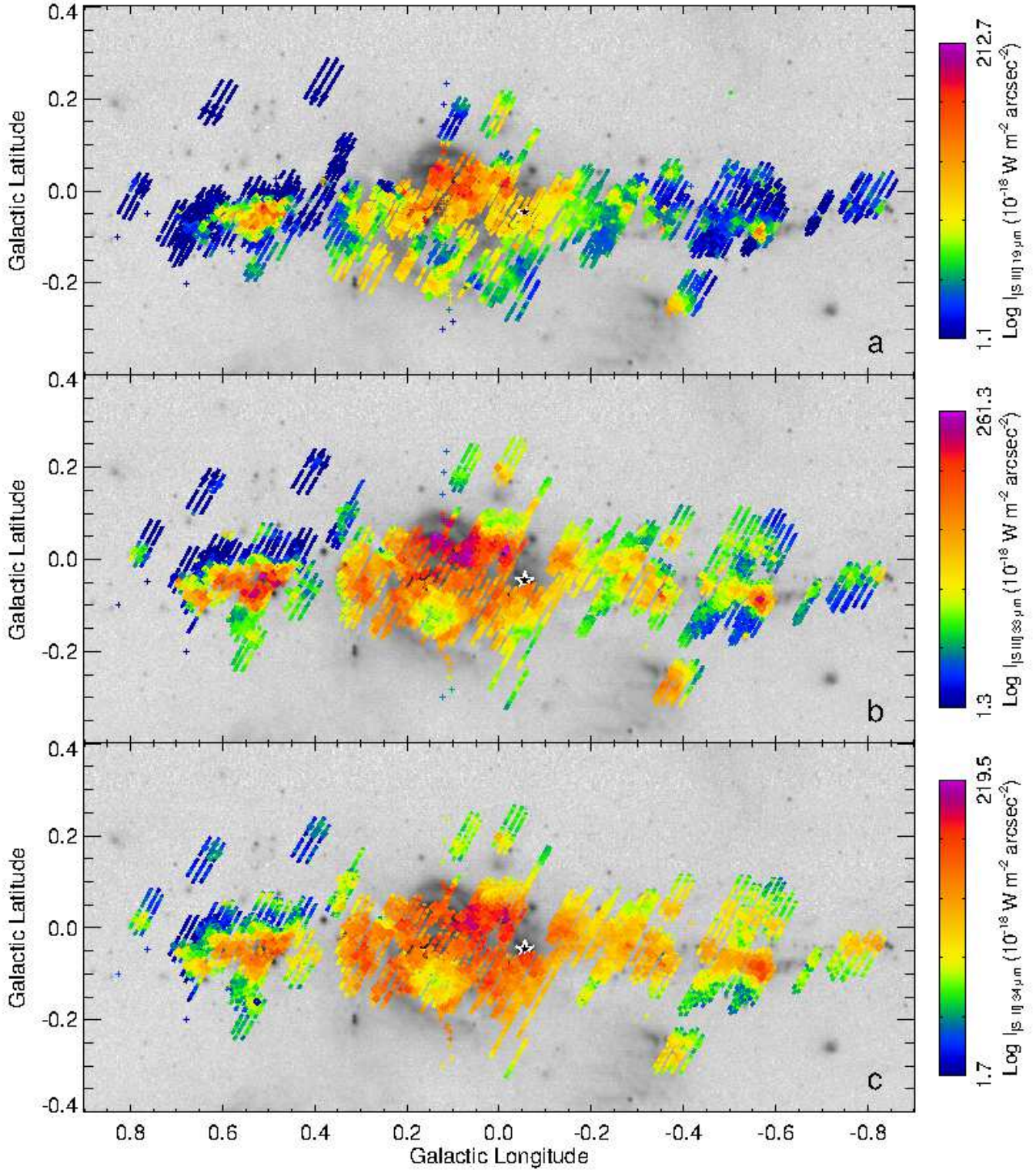}
\caption{Logarithms of the intensities of the observed lines, grayscale from {\it MSX} Band E.
The color scale is given by the bar on the right.
Crosses mark the positions of the SH or LH apertures from Cycle 1 (programs 0018 and 3295)
and diamonds mark the positions of the SH or LH apertures from Cycle 4 (program 40230).
The locations of the LL or SL apertures are drawn as extracted by CUBISM.
(a) The intensity of the [\ion{S}{3}] 19 \micron\ line.
(b) The intensity of the [\ion{S}{3}] 33 \micron\ line.
The positions that have the [\ion{S}{3}] 19 \micron\ line overlapping with the 33 \micron\ line 
show the slit locations from the observing plan AORs; 
the positions that do not overlap are the other order on the LL arrays.
(c) The intensity of the [\ion{Si}{2}] 34 \micron\ line.
%(d) The intensity of the [\ion{Ne}{2}] 12.8 \micron\ line.
%(e) The intensity of the [\ion{Ne}{3}] 15.6 \micron\ line.
%(f) The intensity of the [\ion{S}{3}] 19 \micron\ line.
%Positions colored white are where the [\ion{O}{4}] 26 \micron\ line was not detected. 
%Positions colored dark blue are where the signal/noise for the [\ion{O}{4}] 26 \micron\ line was $< 2$.
%The positions colored magenta at the top of the color scale in panels (e) and (f) may have intensities 
%much larger than $5.6 \times 10^{-18}$ W m$^{-2}$ arcsec$^{-2}$ --- 
%these are the locations of the highly excited positions in Tables 8 and 9.
}
\end{figure*}

%Figure 2 def
\setcounter{figure}{1}
\begin{figure*}
\includegraphics[width=184mm]{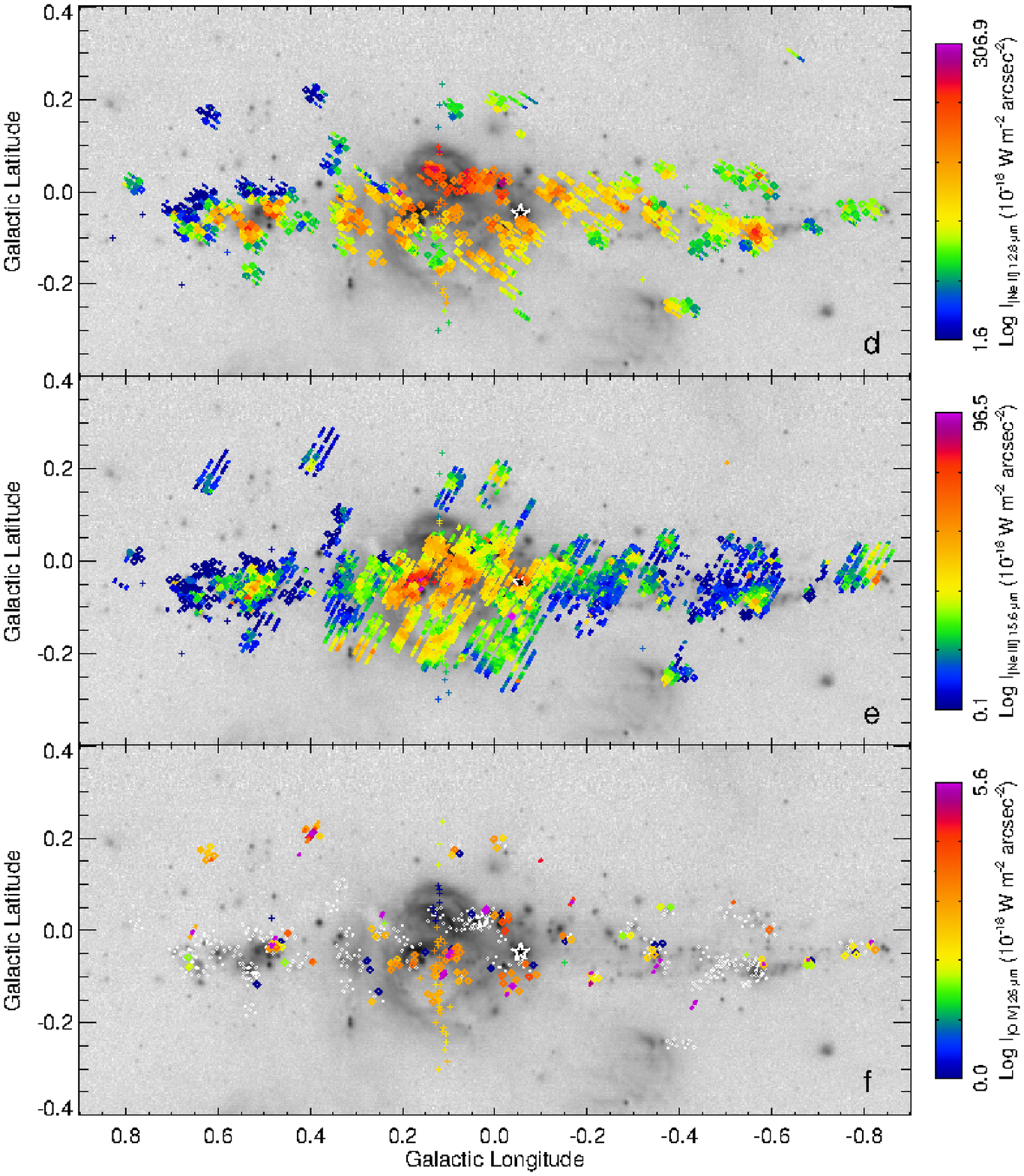}
\caption{{\it Continued.}
%Logarithms of the intensities of the observed lines.
%The color scale is given by the bar on the right.
%Crosses mark the positions of the SH or LH apertures from Cycle 1 (programs 0018 and 3295)
%and diamonds mark the positions of the SH or LH apertures from Cycle 4 (program 40230).
%The locations of the LL or SL apertures are drawn as extracted by CUBISM.
%(a) The intensity of the [\ion{S}{3}] 19 \micron\ line.
%(b) The intensity of the [\ion{S}{3}] 33 \micron\ line.
%The positions that have the [\ion{S}{3}] 19 \micron\ line overlap with the 33 \micron\ line 
%show the slit locations from the observing plan AORs; 
%the positions that do not overlap are the other order on the LL arrays.
%(c) The intensity of the [\ion{Si}{2}] 34 \micron\ line.
(d) The intensity of the [\ion{Ne}{2}] 12.8 \micron\ line.
(e) The intensity of the [\ion{Ne}{3}] 15.6 \micron\ line.
(f) The intensity of the [\ion{O}{4}] 26 \micron\ line.
Positions colored white are the locations of LH observations where the [\ion{O}{4}] 26 \micron\ line was not detected. 
Positions colored dark blue are where the signal/noise for the [\ion{O}{4}] 26 \micron\ line is $< 2$.
Note that for those positions with low S/N or no measurements, 
the $2\sigma$ uncertainties, which are approximately proportional to the continuum intensity, 
are usually similar to the fluxes of the nearby positions with good detections.
The positions colored magenta at the top of the color scale in panels (e) and (f) may have intensities 
much larger than $9.65 \times 10^{-17}$ or $5.6 \times 10^{-18}$ W m$^{-2}$ arcsec$^{-2}$, respectively --- 
these are the locations of the highly excited positions in Tables 8 and 9.
}
\end{figure*}

%Figure 2 ghi
\setcounter{figure}{1}
\begin{figure*}
\includegraphics[width=184mm]{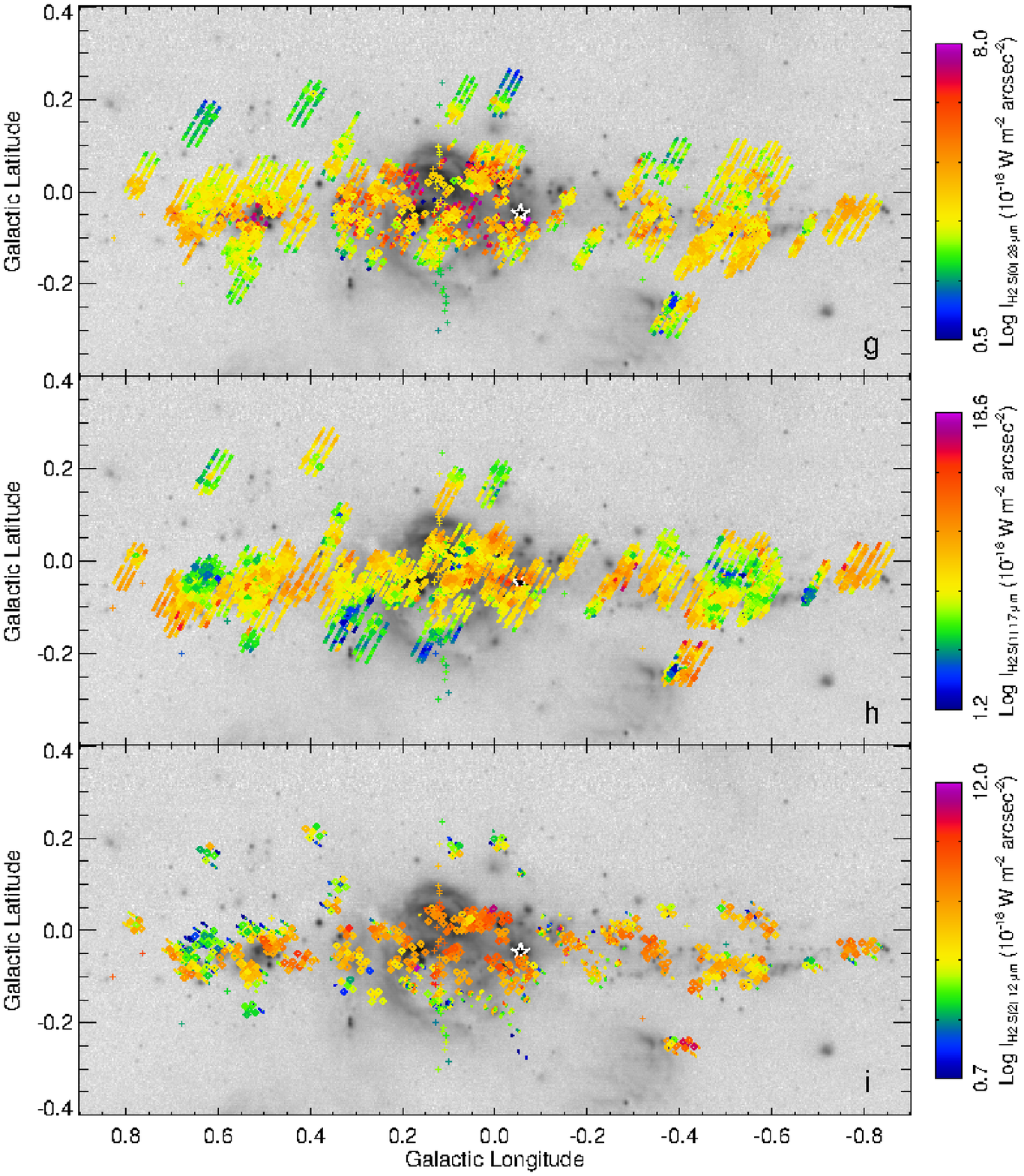}
\caption{{\it Continued.} 
%Logarithms of the intensities of the observed lines.
%The color scale is given by the bar on the right.
%Crosses mark the positions of the SH or LH apertures from Cycle 1 (programs 0018 and 3295)
%and diamonds mark the positions of the SH or LH apertures from Cycle 4 (program 40230).
%The locations of the LL or SL apertures are drawn as extracted by CUBISM.
%(a) The intensity of the [\ion{S}{3}] 19 \micron\ line.
%(b) The intensity of the [\ion{S}{3}] 33 \micron\ line.
%The positions that have the [\ion{S}{3}] 19 \micron\ line overlap with the 33 \micron\ line 
%show the slit locations from the observing plan AORs; 
%the positions that do not overlap are the other order on the LL arrays.
%(c) The intensity of the [\ion{Si}{2}] 34 \micron\ line.
(g) The intensity of the H$_2$ S(0) 28.2 \micron\ line.
(h) The intensity of the H$_2$ S(1) 17.0 \micron\ line.
(i) The intensity of the H$_2$ S(2) 12.28 \micron\ line.
}
\end{figure*}

The observed line intensities are given in Tables 2 -- 7 
for modules SL2 (second order), SL1 (first order), SH, LL2 (second order), LL1 (first order), and LH, 
respectively, and are plotted in Figure 2.
These tables contain the intensities of the common \ion{H}{2} region lines plus [\ion{O}{4}] 26 \micron, 
which is detected widely in the GC although it is not an \ion{H}{2} region line. 
The apertures used in the extraction of the SL or LL spectra are given in the tables.
{ 
The apertures for the CUBISM-extracted spectra for SH and LH are $\sim 51$ 
and $\sim 119$ arcsec$^2$, respectively, 
and the apertures for the programs 0018 and 3295 spectra that were extracted with SMART
by Simpson et al. (2007)  
are assumed to be the IRS's nominal 53.11 and 247.5 arcsec$^2$
for SH and LH, respectively (Houck et al. 2004).
The latter SH and LH line intensities were also corrected for slitloss (telescope and instrument diffraction), 
necessary because SMART assumed the spectra being extracted are from point sources and the GC emission
is extended (the default for CUBISM is to assume extended emission).
}
The units of the intensities from CUBISM (MJy sr$^{-1}$)
were adjusted to be W m$^{-2}$ s$^{-1}$ arcsec$^{-2}$ from W m$^{-2}$ s$^{-1}$ sr$^{-1}$ for the figures
and W m$^{-2}$ s$^{-1}$ sr$^{-1}$ for the online machine readable tables.
So that the final line intensities could be compared, the SH and LH fluxes were also converted 
to W m$^{-2}$ s$^{-1}$ arcsec$^{-2}$ or W m$^{-2}$ s$^{-1}$ sr$^{-1}$.

In addition to the relatively smooth emission from the inner Galaxy, 
there are intriguing serendipitous discoveries in this archived GC data set 
of locations that are so highly excited that the [\ion{Ne}{5}] lines ($IP = 97 - 126$ eV, Table~1) are strong; 
these locations are listed in Table~8. 
These locations also have very strong [\ion{Ne}{3}] 15.6 \micron\ and [\ion{O}{4}] 26 \micron\ if 
the relevant module was observed (in many cases, the module order with the [\ion{Ne}{5}] line 
was the `other' module order from the LL order in the planned AOR and so the observed spectrum 
contains only one order of the nominal LL wavelength range).
These highly excited lines require substantial amounts of ionizing photons with much higher energy  
than are found in OB stars;
such photons are often emitted by the white dwarf stars that ionize 
planetary nebulae or are emitted by active galactic nuclei (e.g., Feuchtgruber et al. 1997),
or are found in the high-energy shocks of supernova remnants (e.g., Sankrit et al. 2014). 
These positions will be discussed further in future papers. 
Probably in the same classes of exciting sources are found an additional number of positions 
that have very strong [\ion{O}{4}] 26 \micron\ but no detected [\ion{Ne}{5}] at either wavelength.
These positions are listed in Table~9.
The references in Tables 8 and 9 are for the radio or Paschen~$\alpha$ sources 
found at approximately the same locations on the sky. 

A few positions observed with the LH module have detected [\ion{S}{1}] 25.25 \micron.
Because sulfur in the interstellar medium (ISM) is either singly ionized (the $IP$ for S$^+$ is 10.4 eV) 
or found in molecules such as SO, 
it is generally thought that this line from neutral sulfur is shock-excited (e.g., Hollenbach \& McKee 1989). 
This line is seen in two orders in the LH module and must be detected in both orders 
to be considered an acceptable measurement. 
These positions and line intensities are given in Table~10. 

Finally, a few positions seen in order 2 of SL have deep ice absorption features at 6.0 and 6.8 \micron\ 
(e.g., Boogert et al. 2015).
 These candidate YSOs are in addition to those described by Immer et al. (2012) and An et al. (2011) 
and are listed in Table 11.

It is seen that the \ion{H}{2} regions marked in Figure 1 
are all detected in the ionized lines of Figure 2, 
although the intensities of the lines in Sgr B2 are substantially lower 
than the line intensities of the other \ion{H}{2} regions. 
This is due to the large extinction towards Sgr B2, which has been well known for a long time.
The other \ion{H}{2} regions appear prominently, particularly the Arched Filaments, 
the small \ion{H}{2} regions at the base of the Arched Filaments (Zhao et al. 1993; Cotera et al. 2000;
{ Dong et al. 2017}), 
the `Sickle' next to the Quintuplet Cluster, the diffuse gas of Sgr A East and
the Bubble Rim, Sgr B2, and Sgr C (see the 20 cm radio survey of Yusef-Zadeh et al. 2004 
for images taken at similar resolutions to the {\it Spitzer} spectra).
Note that the [\ion{Si}{2}] 34 \micron\ line displays a pattern much more similar to that of the \ion{H}{2} region lines,
such as [\ion{S}{3}] 33 \micron, 
than it does to the spatial appearance of the cold molecular clouds that appear in red in Figure~1. 
Although the [\ion{Si}{2}] 34 \micron\ line is often treated as a PDR line (e.g., Kaufman et al. 2006), 
in the GC it is at least as much an \ion{H}{2} region line, and so will be treated as such in the following sections.

{ 
Regarding the high-excitation lines, 
although there appears to be a good correlation of the emission line intensities (Figure~2) 
with star-forming regions (partly due to the selection bias of the {\it Spitzer} observing programs), 
there does not appear to be any correlation of the most highly-excited gas 
with any of the numerous non-thermal filaments detected by Yusef-Zadeh et al. (2004) 
nor with the 6.4 keV X-ray line produced by low-ionization iron.
The latter has been reviewed by Koyama (2018), who concluded that this time-variable line is 
consistent with fluorescence in molecular-cloud iron from an input X-ray flare from Sgr A*.
This line has also been attributed to cosmic-ray ionization by Yusef-Zadeh et al. (2007)
because of its correlation with non-thermal filaments and molecular clouds.
Relativistic cosmic-ray electrons, however, although the source of the non-thermal emission, 
would produce an ionization equilibrium more radical than what is seen here in the GC, 
with the higher ionization-potential [\ion{Ne}{5}] lines stronger than the [\ion{O}{4}] line,
assuming that there is no contribution to either line 
from the ordinary low-excitation \ion{H}{2} regions of the GC.
In fact, the exceptionally highly-excited lines (Tables 8 and 9) occur in decidedly thermal regions,
because they all have counterparts as either thermal radio sources or even Paschen alpha sources 
(references in Tables 8 and 9) where such observations with adequate sensitivity have been performed.
I suggest these are either shocks, or for the compact or symmetric sources, candidate planetary nebulae
(D. An et al., in preparation).
}

\input tab2.tex

\input tab3.tex

\input tab4.tex

\input tab5.tex

\input tab6.tex

\input tab7.tex

\input tab8.tex

\clearpage

\input tab9.tex

\input tab10.tex

\input tab11.tex

\subsection{Line Ratios}

%Figure 3
\begin{figure*}
\includegraphics[width=184mm]{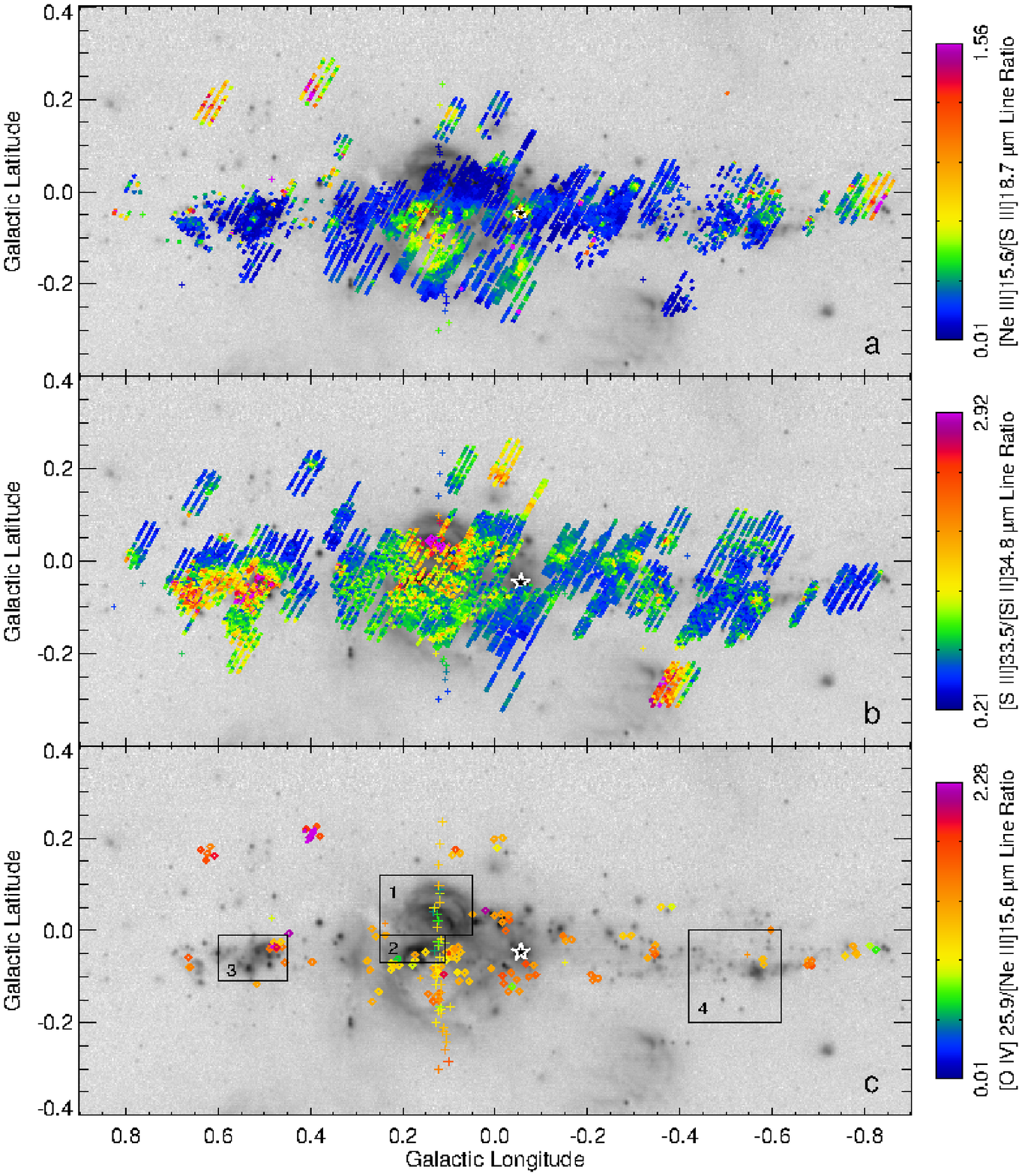}
\caption{Ratios of some of the observed lines, grayscale from {\it MSX} Band E.
The color scale is given by the bar on the right.
Crosses mark the positions of the SH or LH apertures from Cycle 1 (programs 0018 and 3295)
and diamonds mark the positions of the SH or LH apertures from Cycle 4 (program 40230).
The locations of the LL apertures are drawn as extracted by CUBISM.
(a) The ratio of the [\ion{Ne}{3}] 15.6 \micron\ line divided by the [\ion{S}{3}] 18.7 \micron\ line.
(b) The ratio of the [\ion{S}{3}] 33 \micron\ line divided by the [\ion{Si}{2}] 34 \micron\ line.
(c) The ratio of the [\ion{O}{4}] 26 \micron\ line divided by the [\ion{Ne}{3}] 15.6 \micron\ line.
{ The boxes marked `1', `2', `3', and `4' circumscribe the regions known as the GC Filaments, 
the Quintuplet Region, Sgr B1, and Sgr C that are discussed in Sections 3.4 and 4.} 
}
\end{figure*}

Some of the interesting line ratios are plotted in Figure 3.
Although all three line ratios are indicators of excitation (see the required $IP$ in Table~1), 
it is clear that different parts of the GC are affected differently 
by those processes that affect the excitation: the exciting source SED 
and the dilution of the radiation field. 

In Figure 3a the [\ion{Ne}{3}] 15.6/[\ion{S}{3}] 18.7 \micron\ line ratio indicates 
the relative numbers of high energy photons ($> 41$~eV) needed for doubly ionized neon, 
and hence the effective temperatures ($T_{\rm eff}$) of the stars whose SEDs ionize the \ion{H}{2} regions.

In Figure 3b the [\ion{S}{3}] 33/[\ion{Si}{2}] 34 \micron\ line ratio indicates 
the dilution of the radiation field, since the singly and doubly ionized excitation structures 
of both silicon and sulfur are quite similar (see Section 4): 
both elements are at least singly ionized in the non-molecular interstellar gas 
but the $> 16$ or $> 23$ eV photons of a nearby hot star are required to photoionize 
the silicon or sulfur, respectively, in its local \ion{H}{2} region. 
Thus positions colored red or orange in Figure~3b have their exciting stars relatively close by,
whereas positions colored blue or green have very dilute radiation fields, 
indicating that their exciting stars are at some distance. 

Finally, in Figure 3c the [\ion{O}{4}] 26/[\ion{Ne}{3}] 15.6 \micron\ line ratio indicates 
the presence of photons with energies $> 54.9$~eV, as are necessary to produce O$^{3+}$.
In Figure~3c the isolated measurements with exceptionally large ratios, colored red, 
are the candidate planetary nebulae or shocks of Tables 8 and 9; 
however, the majority of observed [\ion{O}{4}] 26 \micron\ lines, colored green through orange, 
are scattered along the Galactic plane with sources of higher ratios at all Galactic longitudes 
(note that the colors represent the logs of the line ratios for panel Figure~3c, unlike 
panels 3a and 3b, where the line ratios have a linear scale).

\subsection{Extinction}

%Figure 4
\begin{figure*}
\includegraphics[width=184mm]{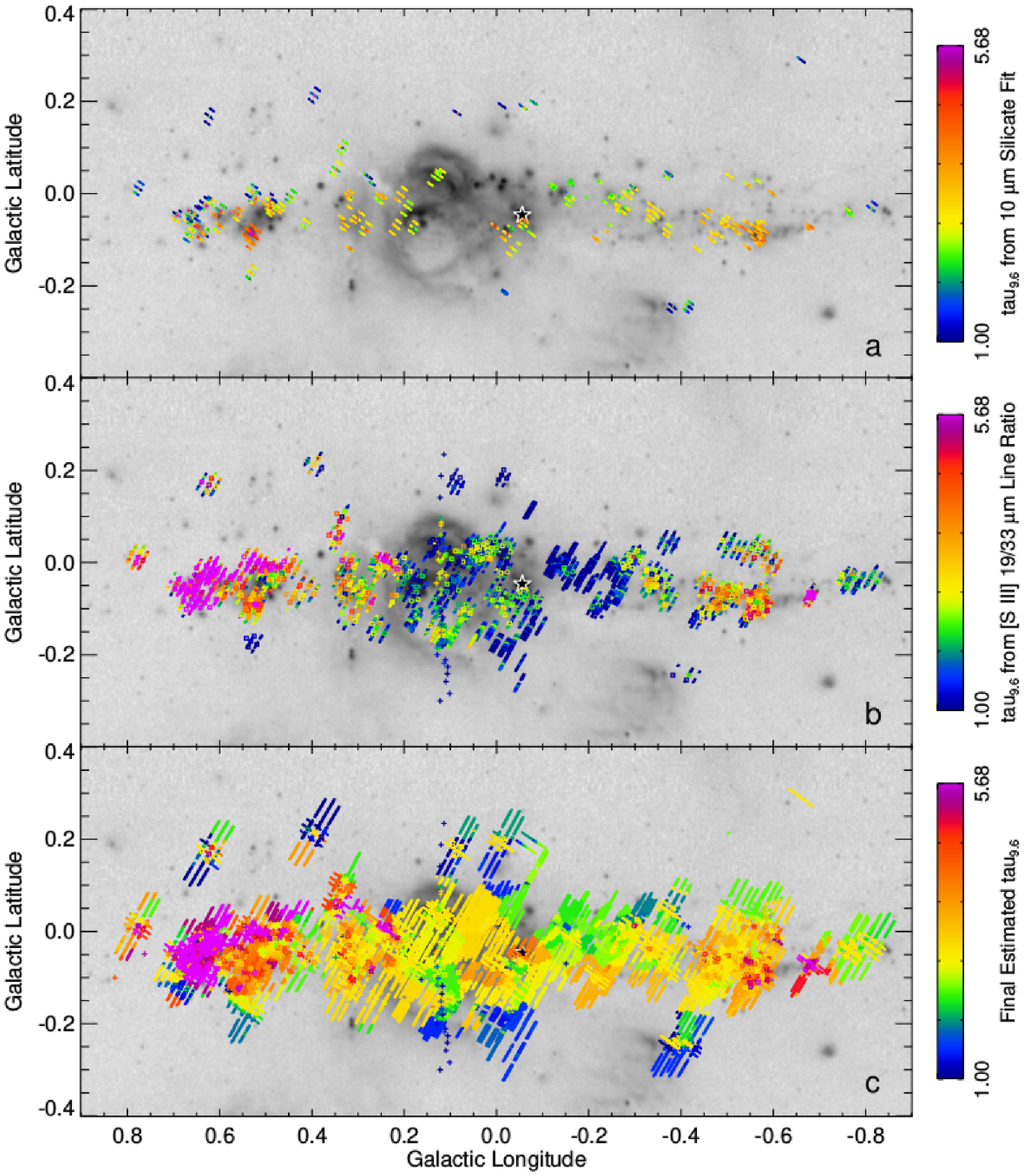}
\caption{Optical depths at 9.6 \micron\ computed by two different methods, grayscale from {\it MSX} Band E. 
The color scale is given by the bar on the right.
Crosses mark the positions of the SH or LH apertures from Cycle 1 (programs 0018 and 3295),
{ squares mark the positions of the SL or LH apertures from Cycle 4 (program 40230), and } 
the locations of the SL or LL apertures are drawn as extracted by CUBISM.
(a) The optical depth at 9.6 \micron\ estimated by fitting the SL spectrum with PAHFIT, see text.
(b) Lower limits to the 9.6 \micron\ extinction computed from the ratios 
of the [\ion{S}{3}] 19 \micron\ lines divided by the [\ion{S}{3}] 33 \micron\ lines.
The plotted extinction is that required to make the ratio equal to at least 0.508, 
the minimum value for $T_e = 6000$~K.
(c) Final estimated extinction from the combination of the two methods, see text.
The values of $\tau_{9.6 \micron}$ in this plot are found in Tables 2 -- 7.
}
\end{figure*}

Although the extinction to the GC has been studied extensively at near-infrared (NIR) wavelengths 
(e.g., { Nishiyama et al. 2008}; Schultheis et al. 2009),
the results are always referenced to visible or K-band extinction, $A_V$ or $A_K$, respectively.
{ For longer wavelengths, Lutz et al. (1996), Nishiyama et al. (2009), and Fritz et al. (2011) 
showed that the steep extinction law seen in the NIR flattens substantially between 3 and 8 \micron\ 
using either hydrogen recombination line ratios measured with ISO or Spitzer IRAC photometry 
(see also Indebetouw et al. 2005 for other Galactic plane sources).} 
For { the longer} MIR wavelengths where almost all the measured lines have wavelengths longer than 10 \micron, 
the GC extinction should be referenced to the optical depth, $\tau_{9.6 \micron}$, 
of the deepest part of the 10 \micron\ silicate feature, 
since it has long been known that the ratio of $\tau_{9.6 \micron}$/$A_V$ 
is a function of position in the Galaxy, and in particular, it is quite different
in the GC itself (Roche \& Aitken 1985; An et al. 2013).

%Figure 5
\begin{figure}
\includegraphics[width=60mm,angle=90,origin=c]{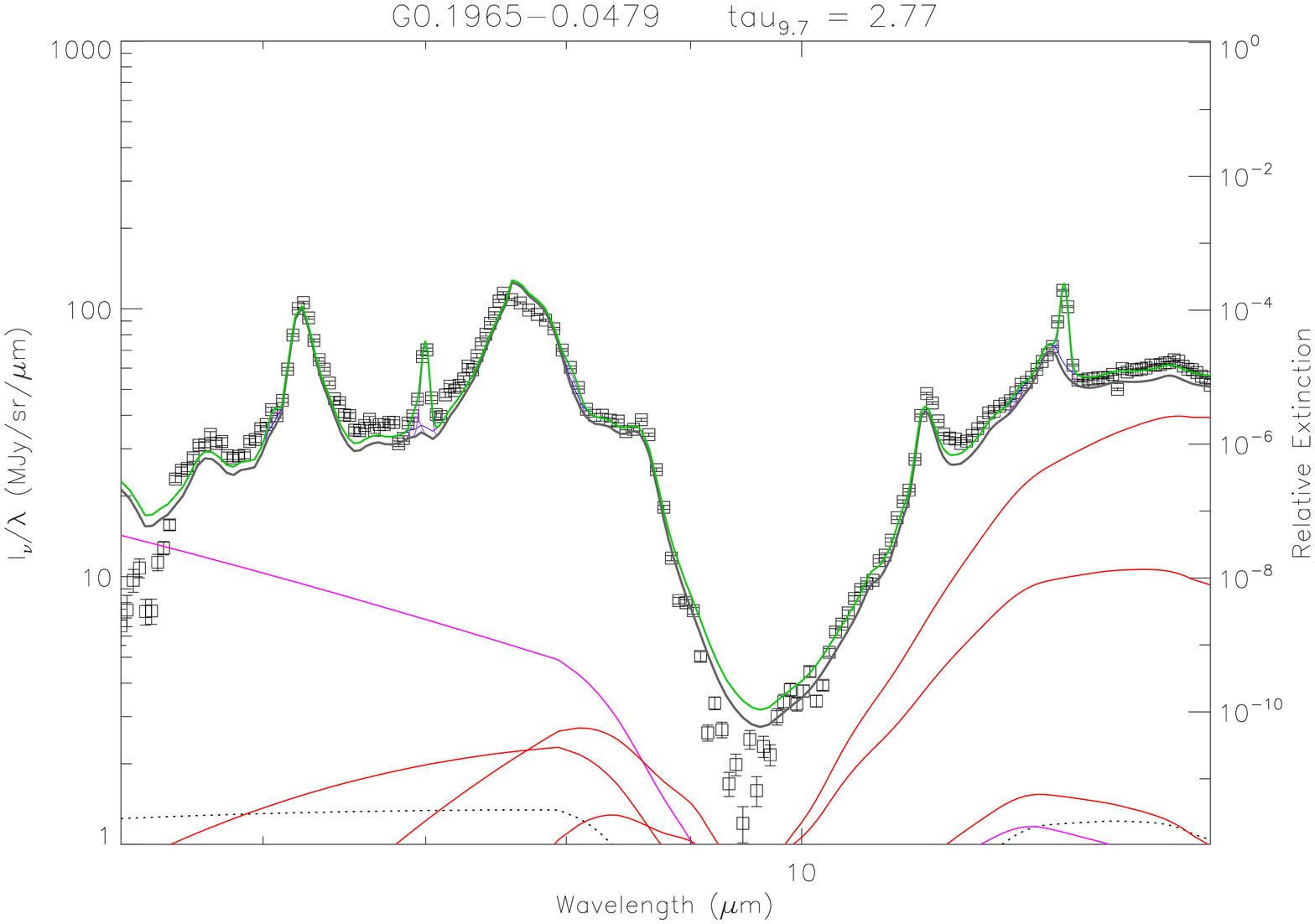} % Figure is landscape with bad bounding box, not eps
\caption{Example of a spectrum where the optical depth at 9.6 \micron\ is estimated 
by fitting the SL spectrum with PAHFIT (Smith et al. 2007b) as modified by Simpson et al. (2012).
The fitted function consists of a template (Simpson et al. 2012) 
for the 6.2, 7.7, 8.6, 11.2, and 12.5 \micron\ PAH features (gray), 
various blackbodies representing the stellar (magenta) and large grain contributions (red), 
additional PAH features (blue), 
atomic and molecular (H$_2$) emission lines (purple), all multiplied by 
the GC extinction law of Chiar \& Tielens (2006). 
The sum of all components is plotted in green; this often obscures the strongest features, 
particularly the purple [\ion{Ar}{2}] 6.98 \micron\ and [\ion{Ne}{2}] 12.8 \micron\ lines. 
The spectrum extracted with CUBISM consists of the black boxes with error bars.
}
\end{figure}

The extinction, as described by $\tau_{9.6 \micron}$, can be computed in two independent ways:

1) The 9.6 \micron\ silicate feature can be modeled and the low-resolution spectra can be fit with 
a combination of this deep absorption feature, plus various black bodies to represent 
the warm dust in the line of sight and the Rayleigh-Jeans tail of the stellar emission, 
and individual broad emission features representing the PAH features that are ubiquitous in the ISM.
A program to fit this combination of features, PAHFIT, was written by Smith et al. (2007b) 
and modified by Simpson et al. (2012) to use an unextincted template for the PAH features 
instead of the multitudinous individual features.
Because there are fewer degrees of freedom, this produces more reliable results for 
the sole absorption feature, the 9.6 \micron\ silicate feature.
Figure~4a shows the values of $\tau_{9.6}$ for the short-low spectra, which cover the wavelength range 
5.2 -- 14.4 \micron. 
The GC extinction law of Chiar \& Tielens (2006) was used.
Figure 5 shows an example of a high signal/noise (S/N) fit with PAHFIT from the Sickle region.
PAHFIT, as described { by} Smith et al. (2007b), is supposed to fit 
the full 5 -- 37 \micron\ low-resolution spectrum; 
however, there are very few positions that have both SL and LL spectra, and these are 
mostly YSOs, which have additional dust extinction from their dusty envelopes and so 
are not representative of the interstellar extinction towards the GC.
Tests were made fitting a few of those positions with both the combination SL--LL wavelength range 
and the SL range as shown here --- the results are not significantly different. 

2) Lower limits on $\tau_{9.6}$ can be computed from the ratio of the [\ion{S}{3}] 18.7/33.5 \micron\ lines 
since the extinction at wavelengths longer than 10 \micron\ is due to the same silicate dust 
grains (Chiar \& Tielens 2006 and references therein). 
The ratio at the lowest density is determined from the atomic data 
and the gas electron temperature, $T_e$ (e.g., Simpson et al. 2007).
At higher densities, the ratio is higher because of collisional de-excitation.
Figure~4b shows the minimum value of $\tau_{9.6}$ needed to make the [\ion{S}{3}] 18.7/33.5 \micron\ line ratio
equal to the minimum ratio, 0.508, predicted for $T_e = 6000$ K 
(effective collision strengths from Grieve et al. 2014, but see the comments in Rubin et al. 2016).

Note that since the deepest part of the 9.6 \micron\ silicate absorption feature is close to zero, 
the contribution from the foreground Zodiacal emission is not negligible.
This was estimated from the four spectra with the smallest integrated flux -- all of these are 
from Program 3121 (PI: K. Kraemer) on the spectra of infrared dark clouds.
These sources are all much fainter than any of the rest of our spectra.
Estimates of the intensity of the Zodiacal emission plus telescope at 9.6 \micron\ computed 
by the {\it Spitzer} software SPOT program 
were not used because these estimates are 
higher than the observed minima in these GC spectra.
The spectra from program 3121 also show low luminosity PAH emission in addition to flux at 9.6 \micron.
This PAH emission, along with some of the continuum, is probably foreground emission 
along the line of sight through the Galactic plane to the GC.
The estimated foreground emission, Zodiacal plus line of sight, was subtracted from 
the SL spectra when fitting the spectra with PAHFIT.

The final estimated extinction is a combination of both methods \#1 and \#2. 
Since the optical depths estimated by method \#2 are only lower limits, 
I increased these values to those estimated by method \#1 for nearby positions.
This was done by making a 6 arcsec grid covering the entire GC and 
then interpolating or extrapolating the observed values of $\tau_{9.6}$ for the grid pixels 
using the Interactive Data Language (IDL) function, GRIDDATA, and the method of `Nearest Neighbor'.
Two grids were made, for method \#1 and for method \#2.
The final extinction grid is the maximum of the method \#2 grid and the method \#1 grid.
Clearly the validity of the results depends on the closeness of each pixel to a location 
where there were either LL observations of both [\ion{s}{3}] lines or PAHFIT computations,
and thus I plot only the regions with observations in Figure~4c.

It is interesting to note that the extinction is fairly uniform across the GC with a value of
$\tau_{9.6} \sim 3$, with the regions with exceptionally higher extinction also being the regions 
with known dense molecular clouds: the `Brick' at G0.26+0.0, Sgr B2 at G0.7$-$0.0, and
the Galactic Center dust ridge described by Immer et al. (2012).
The low extinction regions are mostly at the higher Galactic latitudes, 
where the foreground gas at moderate distances over the Galactic plane 
contributes much more to the line of sight than the gas at the larger distances above 
the Galactic plane of the GC.

The values of the estimated extinction are given in Tables 2 -- 7. 
It should be noted here that the extinction values for those positions in 
the extreme excitation sources of Tables 8 -- 10 may be erroneous --- 
these sources are typically very compact and are not detectable in continuum images.
Thus extinction values estimated from the continuum (method \#1) 
or extinction values estimated from that of their local neighboring positions (method \#2) 
may not be applicable. 
At least some sources may be foreground and actually have much lower optical depths. 
Such sources may be indicated by strong observed [\ion{S}{4}] 10.5 \micron\ line intensities, 
which lines are normally extremely weak or not detectable in regions 
with deep 10 \micron\ silicate absorption.

{ An et al. (2013) also made extinction estimates from the SH spectra of program 40230
by measuring the depths of the 9.6 \micron\ silicate feature from the ratios of the observed intensities 
at approximately 10.24 and 13.9 \micron. 
 These estimates are shown in their figure 4, top.
The estimates of the depth of the 9.6 \micron\ silicate feature in this paper 
for those positions on the sky that have both usable SL spectra and SH spectra 
are in most cases somewhat smaller (the median of the ratio equals 66\%) 
except in the regions of high extinction in Sgr B2, where the extinction estimated in this paper 
is somewhat higher.
At least part of the difference is probably due to the fact that neither wavelength region 
used by An et al. (2013) is free of PAH emission, with the 13.9 \micron\ wavelength region 
in the PAH template (figure 4 of Simpson et al. 2012) that is used in the PAHFIT computation 
having a larger PAH contribution
than the 10.24 \micron\ wavelength region.
Thus after subtraction of the PAH emission, the 9.6 \micron\ silicate feature is not as deep, 
producing a smaller computation of the optical depth by PAHFIT.}

\subsection{Electron Densities and Elemental Abundances}

\input tab12.tex

%Figure 6
\begin{figure*}
\includegraphics[width=184mm]{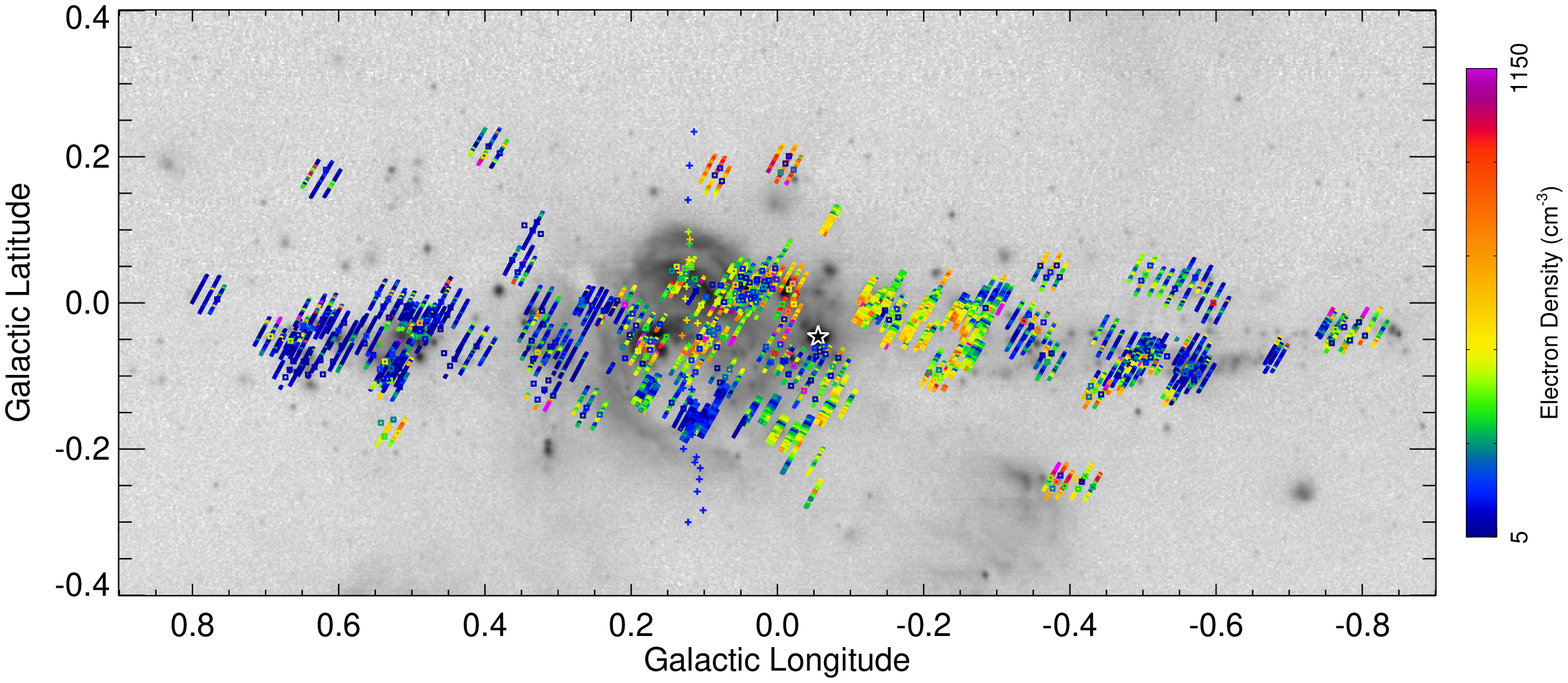}
\caption{Electron densities estimated from the [\ion{S}{3}] 18.7/[\ion{S}{3}] 33 \micron\ line ratios, grayscale from {\it MSX} Band E. 
The color scale is given by the bar on the right.
Crosses mark the positions of the SH or LH apertures from Cycle 1 (programs 3295); 
Square boxes mark the positions of SH/LH apertures from programs 3189 and 40230,
{ and the LL apertures are drawn as extracted with CUBISM}.
}
\end{figure*}

Electron densities, $N_e$, were estimated from the extinction-corrected and co-located 
[\ion{S}{3}] 19/33 \micron\ line intensity ratios 
using the effective collision strengths referenced in Table~1. 
Estimated foreground intensities were first subtracted to compensate for the integrated intensities 
along the lines of sight to the GC; 
this foreground was estimated from the minimum observed intensities 
(3.5 for [\ion{Ne}{2}] 12.8 \micron, 0.2 for [\ion{Ne}{3}] 15.6 \micron, 
0.6 and 3.0 for [\ion{S}{3}] 18.7 \micron, 1.5 and 10.0 for [\ion{S}{3}] 33 \micron,
and 8.0 and 19.0 for [\ion{Si}{2}] 34 \micron, where all estimates 
should be multiplied by $10^{-19}$ W m$^{-2}$ arcsec$^{-2}$ and 
pairs of numbers are for positive and negative Galactic longitudes, respectively).  
The densities estimated from the { observed line ratios}
are plotted in Figure 6, 
along with the densities from Simpson et al. (2007). 
The low resolution lines are preferred for the density computation because 
there are far more observed positions per \ion{H}{2} region 
%(138 to 636 positions per \ion{H}{2} region) 
and because there is no need for correction for differences in aperture size 
for lines at 19 versus 33 \micron. 

Elemental abundances with respect to hydrogen were estimated from the high-resolution line measurements. 
In order to produce sufficiently small errors in the ratios, 
only measurements of the hydrogen line intensities with S/N$ > 4$ were used. 
The measured densities and abundance ratios were then averaged over regions 
that I call the GC Filaments, the Quintuplet Cluster region, Sgr B1, and Sgr C; 
each region should be localized in space and ionized by the stars local to that region. 
{ Outlines of these regions are plotted in Figure~3c.}
A separate average for the Arched Filaments observed by Simpson et al. (2007) 
was also computed  
since the references for the effective collision strengths in Table~1 
have changed substantially since that paper.  
These average densities and ionic abundance ratios are listed in Table~12.
Note that there is very little overlap between the positions measured by Simpson et al. (2007) 
and the positions in the GC Filaments seen in Figure~2f, which are heavily weighted to 
the small \ion{H}{2} regions at the base of the Arched Filaments.

Approximately half of the SL positions with observed [\ion{Ne}{2}] have detectable [\ion{Ar}{2}] 6.98 \micron\ lines.
For these positions, the average [\ion{Ar}{2}] 6.98/[\ion{Ne}{2}] 12.8 \micron\ line ratio equals 0.515.
The positions in the Radio Arc Bubble are excluded from this average 
because they have substantial Ne$^{++}$ (Simpson et al. 2007) 
and probably also substantial Ar$^{++}$, 
making this ratio too dissimilar to that of the rest of the GC.
For $T_e = 6000$~K, this intensity ratio corresponds to an Ar$^{+}$/Ne$^+$ abundance ratio of { 0.032}.

The measured abundances are all higher than those of the Orion Nebula 
[(Ne$^+$ + Ne$^{++}$)/H$^+ = 1.02 \pm 0.03 \times 10^{-4}$ and 
(S$^{++}$ + S$^{3+}$)/H$^+ = 9.6 \pm 0.3 \times { 10^{-6}}$, Rubin et al. 2016]; 
this is expected considering the known abundance gradients in the Galaxy 
(e.g., Mart\'in-Hern\'andez et al. 2002; Rudolph et al. 2006).  
Since it is essential, in a comparison of abundances, that all abundances 
are computed using the same atomic physics (e.g., Table~1), 
it is beyond the scope of this paper to recompute the Galactic abundance gradients 
using these new results. 

\section{Discussion}

\subsection{Comparison with Models}

By comparing the line flux ratios to ratios produced by \ion{H}{2} region models,
one can gain insight into the gas chemical composition 
and the nature of the sources ionizing the gas. 
Such models need to cover a range of gas densities and geometries, 
gas compositions, ionizing source SEDs, etc.
At one extreme, one can employ the million model grid of 
Morisset, Delgado-Inglada, \& Flores-Fajardo (2015) with its access via a data-base query language, 
and at the other extreme, one can compute detailed models specific to each source,
such as the models of the Galactic \ion{H}{2} regions W43 and G333.6$-$0.2 of Simpson et al. (2004).
In general, large grids are useful to get a first guess of the range of model parameters, 
which are then refined with detailed modeling. 
Because of the extreme nature of the GC and because it has already been seen that 
there are difficulties in reproducing the observed combination 
of both low and high excitation gas (e.g., Simpson et al. 2007), 
the pre-computed models from the extant databases are not adequate to represent 
the observations in the previous section.
I here expand the model parameter space from that of Simpson et al. (2007) 
to demonstrate that X-rays in addition to the SEDs of OB stars 
are required to fit the observed line intensities.

I have modeled the ionization structure of the GC gas 
with a relatively coarse grid of model \ion{H}{2} regions 
computed with the code { Cloudy 17.00 (Ferland et al. 2017)}.
Because of the low gas density in the GC, as seen in the previous section, 
the models all have low hydrogen density ($N_{\rm H} = N_p = 100$~cm$^{-3}$)
but high photon input, to correspond to the massive clusters ionizing the gas.
The parameters that are varied to make a grid of models 
consist of the shapes of the ionizing SEDs, which determine the amounts of 
the more highly excited states of oxygen and neon, 
and the dilution of the radiation field that so strongly affects 
the low-ionization states of silicon and sulfur.

The radiation field intensity is usually described by 
the ionization parameter, $U$, which is defined as the ratio of 
the photon density divided by $N_e$ at the outer edge 
of the fully ionized gas sphere of radius, $R_S$ (e.g., Mathis 2000).
The low values of $U$ needed to produce the low observed [\ion{S}{3}] 33/[\ion{Si}{2}] 34 \micron\ ratio (Figure~3b)
are found only for \ion{H}{2} regions at extremely large distances from their exciting stars 
(thus producing very dilute radiation fields) or 
for \ion{H}{2} regions with very low average densities, much lower than the densities 
estimated from the [\ion{S}{3}] line ratios. 
Here I define the filling factor $f$ as the fraction of the volume with clumps 
of proton density $N_p$ such that the average proton density equals $fN_p$.
With this definition, $R_S$ and $U$ are given by 
\begin{equation}
R_S = {\left( {\frac{3 N_{\rm LyC}}{4 \pi f N_{\rm H}^2 (N_e/N_p) \alpha_{\rm B} F_{\rm He}}} \right)}^{\frac{1}{3}}
\end{equation}
and 
\begin{equation}
U = \frac{N_{\rm LyC}}{4 \pi R_S^2 c} \frac{1}{N_{\rm H}(N_e/N_p)} \approx 
\left[ \frac{N_{\rm LyC}}{36 \pi c^3} f^2 \alpha_{\rm B}^2 F_{\rm He}^2 \frac{N_p}{N_e}N_{\rm H} \right]^{\frac{1}{3}}, 
\end{equation}
where 
$F_{\rm He} = 1/(1+f_i <{\rm He}^+/({\rm H}^+ + {\rm He}^+>)$ (Rubin 1968 as written by Simpson \& Rubin 1990),
$f_i$ is the fraction of helium recombination photons to excited states that are energetic enough
to ionize hydrogen ($f_i \sim 0.65$),
$N_{\rm LyC}$ is the number of photons emitted per second in the Lyman continuum 
that can ionize hydrogen, 
$N_e$ is the electron density, $c$ is the speed of light, 
 and $\alpha_{\rm B}$ is the total recombination rate to the second level of hydrogen.

For this grid I use SEDs from the compilation of SEDs with Solar abundances from Starburst99, 
most recently described by Leitherer et al. (2014), with individual 
O-star atmosphere models from Leitherer et al. (2010).
{ These input SEDs were computed using the Starburst99 default parameters for an instantaneous burst of star formation.}
In Starburst99, an initial mass function for a massive stellar cluster is assumed, 
the cluster stars of varying masses are evolved following the evolutionary tracks for each star, 
and a composite SED for the whole cluster is calculated using the stellar types that the stars have for the given cluster age.
In Cloudy, the SED is input by calling for it by atmospheric model type, and 
for Starburst99, by model age, expressed as the logarithm of the age. 
The ages used for this grid are $10^{6.0}$, $10^{6.2}$, $10^{6.3}$, $10^{6.4}$, $10^{6.5}$, $10^{6.6}$, $10^{6.65}$, 
and $10^{6.7}$ years; these ages produced line intensity ratios 
that bracket the observed ratios. 

The 2014 version of Starburst99 used the `Geneva' evolutionary tracks of Ekstr\"om et al. (2012)  
for stellar models with both zero rotation and rotation with velocities of 40\%\ 
of the break-up velocity.
The effect of the higher rotational velocity is to produce SEDs that are bluer and more luminous 
due to more mixing in the stellar core. 
The result is that the cluster SED for a given age has more high energy photons 
than the SED for zero rotation, or conversely, if two SEDs must be essentially identical 
to produce an \ion{H}{2} region model that best fits the observations, 
the age of the Starburst99 SED with rotation is significantly older than the age 
of the Starburst99 model without rotation (Leitherer et al. 2014). 
I will show that such models with rotation require ages that are almost certainly too old 
for fitting the spectra of the diffuse gas of the GC.

In Simpson et al. (2007) we attempted to model the gas of the Radio Arc Bubble 
with \ion{H}{2} region SEDs consisting of multiple component blackbodies, with the higher temperature 
blackbodies having temperatures of either $10^5$ or $10^6$~K.
None of these models was satisfactory. 
In retrospect, as a result of computing the models in this paper, I conclude that the reason 
is surely the use of blackbodies to represent the stellar spectra --- 
such SEDs all have too much flux in the 40 -- 54 eV range compared to the 54 -- 77 eV range 
and so produce too much Ne$^{++}$ compared to O$^{+3}$. 

The models here use stellar SEDs from Starburst99 plus blackbodies with temperatures 
of either $10^6$ or $10^{6.5}$~K and blackbody luminosities ranging from $10^{37}$ to $10^{39}$ erg s$^{-1}$
in increments of $10^{0.5}$;
these models adequately cover the range of line ratios observed by Simpson et al. (2007). 
The SEDs used in this paper are shown in Figure 7.
As mentioned above, the SEDs for the Starburst99 model with age $= 10^{6.7}$~yrs and 
rotation 40\% of breakup have much more flux than SEDs of the same age but zero rotation. 
%Fig. 12 from Simpson et al. ??
Blackbodies were chosen only as a way of adding a smooth, well-defined component to the SED 
between 54 and $\sim 100$ eV; 
other spectral shapes may be more appropriate, particularly a non-thermal shape 
if the actual X-ray emission is optically thin.

%Figure 7
\begin{figure*}
\includegraphics[width=165mm]{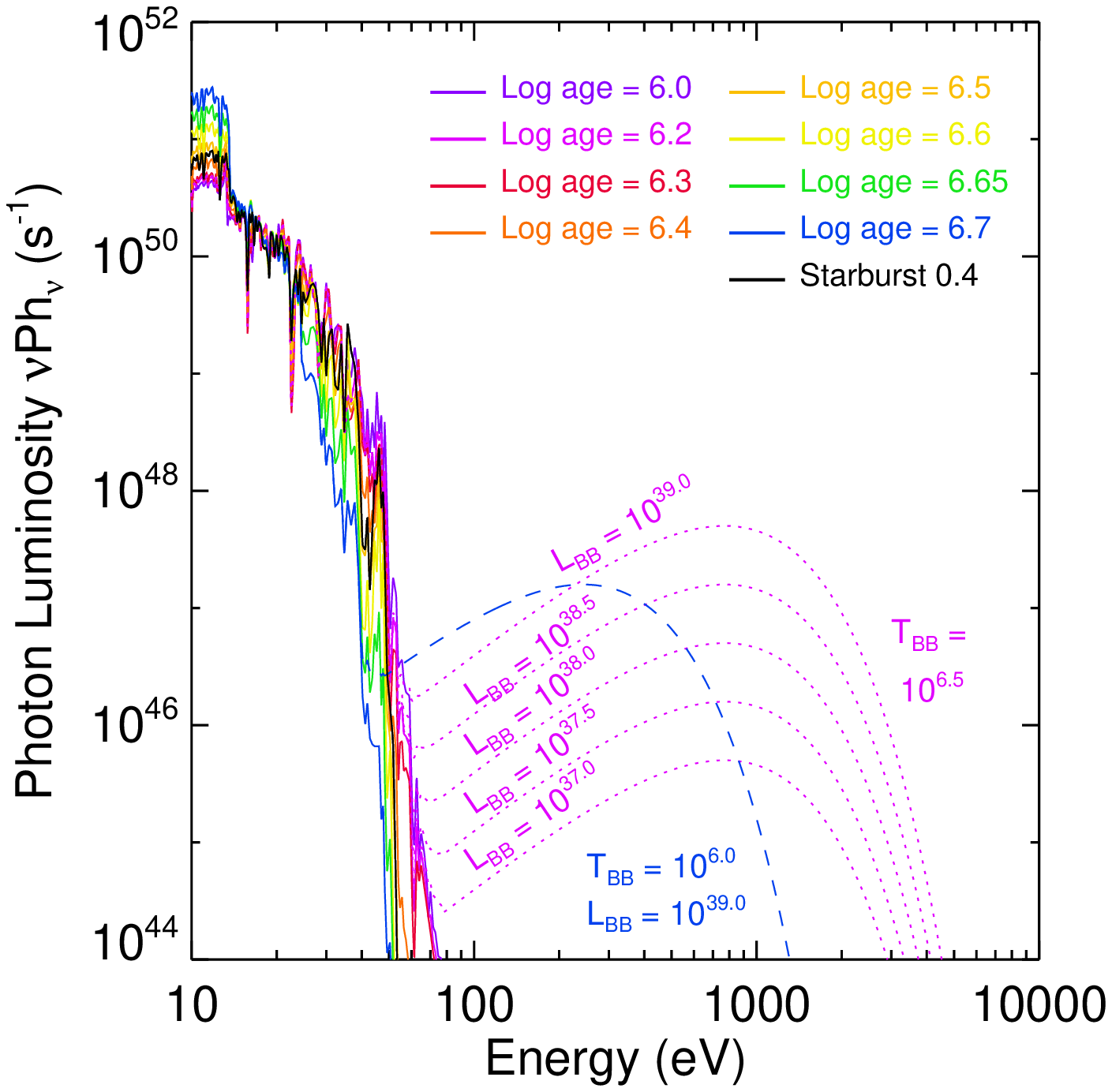}
\caption{Spectral energy distributions from Starburst99 for various ages with 
additional fluxes from blackbodies 
with the given luminosities (labels in erg s$^{-1}$) and temperatures (in K).
The numbers of photons emitted by the central source, $Ph_\nu$, are multiplied 
by the frequency, $\nu$.  
All the Starburst99 models drawn with colored lines have zero rotation velocity;
the black line is the Starburst99 model with rotation 0.4 times the break-up velocity 
and log age = 6.7.
Note that the SED with $T_{\rm BB} = 10^{6.0}$~K and $L_{\rm BB} = 10^{39.0}$~erg s$^{-1}$ 
has substantially more flux in the important energy range of 41--100 eV than 
the SED  with $T_{\rm BB} = 10^{6.5}$~K and $L_{\rm BB} = 10^{39.0}$~erg s$^{-1}$.
}
\end{figure*}

The abundances used in the models are my best estimate of the abundances 
of the atomic gas component of the GC.
{ 
With respect to hydrogen, these are, for He, C, N, O, Ne, Si, S, Ar, and Fe: 
0.095, 5.13e-4, 1.16e-4, 6.84e-4, 1.74e-4, 2.40e-5, 1.90e-5, 6.20e-6, and 2.6e-6, respectively.
The abundances of Ne/H, (S$^{++}$ + S$^{3+}$)/H$^+$, and Si$^+$/H$^+$ are the averages of the abundances in Table~12.
Ionization correction factors (icf) were computed for Si$^+$/H$^+$ (0.56)
and Ar$^+$/Ne$^+$ (0.90) from the fully ionized regions of the models computed herein. 
For S/H there is a 15.6\% addition to the (S$^{++}$ + S$^{3+}$)/H$^+$ ratio for S$^+$ (Rubin et al. 2016).
Since both Ne/H and S/H are factors of 1.71 times the abundances used by Rubin et al. (2016) 
to describe the Orion Nebula, 
I multiplied their Orion Nebula abundances for C, N, O, and Fe by this same factor 
to get the GC abundances in this paper (omitting the effects of the likely larger N/H abundance gradient,
e.g., Rudolph et al. 2006).
}
The other elements have the abundances used for the Cloudy \ion{H}{2} region mix, 
which are essentially those of the Orion Nebula and so are probably of too low abundance for the GC.
However, the elements in this mix all have very low abundances compared to the elements 
listed above and so contribute very little to the cooling of the \ion{H}{2} region gas. 

The other chief parameters for the grid are the electron and photon densities. 
Because the observed line ratios indicate that the electron density is never high 
(except in a few of the high excitation sources of Tables 8 and 9), 
densities of $N_p = 100$~cm$^{-3}$ and cluster photon luminosities 
$N_{\rm LyC} = 10^{50}$ photons s$^{-1}$ were used in the models.
The average gas densities were varied by using a range of filling factors $f$ of 1.0, 0.31623, 0.1, 
0.031623, 0.01, 0.0031623, and 0.001. 
The photon densities at the inner edges of the \ion{H}{2} regions were varied by 
using inner \ion{H}{2} region radii, $R_{\rm inner}$, of 1.0, 3.1623, and 10 pc; 
a few models were also computed with inner radii of 31.623 and 100 pc but
these models have such low $U$ that their predicted line ratios are outside the observed range 
($\log U \lesssim -3.5$).

Given the constant density and filling factor with distance from the exciting star cluster, 
the integrated line fluxes should scale with the ionizing luminosities $N_{\rm LyC}$,
and adjustments can be made for changes in $N_e$ keeping $U$ constant 
according to equations 1 and 2. 
This is not exact, as changing the density changes the cooling rates, and hence $T_e$, 
thus affecting $R_S$ and $U$ through the $\sim T^{-0.8}$ temperature dependence of $\alpha_{\rm B}$.
MIR line emissivities are relatively insensitive to $T_e$ (proportional to $\sim T_e^{-0.3}$)
and the predicted line ratios of the MIR forbidden lines are quite insensitive. 
Changing the input $R_{\rm inner}$ does not significantly change $U$ or $R_S$ until 
$R_{\rm inner}$ is quite a bit larger than 10 pc.

%Figure 8
\begin{figure*}
\includegraphics[width=165mm]{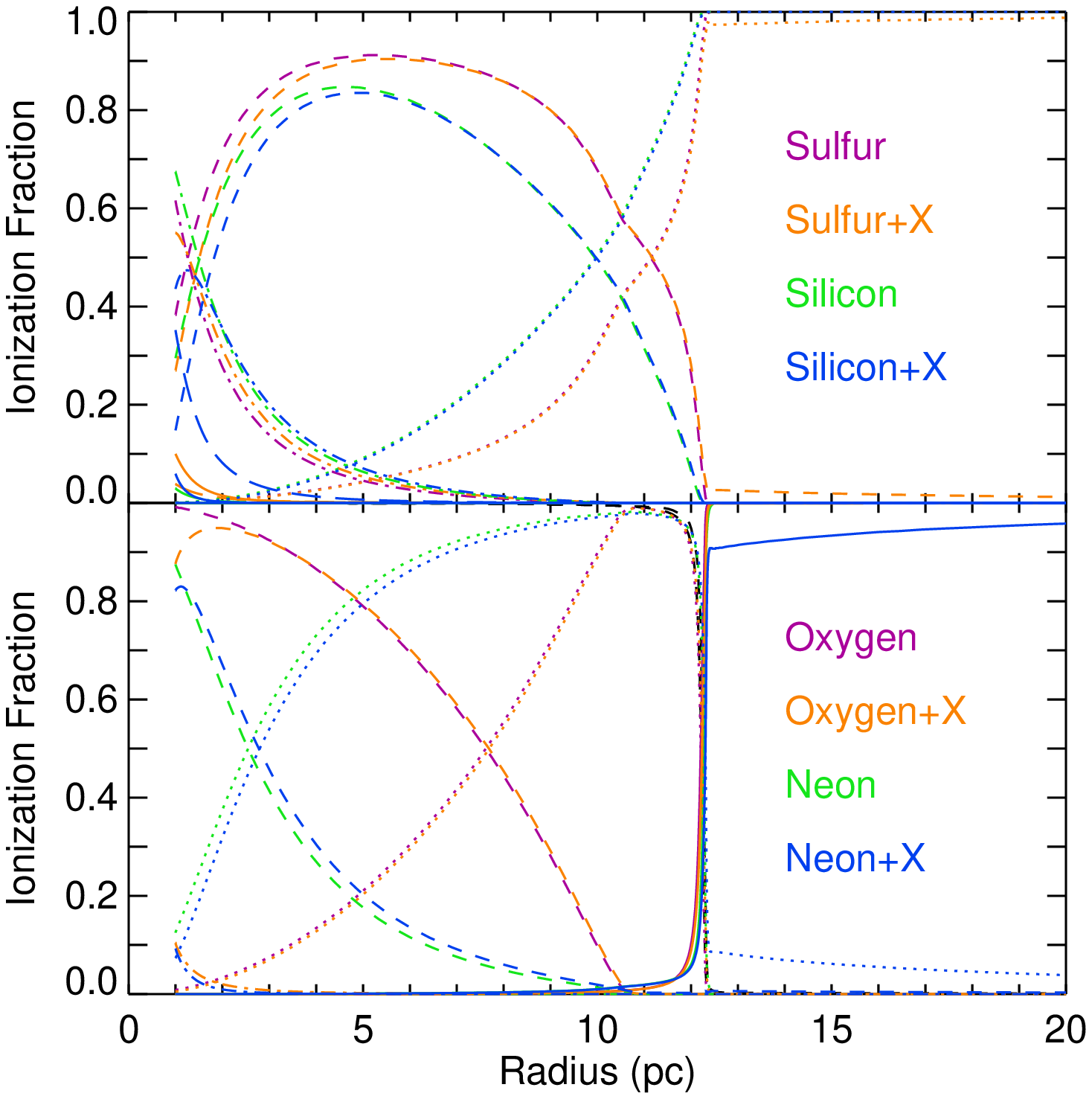}
\caption{
Ionization structures for different elements for models with Starburst99 SEDs for zero rotation and age 6.6 Myr.
The models are {\tt starburst0\_66\_0001} and {\tt starburst0\_66\_6570} and have inner radii of 1.0 pc, Str\"omgren radii $\sim 14$~pc, and filling factor equal to 0.1.
The different elements and models with or without X-rays are indicated by color, as shown in the figure.
The bottom panel shows the ionization of oxygen and neon and the top panel shows the ionization of silicon and sulfur.
For all elements, the neutral ionization stage is plotted with a solid line, singly ionized is dotted, doubly ionized is dashed, triply ionized is dash-dot, quadruply ionized is long dashes, and five times ionized is plotted with a solid line again.
}
\end{figure*}

Examples of the ionization structures of models with and without additional X-rays are given 
in Figure 8.
In particular, the locations of gas with doubly ionized silicon and sulfur are seen to be 
quite similar and unlike either oxygen or neon.
For this reason the easily observed ratio of doubly ionized sulfur to singly ionized silicon 
([\ion{S}{3}] 33 \micron/[\ion{Si}{2}] 34 \micron) 
is an excellent indicator of the local ionization parameter $U$.

%Figure 9
\begin{figure*}
\includegraphics[width=160mm]{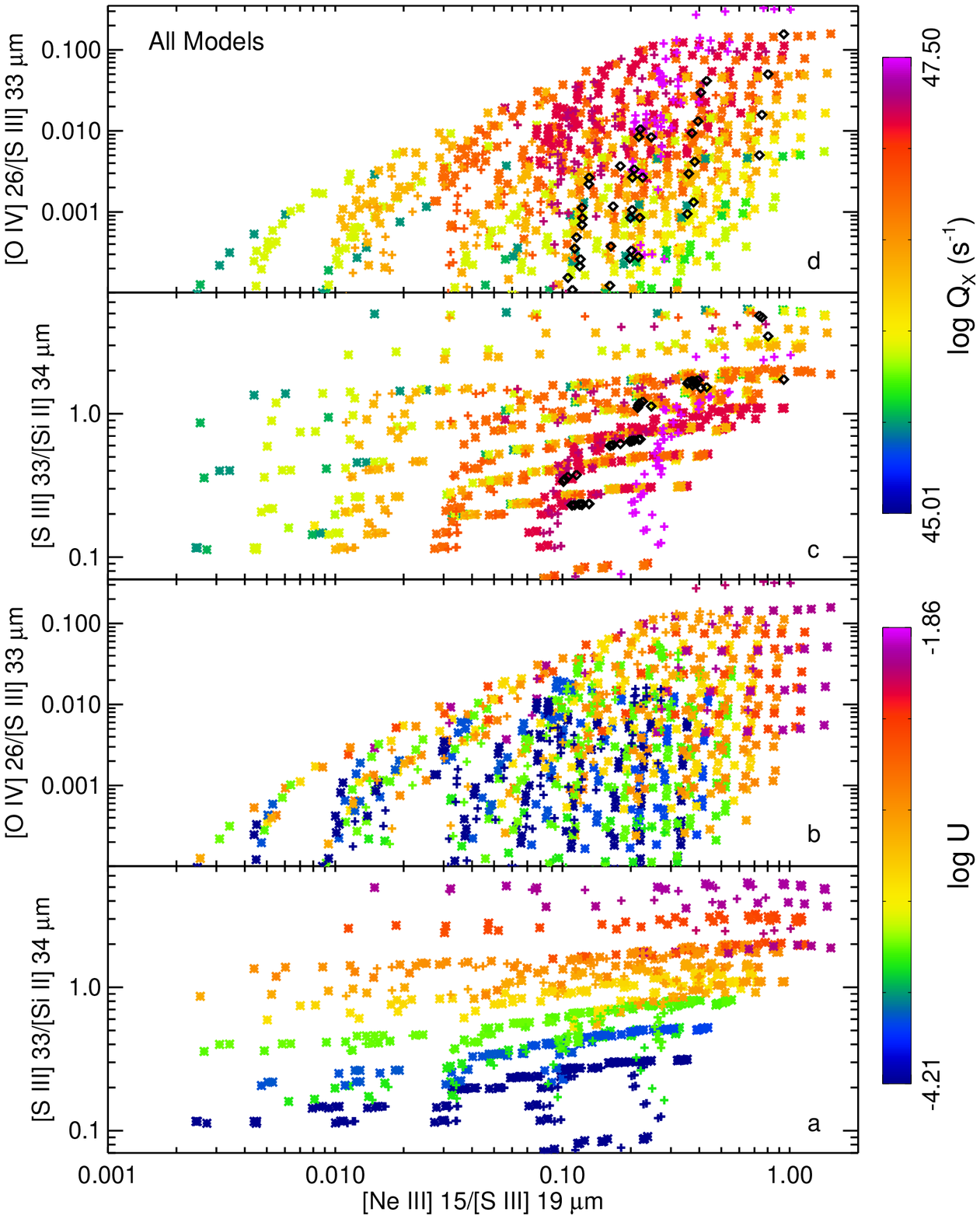}
\caption{
Plots of the ratios of the lines observed with {\it Spitzer} IRS (Section 3) computed from all the models 
in the grid of \ion{H}{2} regions.
The points marked with asterisks include ionizing fluxes by blackbodies with $T = 10^{6.5}$~K 
and the points marked with plus signs include ionizing fluxes by blackbodies with $T = 10^{6.0}$~K.
The points marked with black diamonds in panels (c) and (d) employ the Starburst99 SED 
for rotation 0.4 times the break-up velocity and log age = 6.7.
The colors of the data points in panels (a) and (b) are functions of the log of the ionization parameter $U$
and the colors in panels (c) and (d) are functions of the numbers of stellar photons with energies 
between 54.4 and 280 eV, here written as $Q_X$. 
Such photons from GC sources are too soft to be detectable by {\it Chandra} 
but have the right energy range to ionize the higher-excitation ionization states reported here.
It is seen that $Q_X$ has an especially strong influence on the [\ion{O}{4}] 26/[\ion{S}{3}] 33 \micron\ line ratios 
and that $U$ is the main parameter influencing the [\ion{S}{3}] 33/[\ion{Si}{2}] 34 \micron\ line ratios, 
as expected.
}
\end{figure*}

The line ratios predicted by this grid of \ion{H}{2} region models, seen in Figure~9,  
can be compared to the observed ratios in the GC. 
Here I subdivide the area of the GC to analyze specific \ion{H}{2} regions of interest: 
the regions ionized by the Quintuplet or the Arches Clusters, Sgr B1, and Sgr C.
Obviously, Sgr B2 would be of great interest; however, the extinction is so large 
towards Sgr B2 that the {\it Spitzer} IRS data set includes only a few sources (mostly candidate YSOs) 
and thus the region was not well enough observed by the IRS (Figure~2).

The models and data for the four regions are plotted in Figures 10 to 13, 
and summaries of the best-fitting models are given in Table 13.
For each region, the line intensities divided by the [\ion{S}{3}] intensity 
observed in the same {\it Spitzer} IRS module and corrected for extinction 
were averaged, with the low-resolution modules and high-resolution modules 
computed separately, since there are far more low-resolution spectra 
but only the high-resolution spectra include measurable [\ion{O}{4}] 26 \micron\ line intensities
(not including the shocked regions of Tables 8 and 9). 
Then one by one the log of these intensity ratios with respect to a sulfur line was compared to the logs 
of the same ratios of each model, and the sum of the squares of the differences (`$chisq$') was computed:
\begin{equation}
chisq = \sum_{i} \left[ \log (\frac{x_i}{y_i}) - \log (\frac{x^\prime_i}{y^\prime_i}) \right]^2 ,
\end{equation}
where $x_i$ is one of the observed [\ion{Ne}{3}] 15, [\ion{Si}{2}] 34, or [\ion{O}{4}] 26 \micron\ line intensities, 
$y_i$ is the [\ion{S}{3}] line flux observed in the same IRS module, 
and $x^\prime_i$ and $y^\prime_i$ are the same lines as computed by one of the Cloudy models.

The model parameters were not iterated, and since the spacing of the parameters of the grid is 
not small (0.5 dex), the models with minimum $chisq$ give only an indication 
of what is needed to produce a better fit.
The results of three models from each X-ray SED group are given in Table 13 -- 
although there are differences in $chisq$, these models are probably equally good fits 
and simply show the ranges of acceptable parameters. 

\input tab13.tex

Table 13 contains the numbers of ionizing photons, $N_{\rm LyC}$, for each of the \ion{H}{2} regions.
These were estimated from the single-dish radio continuum 
measurements, $S_\nu$, of Altenhoff et al. (1978), Downes et al. (1980), Reifenstein et al. (1970), and 
Wilson et al. (1970) at $\sim 5$~GHz 
using the relation of Rubin (1968) as written by Simpson \& Rubin (1990) for an assumed $T_e = 6000$~K
(Simpson et al. 2007).
It is especially important to use single-dish radio telescope measurements for these estimates 
because the total numbers of ionizing photons are needed for computing scale factors for the models, 
which were all computed for $10^{50}$ photons s$^{-1}$, 
and interferometers lose flux owing to their lack of dishes with almost zero spacing. 

The estimates of the photon numbers for the Quintuplet Cluster region and the Arched Filaments 
are particularly uncertain because of the large contribution to the radio fluxes from 
the non-thermal Radio Arc (e.g., Yusef-Zadeh \& Morris 1987).
However, the values in Table~13 must be underestimates since neither cluster 
is embedded in its natal molecular cloud, thereby allowing a sizable fraction of ionizing photons 
to escape the region, and the $N_{\rm LyC}$ vs. $S_\nu$ relation assumes that 
the \ion{H}{2} region producing the radio continuum is ionization-bounded in all directions (Rubin 1968).
In fact, Figer et al. (1999, 2002) estimated ionizing fluxes for both the Quintuplet and Arches clusters 
of close to or more than $10^{51}$ photons s$^{-1}$ from the numbers of OB and Wolf-Rayet (WR) stars.
In summary, to compare the X-ray fluxes from the best fit models to X-ray observations of the GC, 
one should multiply the X-ray $L_{\rm BB}$ by the observed $N_{\rm LyC}$ in Table~13
and divide by the $N_{\rm LyC}$ used in the models, which was $N_{\rm LyC} = 10^{50}$ photons s$^{-1}$.

\subsection{Arched Filaments}

%Figure 10
\begin{figure}
\includegraphics[width=84mm]{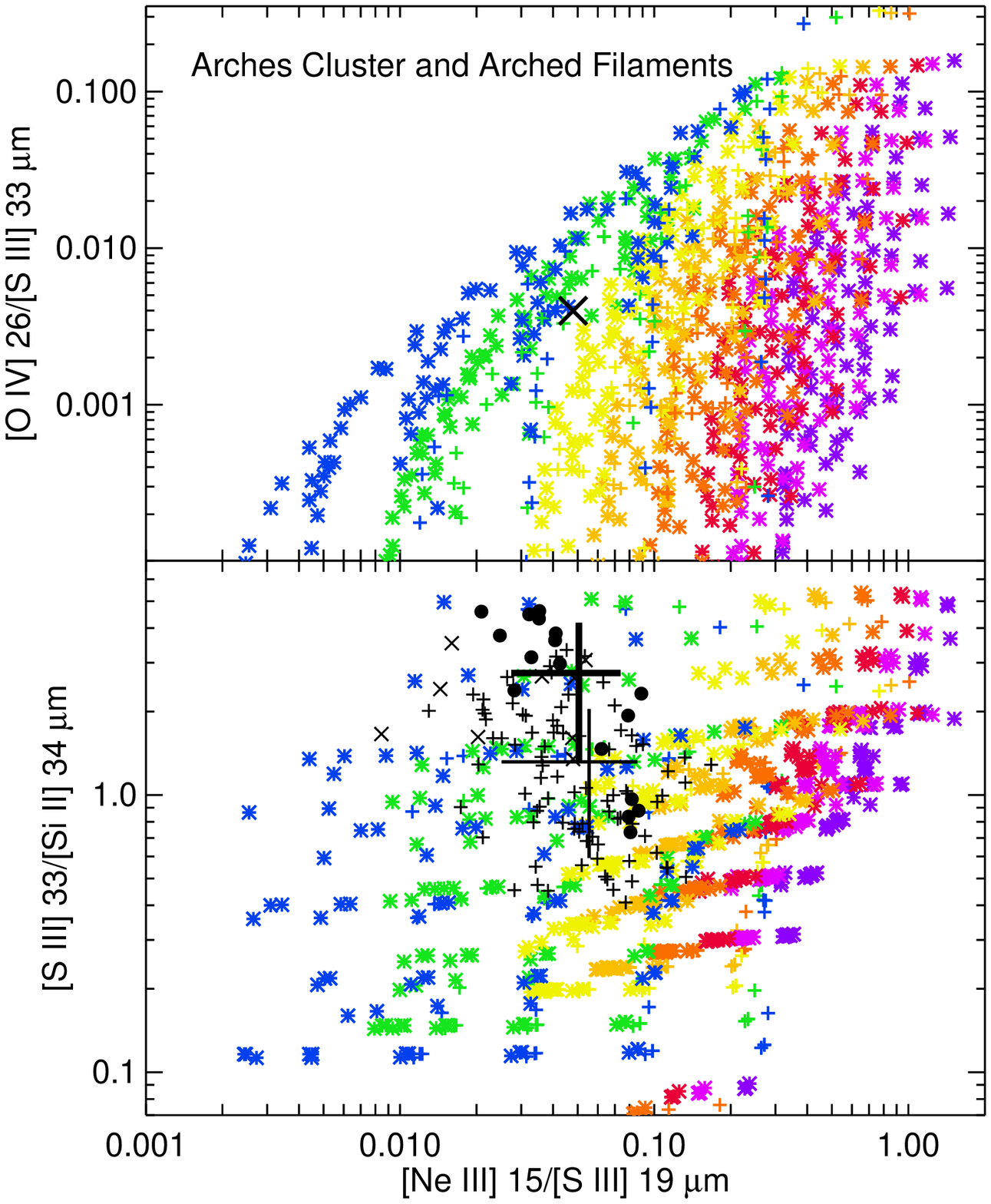}
\caption{
Plots of the ratios of the lines observed with {\it Spitzer} IRS (Section 3) computed from all the models 
with zero rotation 
in the grid of \ion{H}{2} regions and the ratios observed in the Arched Filaments.
The points marked with asterisks include ionizing fluxes by blackbodies with $T = 10^{6.5}$~K 
and the points marked with plus signs include ionizing fluxes by blackbodies with $T = 10^{6.0}$~K.
The colors of the points representing the models in both panels are functions of 
the ages of the zero rotation models from Starburst99: purple is log age = 6.0, 
magenta is 6.2, red is 6.3, red-orange is 6.4, yellow-orange is 6.5, 
yellow is 6.6, green is 6.65, and blue is 6.7, where the age is in yrs.
For the locations of the points representing the models computed with Starburst99 SEDs 
with rotation velocity 0.4 times the break-up velocity and log age = 6.7, see Figure~9.
In the lower panel the small black crosses are ratios of lines observed with the LL module 
and in both panels the black dots are the ratios of lines observed with the SH and LH modules 
in program 40230.
The black X's show the ratios of lines observed with the SH and LH modules 
in program 3295 (Simpson et al. 2007) { or program 0018}.
The large black crosses plot the averages of the observed ratios and the error bars their 
standard deviations, with the thicker error bars from the high resolution modules.
}
\end{figure}

The line ratios from the Arched Filament positions used in this section are the combination 
of the Arched Filament positions of Simpson et al. (2007) 
and a subset of the GC Filaments, described in Table~12, with Galactic longitudes between 0.10 and 0.25, 
and are plotted in Figure~10.
The Arches Cluster is assumed to be the source of the ionizing photons for the Filaments 
along Galactic longitude 0.15 -- the high [\ion{S}{3}] 33/[\ion{Si}{2}] 34 \micron\ ratio in this region (Figure~3b)
shows that the exciting stars are near by, providing the undiluted radiation field 
(high ionization parameter) needed to produce this ratio. 

The average model ages are $10^{6.6}$ to  $10^{6.7}$ yr with average $chisq$ = { 0.07 and 0.05 
for the models with X-ray $T_{\rm BB} = 10^{6.5}$~K and $L_{\rm BB} = 10^{38.0}$ erg s$^{-1}$
or $L_{\rm BB} = 10^{38.5}$ erg s$^{-1}$ 
or X-ray $T_{\rm BB} = 10^{6.0}$~K and $L_{\rm BB} = 10^{37.0}$ or $10^{37.5}$ erg s$^{-1}$, respectively.}
%One of the reasons for the large $chisq$ (indicating poor fits) are the large differences  
%between the observed [\ion{S}{3}] 33/[\ion{Si}{2}] 34 \micron\ and [\ion{O}{4}] 26/[\ion{S}{3}] 33 \micron\ line ratios
%of the different IRS modules (Figure~10). 

The fitted SEDs all have ages in the range of $4 - 5 \times 10^6$ yrs; 
a substantially younger age does not produce any reasonable fit.
On the other hand, going back to Figure~9, notice that the black diamonds that mark the Starburst99 SEDs 
for rotation of 0.4 times the break-up velocity all have 
much higher [\ion{Ne}{3}] 15/[\ion{S}{3}] 19 \micron\ line ratios,
even though they all also have ages $\sim 5 \times 10^6$ yrs. 
To get models using the Starburst99 SEDs for the 0.4 times break-up sequence 
that also match the observed [\ion{Ne}{3}] 15/[\ion{S}{3}] 19 \micron\ line ratios for the Arched Filaments, 
one would need substantially older cluster ages than $5 \times 10^6$ yrs.
Such long ages are in conflict with { ages estimated by other means, for example, the 
$3.5 \pm 0.7$ Myr for the Arches Cluster (Schneider et al. 2014).} 
Moreover, long ages are not reasonable considering the short orbital time for clusters in the GC 
and the subsequent loss of stars due to tidal interactions (e.g., Portegies Zwart et al. 2002;
{ Habibi et al. 2014}).
For these reasons, that the Starburst99 SEDs computed for stellar models with rotation cannot produce reasonable fits, they are not discussed further in this paper.

The GC filaments of { Table 12} include the small \ion{H}{2} regions 
at the base of the Filaments near Galactic longitude 0.05.  
These \ion{H}{2} regions also have hot stars (Cotera et al. 1999) 
and so should be analyzed separately from the Arched Filaments. 
Moreover, any models should use individual stellar SEDs instead of the composite cluster SED 
of Starburst99. Such models will be discussed in a later paper. 
{ However, if the data for only the small \ion{H}{2} regions
are compared to the models, the best fits have similar ages to those of the Arched Filaments 
in Table 13 but require slightly more X-rays, possibly owing to their location closer to Sgr A*, 
and a lower $U$.
The similarity in inferred ages to those of the Arches Cluster stars 
could be a possible indication that these stars were not formed independently from 
the Arches Cluster but were originally in the Cluster and have been stripped away 
by the tidal field of the GC (e.g., Habibi et al. 2014).
}

Note that there { is only one value } plotted in the top panel of Figure 10 ---
the reason is that the [\ion{O}{4}] 26 \micron\ line is required to have a signal/noise ratio of at least 2 
(the observed S/N for the { plotted point is 2.5}).
Although all positions in the Arched Filaments appear to have detectable [\ion{O}{4}] 26 \micron\ lines, 
the other positions are in the high-continuum filaments, 
with the result that the noise from uncorrectable rogue pixels is also high, leading to poor S/N 
in the measurement of the faint [\ion{O}{4}] 26 \micron\ line (Simpson et al. 2007).
Thus the estimate of the X-ray luminosity for the Arched Filaments is the least certain of
all four \ion{H}{2} region estimates. 

Erickson et al. (1991), Colgan et al. (1996), and Cotera et al. (2005) 
found that the excitation of the Filaments is consistent with the exciting source being 
located at some distance from the Filaments, such as the location of the Arches Cluster.
Simpson et al. (2007) also found that that the excitation of the `W' Filament is 
much lower than that of the `E2' Filament, again in accord with the supposition 
that the Filaments are ionized by the Arches Cluster.
Unfortunately, the new IRS fluxes reported in this paper do not include 
much additional coverage of the E2 Filament and none of the W Filament beyond 
that described by Simpson et al. (2007), as seen in Figure 2.
The most likely configuration of the Filaments with respect to the Arches Cluster 
is that the Filaments are the lit-up edges of molecular clouds at 
distances of $\sim 10 - 20$~pc from the Arches Cluster (Lang et al. 2001; Yasuda et al. 2009).

On the other hand, Hankins et al. (2017) mapped the Arched Filaments in several MIR bands with 
the Faint Object InfraRed CAmera for the {\it SOFIA} Telescope (FORCAST). 
They found that if they assumed standard size interstellar dust grains ($\sim 0.1$~\micron), 
the heating source of the dust would have to be much closer to the Filaments 
than the distances observed for the Arches Cluster on the sky. 
Either the heating sources would have to be more localized in the Filaments 
or the dust grains would have to be significantly smaller, $\sim 0.01$~\micron\ 
if the heating source is required to be the Arches Cluster.
{ They suggest that the presence of} such small grains would indicate substantial dust processing in the GC.
{ Small grains could result from processes such as shattering by supernovae (e.g., Andersen et al. 2011) 
or shattering by high grain velocities due to the turbulence in the GC molecular clouds (Hankins et al. 2017; Hirashita \& Yan 2009).}

\subsection{Quintuplet Cluster Region}

%Figure 11
\begin{figure}
\includegraphics[width=84mm]{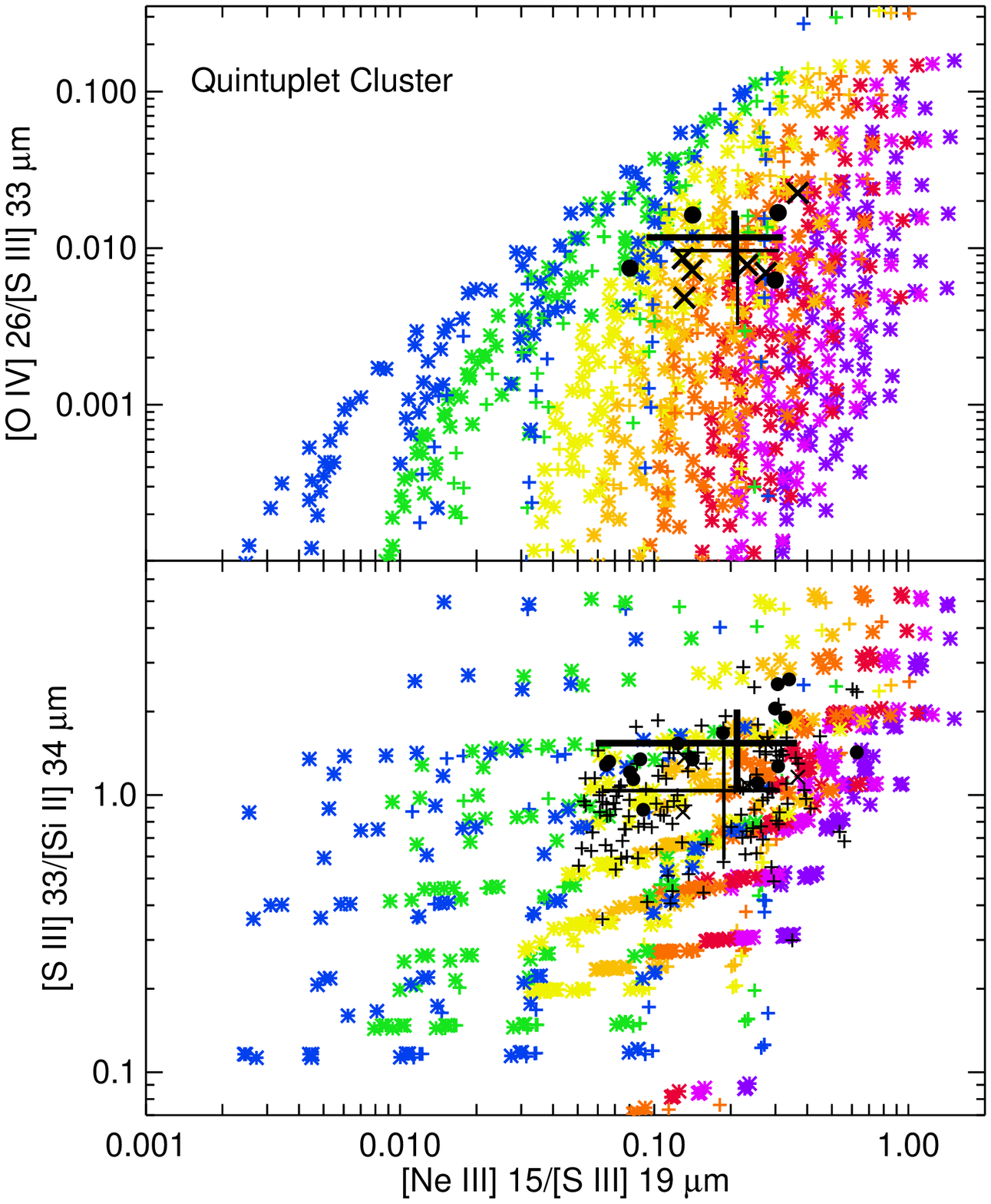}
\caption{
Plots of the ratios of the lines observed with {\it Spitzer} IRS (Section 3) computed from all the models 
with zero rotation 
in the grid of \ion{H}{2} regions and the ratios observed in the Quintuplet Cluster region.
The colored points marked with asterisks include ionizing fluxes by blackbodies with $T = 10^{6.5}$~K 
and the colored points marked with plus signs include ionizing fluxes by blackbodies with $T = 10^{6.0}$~K.
The colors of the points representing the models and the black symbols in both panels,
representing the line ratios observed in the Quintuplet Cluster region,  
are described in Figure 10.
}
\end{figure}

The region surrounding the Quintuplet Cluster (G0.1659$-$0.0656), 
$0.11 < l < 0.25$ and $-0.07 < b < -0.01$ in Galactic coordinates ($l,b$), 
was chosen to specifically avoid the Radio Arc Bubble, which is known to 
contain shocks (Simpson et al. 2007).
Previous observations (Rodr\'{\i}guez-Fern\'andez et al. 2001; Simpson et al. 2007) 
have demonstrated that the gas excitation decreases uniformly with distance from the Quintuplet Cluster,
from which I infer that this massive cluster is the ionizing source. 
Modeling the Quintuplet Cluster region (Figure~11), I find that 
the average Starburst99 model age is $10^{6.4}$ yr for the models with $T_{\rm BB} = 10^{6.5}$ erg s$^{-1}$
and $10^{6.5}$ yr for the models with $T_{\rm BB} = 10^{6.0}$ erg s$^{-1}$, 
significantly younger than the average Starburst99 model age of { $10^{6.66}$} yr 
needed for the Arches Cluster.

The Quintuplet Cluster, like the Arches Cluster, contains numerous WR stars, 
indicating that there has been considerable evolution away from the main sequence 
for its most massive stars. 
This change in the individual stellar SEDs owing to such evolution is included 
in the composite SED of Starburst99 (Leitherer et al. 2014).
Schneider et al. (2014) modeled the change in the stellar mass function 
for these two clusters taking into account that most massive stars are found in binary systems 
(see, e.g., Sana et al. 2014)
and exchange mass, thereby changing their spectral types and the apparent mass function of the cluster.
By fitting the observed stellar mass functions to their population synthesis models, 
they find { an age for the Quintuplet Cluster of $4.8 \pm 1.1$ Myr, older than their $3.5$ Myr age of the Arches Cluster}.
Although the uncertainties are sizable, the turn-off point from the main sequence 
is well determined in both clusters; thus they find there is no uncertainty 
in their determination that the Quintuplet Cluster is older than the Arches Cluster.
Another method of determining ages of stellar clusters is by isochrone fitting; 
using this method Liermann et al. (2014 and references therein) estimate an age of $3.0 \pm 0.5$ Myr 
for the OB stars in the Quintuplet Cluster. 
This should not be surprising, according to Schneider et al. (2014), 
since the most massive stars in the Quintuplet Cluster are effectively rejuvenated 
by receiving mass transferred from their binary companions. 

However, I find from the {\it Spitzer} IRS spectra of the [\ion{Ne}{3}]/[\ion{S}{3}] ratios 
(Figures 10 and 11) that the Quintuplet Cluster
{ Region is ionized by higher energy photons than the Arched Filaments. }
This { conclusion} is not new, if one looks back at previous observations of the [\ion{O}{3}], [\ion{S}{3}], and [\ion{Si}{2}] lines 
taken from the Kuiper Airborne Observatory (KAO, Colgan et al. 1996; Simpson et al. 1997) 
and the [\ion{O}{3}] 88 and [\ion{N}{2}] 122 \micron\ lines measured by Yasuda et al. (2009) with {\it AKARI}.
In the KAO spectra the O$^{++}$/S$^{++}$ ratios were measured to be higher in the Sickle than in the Arched Filaments, 
indicating ionization by higher temperature stellar SEDs in the former region.
{ Given the assumption that the two clusters 
are the sources of the photons ionizing their respective regions, 
one would normally conclude, therefore, that the Quintuplet Cluster 
must contain hotter stars than the Arches Cluster. 
It is important to note that both }
clusters are massive enough that they should not be affected by small numbers of stars 
producing uneven initial mass functions. 

{ On the other hand, if the Quintuplet Cluster is indeed older than the Arches Cluster 
(e.g., Schneider et al. 2014),
then there are additional high energy photons in the region of the Quintuplet Cluster 
that are contributing to the ionization of the Sickle, 
photons that are not included in the composite SEDs that are produced by Starburst99.}
One possibility is that the stars of the Quintuplet Cluster, unlike the Arches Cluster, 
rotate with a high velocity, thereby making the models computed with Starburst99 SEDs 
with rotation 0.4 times the breakup velocity applicable. 
Another possibility is that the actual SED includes energetic photons from some 
transient event, such as a supernova or a mass-loss episode from 
the Quintuplet Cluster's Pistol Star.
(I note, however, that the Pistol Nebula is not as highly ionized as the rest of the Sickle 
--- Simpson et al. 1997 --- thereby making the latter suggestion less likely.) 
In fact, Ponti et al. (2015) suggest that the Radio Arc Bubble is the result of multiple 
supernova explosions from Quintuplet Cluster stars, occurring over a range of time 
as the cluster moved from lower to higher Galactic longitudes (Stolte et al. 2014),
and Heard \& Warwick (2013) suggest that there is a supernova remnant at G0.13-0.12.

\subsection{Sgr B1}

%Figure 12
\begin{figure}
\includegraphics[width=84mm]{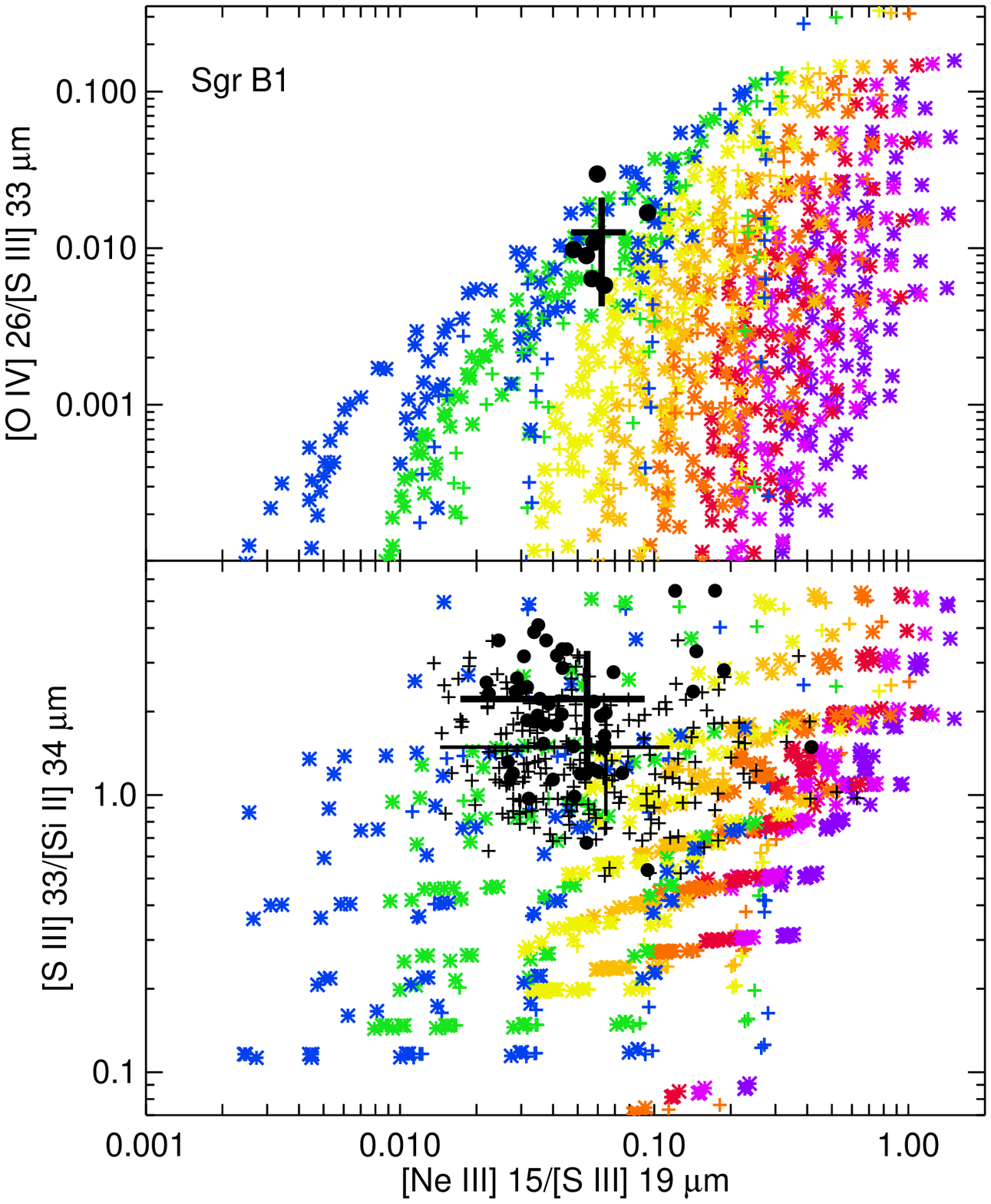}
\caption{
Plots of the ratios of the lines observed with {\it Spitzer} IRS (Section 3) computed from all the models 
with zero rotation 
in the grid of \ion{H}{2} regions and the ratios observed in Sgr B1.
The points marked with asterisks include ionizing fluxes by blackbodies with $T = 10^{6.5}$~K 
and the points marked with plus signs include ionizing fluxes by blackbodies with $T = 10^{6.0}$~K.
The colors of the points representing the models and the black symbols  in both panels,
representing the line ratios observed in Sgr B1,  
 are described in Figure 10.
}
\end{figure}

In Galactic coordinates, Sgr B1 is defined to be located in the region $0.45 < l < 0.60$ and $-0.11 < b < -0.01$. 
The line ratios observed in Sgr B1 are plotted in Figure~12 
and the best fit models are tabulated in Table~13.

Sgr B1 appears to have two components, `East' and `West', 
where the West component has relatively less absorption from both \ion{H}{1} (Lang et al. 2010) 
and formaldehyde (Mehringer et al. 1995) than the East component.
The measurements of the MIR dust extinction also show that the eastern part of the source 
has higher $\tau_{9.6 \micron}$ than the western part of Sgr B1 (Figure~4).

Although Sgr B2 is thought to be extremely young because of the multiple compact components 
in its radio thermal continuum emission 
(e.g., De Pree et al. 2015 and references therein)
{ and its numerous hot molecular cores (e.g., Vogel et al. 1987; Etxaluze et al. 2013; Schmiedeke 2016)}, 
this is not true for Sgr B1. 
The radio continuum of Sgr B1 mostly appears to consist of extended ridges and shell-like structures 
(Mehringer et al. 1992). 
The measurement of the density of Sgr B1 equal to { $\sim 290$} cm$^{-3}$ (Table 12) 
is in accord with this finding that the gas has no { truly} compact components. 
Such morphology in an \ion{H}{2} region can arise only as the result of strong stellar winds dispersing the gas.
%In fact, Sgr B1 is known to have both an O4-6I and three WR WN7--8 stars (Mauerhan et al. 2010),  
{ Such strong winds might originate in stars like the O4-6I and three WR WN7--8 stars 
found in Sgr B1 (Mauerhan et al. 2010),  
where the O star is located in a cavity in the dust at G0.52$-$0.046 (Ponti et al. 2015) 
and one of the WR stars is in a possible bubble
(although on the other hand, 
these stars might actually be unrelated to Sgr B1 and instead be some of the stars 
that were stripped from the Arches or Quintuplet Clusters and have drifted many pc along the 
plane of the Galaxy, Habibi et al. 2014). 
However, there might be another indicator of significant age in that}
 Nobukawa et al. (2008) and Ponti et al. (2015) suggest that there is a supernova remnant 
at G0.42$-$0.04 in Sgr B1 (an area not covered by the IRS),  
which they detect in X-ray observations with {\it Suzaku} and {\it XMM-Newton}, respectively. 

The models of Table 13 require Starburst99 SEDs with ages of $10^{6.65}$ to $10^{6.7}$ yrs.  
However, in contrast to the above suggestions of substantial age for Sgr B1, 
I note that it does contain YSOs (e.g., An et al. 2011) and 
OH and H$_2$O masers (Mehringer et al. 1993), 
both indicators that star formation is currently occurring. 

{ To summarize, the status of Sgr B1 still presents a major puzzle to the theory of star formation 
in the orbital streams around the gravitational center of the Galaxy at Sgr A 
proposed by Kruijssen et al. (2015).
In this theory molecular clouds are compressed and stars form as a result of close passage to Sgr A*  
with Sgr B2 following Sgr B1 around an extreme of the orbit as viewed from the Earth,
Sgr B2 still on the front side of the orbit relative to the Earth but Sgr B1 already on the back side.
From their orbital positions, Sgr B2 would have an age of $\sim 0.7$ Myr and Sgr B1 would have an age
of $\sim 1.5$ Myr (Barnes et al. 2017).
The presence of YSOs in Sgr B1 would be compatible with this age although the $\sim 5$ Myr age 
inferred from the best fitting models along with the dispersed appearance of the ionized gas in Figure 1 
indicate a significantly larger age.

However, the location of Sgr B1 is not exactly on this orbital path 
as seen in figure 4 of Barnes et al. (2017),
and there is substantially more extinction towards the supposedly closer Sgr B2 than Sgr B1. 
This extra optical depth is seen in the extinction map of Figure~4 
and in both formaldehyde absorption by Mehringer et al. (1995) 
and in neutral hydrogen absorption by Lang et al. (2010).
For the age difference between Sgr B2 and Sgr B1 to be as large as the 4 Myr that 
the combination of orbital kinematics and the ages inferred from the models suggest, 
Sgr B1 would have to be much further along in its orbit and there would have to 
be substantial line-of-sight separation between the two \ion{H}{2} regions. 
On the other hand, the velocity structure of both Sgr B2 and Sgr B1 seems to indicate 
that they are physically connected,  
with both \ion{H}{2} regions consisting of multiple sources at different distances along the line of sight 
interspersed with dense molecular cloud material 
(Mehringer et al. 1995; Lang et al. 2010).
Clearly, this region has no simple description, and given the measured velocities,
there does not appear to be any good evidence 
that Sgr B1 lies on the back side of the orbiting molecular cloud streams.
}

Sgr B1 will be discussed in more detail in a later paper (J. Simpson et al., in preparation).

\subsection{Sgr C}

%Figure 13
\begin{figure}
\includegraphics[width=84mm]{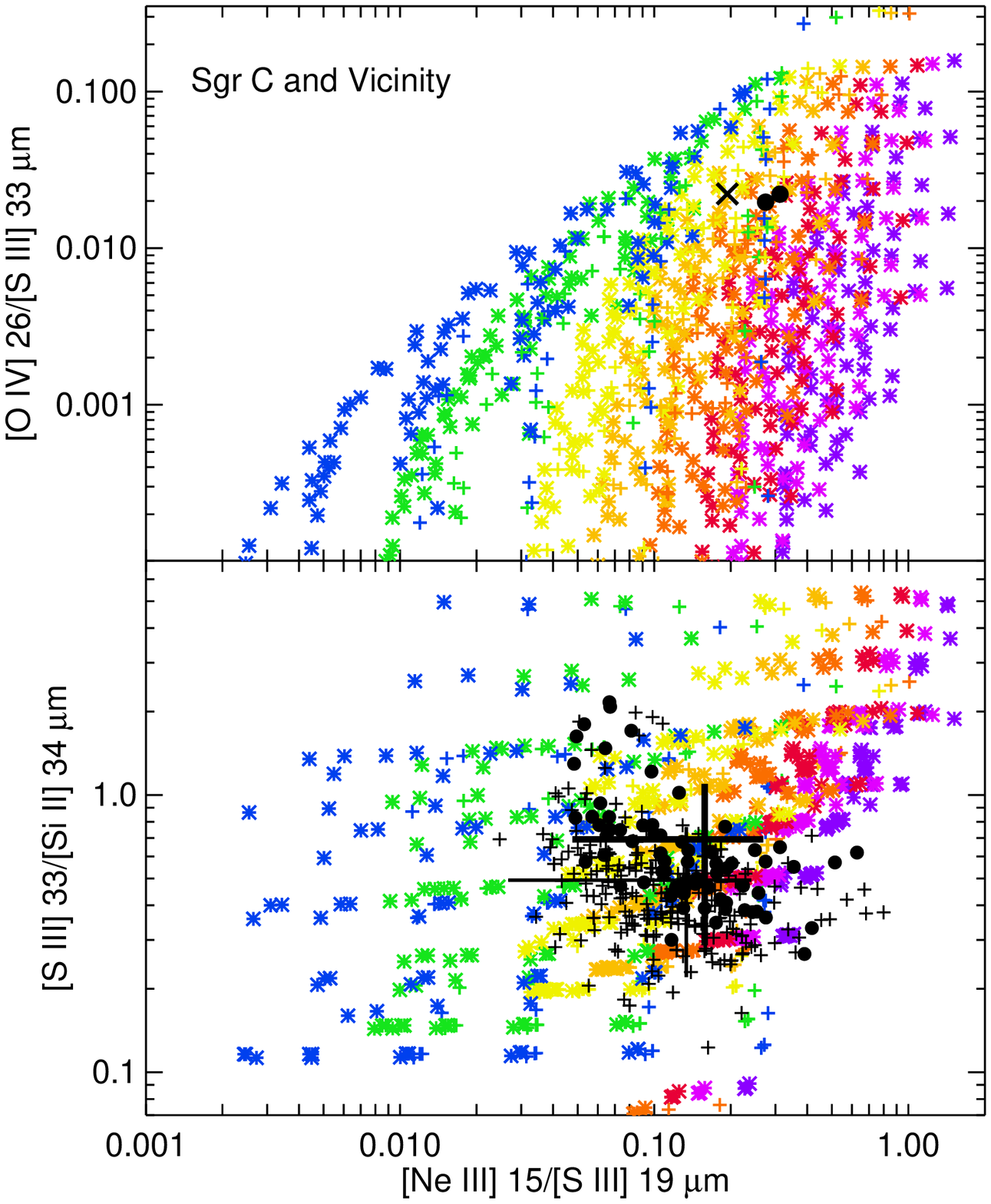}
\caption{
Plots of the ratios of the lines observed with {\it Spitzer} IRS (Section 3) computed from all the models 
with zero rotation 
in the grid of \ion{H}{2} regions and the ratios observed in Sgr C.
The points marked with asterisks include ionizing fluxes by blackbodies with $T = 10^{6.5}$~K 
and the points marked with plus signs include ionizing fluxes by blackbodies with $T = 10^{6.0}$~K.
The colors of the points representing the models and the black symbols in both panels, 
representing the line ratios observed in Sgr C,  
are described in Figure 10.
}
\end{figure}

Sgr C, G359.43$-$0.08, 
is the least luminous of the bright \ion{H}{2} regions discussed in this paper.
In addition to the thermal continuum of the \ion{H}{2} region, there appears 
to be an associated non-thermal filament, G359.45$-$0.06, reminiscent of the radio arc 
immediately adjacent and partially superposed on the Sickle and Arched Filaments 
(Liszt 1985; Liszt \& Spiker 1995).  
Tsuru et al. (2009) suggested that a nearby X-ray source, G359.41$-$0.12, is a supernova remnant 
with an outflow in the direction of Sgr C,
although Ponti et al. (2015) find additional structures with {\it XMM-Newton}
and suggest that it is part of the Galactic Center Lobe. 
Chuard et al. (2017) also find additional components and suggest that G359.41$-$0.12 
may include multiple supernova remnants. 

The line ratios observed in Sgr C are plotted in Figure~13 
and the best fit models are tabulated in Table~13.
An age of { $\sim 10^{6.6}$} yrs is estimated for Sgr C ---  
this is not unreasonable given its position at negative Galactic longitudes 
in the models described in the Introduction. 
{ However, there is also evidence of ongoing star formation in the presence 
of a YSO with an `extended green object' outflow (Kendrew et al. 2013) 
and the possible YSOs detected by An et al. (2011).
Thus this source appears to be similar to Sgr B1 with signs of both current star formation 
along with indications of significantly evolved stars. }

\subsection{Other Sources}

In summary, there appears to be extensive low-excitation gas in the GC, 
with a few regions of high-excitation gas that are confined to either known \ion{H}{2} regions 
or the high-excitation sources of Tables 8 and 9. 
There are two \ion{H}{2} regions seen in Figure~3b that are not otherwise discussed in this paper: 
G0.00+0.20 and G359.25$-$0.25. 
The former is probably the foreground \ion{H}{2} region S17 (Langer et al. 2017) 
and the latter may also be foreground of the GC, given its relatively low $\tau_{9.6 \micron}$.

\subsection{\ion{H}{2} Regions and the Warm Ionized Medium of Galaxies}

Extragalactic astronomers are often urged to study the Milky Way's GC because 
its relative nearness allows one to distinguish different phenomena that 
are confused in the effectively large apertures used on external galaxies.
Figures 2 and 3 show that the GC has individual \ion{H}{2} regions/star-forming regions 
widely spaced and surrounded by regions of lower density gas. 
The locations of the very low-excitation gas are seen in Figure~3b, 
where the [\ion{S}{3}] 33/[\ion{Si}{2}] 34 \micron\ ratio is  
proportional to the ratio of S$^{++}$/Si$^+$,  
which is the equivalent of the S$^{++}$/S$^+$ ratio (see Figure~8).
This low ratio is indicative of a low ionization parameter or a very dilute radiation field. 

Such dilute radiation fields occur when ionizing photons escape from density-bounded \ion{H}{2} regions, 
sometimes traveling great distances to high galactic latitudes (see Haffner et al. 2009 for a review). 
The gas ionized by this dilute radiation field is known as the warm ionized medium (WIM). 
Many galaxies have over half of their H$\alpha$ emission originating in the WIM (Oey et al. 2007). 
Usually, visible observations of the WIM largely compare the lines of H$\alpha$, 
[\ion{Ne}{2}], and [\ion{S}{2}] to [\ion{O}{3}], although 
another possibility is the [\ion{Ne}{3}] 3869/[\ion{O}{2}] 3727 \AA\ ratio 
for low-extinction galaxies (Levesque \& Richardson 2014).

To study the sources of the photons ionizing high-galactic latitude gas, 
Pellegrini et al. (2012) devised a technique they call `ionization-parameter mapping',
wherein they form images of extra-galactic \ion{H}{2} regions in the [\ion{S}{2}]/[\ion{O}{3}] line ratios.
They notice that sources with noticeable [\ion{S}{2}]/[\ion{O}{3}] halos must be optically thick 
in the Lyman continuum, such that no ionizing photons escape outside of the \ion{H}{2} region. 
In contrast, \ion{H}{2} regions with no noticeable [\ion{S}{2}] must be optically thin (i.e., density bounded),
and ionizing photons do escape. 
Blister \ion{H}{2} regions are optically thick (ionization-bounded) on one side and density-bounded on the other. 
Similarly, the escape of ionizing radiation in starburst galaxies was studied by Zastrow et al. (2013),
who find, from their observations of [\ion{S}{3}], [\ion{S}{2}], and H$\alpha$, that 
the star formation history and galaxy morphology are both important, with the radiation 
mostly escaping through ionization cones produced by feedback from stellar winds and supernovae.

Although these visible wavelength lines cannot be detected in the GC, owing to the GC's extinction, 
the two bright [\ion{S}{3}] 33 and [\ion{Si}{2}] 34 \micron\ lines can be used to test 
the same questions regarding the WIM in the Milky Way's GC. 
In Figure~3b it is seen that the [\ion{S}{3}] 33/[\ion{Si}{2}] 34 \micron\ line ratio 
is high only in the region of the Arched Filaments and in Sgr B1 and B2, 
but the ratio is substantially lower in the regions to the west of Sgr B and 
all around Sgr C. 
Even though the mapping is incomplete to the east of Sgr B, 
it is not to the west, where the two \ion{H}{2} regions 
must be ionization bounded, possibly by the thick molecular clouds 
of the Dust Ridge, seen in Figure~1.
On the other hand, 
I have already noted that the numbers of ionizing photons estimated from the radio luminosities 
are much smaller than the numbers of photons estimated by counting stars 
for the Arches and Quintuplet Clusters.
Since these clusters are well separated from their natal molecular clouds (e.g., Stolte et al. 2014), 
a significant fraction of their ionizing photons probably do escape to the surrounding 
ISM above and below the Galactic plane.

\subsection{X-rays in the Galactic Center}

The high-excitation [\ion{O}{4}] 26 \micron\ line was observed 
in a number of starburst galaxies by Lutz et al. (1998) using {\it ISO} 
and Dale et al. (2009) using {\it Spitzer}'s IRS.
Lutz et al. (1998) discussed several possible sources of the $> 54$~eV photons needed to ionize O$^{++}$ to O$^{3+}$, 
including weak AGNs, low-metallicity OB stars and WR stars, 
and ionizing shocks in starburst outflows. 
They concluded the last is the most likely source of the O$^{3+}$.
In addition, Schaerer \& Stasi\'nska (1999) pointed out that the [\ion{O}{4}] 26 \micron\ line 
could be easily formed by photoionization in `Wolf-Rayet galaxies'; however, their examples 
are all low metallicity, which the GC is not. 
Dale et al. (2009) and An et al. (2013) compared the [\ion{O}{4}] 26 \micron\ lines 
to other lines observed with the IRS and concluded 
that strong [\ion{O}{4}] 26 \micron\ lines are most commonly found in AGN. 
In particular, An et al. (2013) observed that the line ratios found in the GC 
are not compatible with those found in LINER galaxies or AGN; 
they preferred the starburst designation for the GC.
However, there is still the problem that the high excitation [\ion{O}{4}] lines 
are not seen in any of the giant \ion{H}{2} regions that are not low metallicity; 
there must be some way to ionize the gas in addition to the shocks in starburst outflows.

Correcting for the larger numbers of ionizing photons of Table~13, 
I estimate from the best-fit models that 
the X-ray luminosities for the four regions are 
%$10^{38.6}$, $10^{38.6}$, $10^{38.7}$, and $10^{38.7}$ erg s$^{-1}$ for 
{ $10^{38.9}$, $10^{38.7}$}, $10^{38.7}$, and $10^{38.7}$ erg s$^{-1}$ for 
the Quintuplet Cluster region, the Arched Filaments, Sgr B1, and Sgr C, respectively.
This distribution is surprisingly flat across the GC --- I conclude that the source of 
the 54 -- 77 eV photons ionizing O$^{3+}$ is not the Sgr A* black hole. 
The models, of course, do not have exactly the same $N_{\rm Lyc}$ and density as the estimates 
in Tables 13 and 12, respectively; however, if models are run with values of $N_{\rm Lyc}$ and $N_p$ 
from the observations and $R_{\rm inner}$ (for $R_S$) and filling factor $f$ adjusted using Equation (2) 
to produce the same value of $U$, 
the new models produce almost the same forbidden line ratios and almost the same $chisq$
if the X-ray luminosity is scaled by the same factor as $N_{\rm Lyc}$.

After adjusting for the actual numbers of ionizing photons, then, 
the best-fitting models for the Arched Filaments, Sgr B1, and Sgr C 
all have ages of $10^{6.6}$ to $10^{6.7}$ yrs and X-ray luminosities of $\sim 10^{38.7}$ 
{ to $\sim 10^{38.9}$} erg s$^{-1}$.
For producing the observed triply ionized oxygen, only the energy range of $\sim 55 - 100$ eV is important.
Integrating the $10^{6.6}$ yr SED with X-ray blackbody $T_{\rm BB} = 10^{6.5}$~K and
luminosity $10^{38.5}$ erg s$^{-1}$, 
I find that the integrated luminosity for this energy range is $5.7 \times 10^{35}$ erg s$^{-1}$.
These extreme ultraviolet (EUV) photons are completely absorbed by hydrogen gas in the intervening ISM 
and cannot be detected from Earth.
In the models they have been represented by the Rayleigh-Jeans tail of the black-body function 
of the added X-rays (Figure 7), 
but there is no reliable way to extrapolate this SED to energies observable with {\it Chandra}
or some other X-ray telescope;
bremsstrahlung SEDs, for example, could have a much lower integrated luminosity 
than the blackbody SEDs of our models.
In fact, a thermal emission model could be tweaked to fit the observed 3 -- 10 keV {\it Suzaku} observations 
of the Arches Cluster (Tsujimoto et al. 2007).

The source of the diffuse X-ray emission seen in the GC in both soft 
and hard X-ray bands has no conclusive identification
(e.g., Ponti et al. 2015). 
For example, Yusef-Zadeh et al. (2002) suggested that low-energy cosmic ray electrons 
interact with the gas, thereby producing the diffuse bremsstrahlung X-ray emission as well as 
fluorescent line emission, 
whereas Muno et al. (2004) and Park et al. (2004) suggest the energy source 
of the softer component of the emission is shocks from supernovae or the winds of O and WR stars. 

On the other hand, as spatial resolution and sensitivity due to deep integrations improved, 
more and more of the apparently diffuse hard X-ray emission has been resolved into point sources: 
Muno et al. (2009) found 9017 point sources with {\it Chandra} in the 2 -- 8 keV range 
and Hong et al. (2016) found 77 in the harder 3 -- 79 keV X-ray range with {\it NuSTAR}.
Most of these point sources are thought to be accreting white dwarfs (e.g., cataclysmic variables, CV,  
Muno et al. 2006; see also the review of Mukai 2017) or binaries with active coronae.
Further observations of CVs with {\it Chandra}, {\it Swift}, and {\it Suzaku}  
indicate that they could indeed make up a significant contribution 
to the diffuse hard X-ray emission from the Galactic Ridge (e.g., Revnivtsev et al. 2009; Reis et al. 2013)
but not to the central GC itself (Nobukawa et al. 2016).
Not all of the {\it Chandra} point sources are CVs, however. 
Thanks to the high spatial resolution of {\it Chandra}, NIR counterparts have been identified 
for some of these point sources, and some of these have been determined 
to be O supergiants or WR stars by NIR spectroscopy (Mauerhan et al. 2010).

I conclude that the EUV photons that triply ionize oxygen 
to produce the observed [\ion{O}{4}] 26 \micron\ line 
are probably the low-energy tails of the X-ray spectrum produced by the known sources in the GC: 
accreting white dwarfs, hot O and WR stars with winds, and supernova remnants. 
This would explain why the [\ion{O}{4}] line intensities are distributed widely across the GC --- 
they are correlated with the mass distribution of the inner parts of the Galaxy
and not with the locations of the star forming regions such as Sgr B and Sgr C.

\section{Summary and Conclusions}

Spectra taken of the Galactic Center region with all four modules of the {\it Spitzer} IRS were 
downloaded from the {\it Spitzer} Heritage Archive and analyzed with  
SMART { and/or} CUBISM.
% (high resolution or low resolution modules, respectively).
From these spectra, intensities were measured over a large fraction of the GC of the lines 
H$_2$ S(0), S(1), S(2), and S(7), \ion{H}{1} 7--6 12.37 \micron, 
[\ion{O}{4}] 26 \micron, [\ion{Ne}{2}] 12.8 \micron, [\ion{Ne}{3}] 15.6 \micron, [\ion{Si}{2}] 34 \micron, 
[\ion{S}{1}] 25.3 \micron, [\ion{S}{3}] 18.7 and 33 \micron, [\ion{S}{4}] 10.5 \micron, [\ion{Cl}{2}] 14.4 \micron, 
[\ion{Ar}{2}] 6.98 \micron, [\ion{Fe}{2}] 26 \micron, and [\ion{Fe}{3}] 23 \micron.
Locally strong intensities were measured for the lines 
[\ion{Fe}{2}] 5.34 \micron\ and [\ion{S}{4}] 10.5 \micron\ 
but for most spectra only upper limits could be obtained for these lines
(in particular, the gas with strong [\ion{S}{4}] 10.5 \micron\ intensities may be foreground to the GC).
Extinction was estimated from the ratios of the [\ion{S}{3}] 18.7 and 33 \micron\ lines 
and from the depths of the 9.6 \micron\ silicate absorption features 
seen in the short-low module spectra. 
In general, it is seen that the [\ion{Ne}{3}] 15.6/[\ion{S}{3}] 18.7 \micron\ line ratio 
is fairly low, indicating ionization by late O stars.  
From this one can infer either a dearth of the highest mass main-sequence stars 
and/or substantial evolution away from the main sequence.  
All together, after rejection of those spectra affected by saturation, 
intensities measured from { 47,469} spectra are published in this paper (Tables 2 -- 7).

Serendipitously, seventeen of the locations on the sky, almost all very compact, show high excitation 
characteristic of planetary nebulae or high-velocity shocks (Tables 8 and 9). 
In addition to significantly stronger [\ion{O}{4}] 26 \micron\ line emission compared to 
their surroundings, these regions 
sometimes exhibit emission in the [\ion{Ne}{5}] 14.3 and 24.3 \micron\ lines.
{ I suggest that the high excitation sources that have compact radio counterparts 
be investigated as candidate planetary nebulae;
the shocked sources may be indicators of outflows from so-far undetected, isolated massive hot stars 
that have been modeled as escapees from the Arches or Quintuplet Clusters by Habibi et al. (2014).}
At the other extreme, emission from low-velocity shocks, 
characterized by emission in the [\ion{S}{1}] 25.3 \micron\ line, 
was detected in { 24} locations on the sky (Table 10).
Twenty-three candidate YSOs or sources behind dense molecular clouds 
with ice absorption features at 6.0 and 6.8 \micron\ were also detected (Table 11).

Electron densities and ionic abundances were estimated from the line intensities
for the GC \ion{H}{2} regions usually named Sgr B1, Sgr C, the Arched Filaments, and 
the Sickle, here called the Quintuplet Region to include somewhat more area on the sky.
{ The average densities ranged from 270 to 310 cm$^{-3}$.}
%The lowest average densities, 90 and 200 cm$^{-3}$, were estimated for Sgr B1 and Sgr C, respectively. 
I conclude that the hot stars in these \ion{H}{2} regions are not 
ionizing gas from local molecular clouds but that the natal molecular { clouds have} already been dispersed 
by the strong winds from O and WR stars, and possibly a supernova in Sgr B1.
{ Other possibilities include the idea that 
the ionized gas consists of the lit-up edges of dense molecular clouds,  
particularly in those regions close to Sgr A (e.g., Lang et al. 2001).}
%Average electron densities are higher for the Quintuplet Region and the Arched Filaments,
%including the base of the Filaments towards Sgr A, here called the GC Filaments:
%these average densities are estimated as 390, 600, and 420 cm$^{-3}$, respectively. 
%It is suggested that this gas consists of the lit-up edges of the dense molecular clouds 
%close to Sgr A (e.g., Lang et al. 2001). 

The computed ionic abundances range from 
{ 1.55 -- $1.97 \times 10^{-4}$ for the (Ne$^+$ + Ne$^{++}$)/H$^+$ ratio 
and 0.84 -- $2.09 \times 10^{-5}$ for the (S$^{++}$ + S$^{3+}$)/H$^+$ ratio.} 
Although the former ratio is a good approximation to the Ne/H ratio, 
there being very little neutral or triply ionized neon in \ion{H}{2} regions, 
the latter ratio needs some sort of correction for singly and quadruply ionized sulfur. 
S$^+$ is especially prevalent in regions with very low ionization parameters 
and is probably the reason why the measurement for Sgr C diverges so much 
from that of the other \ion{H}{2} regions (Table 12, Figure~3b). 
The Ar$^+$/Ne$^+$ abundance ratio was measured to be { $\sim 0.032$}; 
since the median of the ratio of the ionization fractions of Ar$^+$ and Ne$^+$ 
is measured to be 0.90 for the computed models, 
the inferred Ar/Ne abundance ratio is $\sim 10$\% larger, or { $\sim 0.036$}.

I conclude from these measurements that the Galactic Center,
{ with Ne/H, S/H, and Ar/H ratios of $\sim 1.7 \times 10^{-4}$, $\sim 1.9 \times 10^{-5}$, and $\sim 6.2 \times 10^{-6}$, respectively,} 
has { abundances somewhat higher than Solar abundances}  
(Ekstr\"om et al. 2012 assumed $1.29 \times 10^{-4}$ for Ne/H, which, { like Ar/H}, cannot be measured in the Sun, 
and Scott et al. 2015 inferred $1.35 \times 10^{-5}$ for S/H 
from their analysis of Solar spectra).
The average measurements of Si$^+$/H$^+$ range from { 1.17 to $1.46 \times 10^{-5}$};
{ after correcting for higher ionization stages, 
the average value of gas-phase Si/H equals $2.4 \times 10^{-5}$,
whereas the Solar value of Si/H equals $3.3 \times 10^{-5}$ (Scott et al. 2015).
Thus, even though the estimated gas-phase abundance of Si/H is still below that of the Solar Si/H ratio, 
it is }
much more than what is often assumed for the abundances used in modeling \ion{H}{2} regions or PDRs. 

Because the observed [\ion{Ne}{3}] 15.6/[\ion{S}{3}] 18.7 \micron\ line ratios are 
very sensitive to the input SEDs of the ionizing stars 
and the [\ion{S}{3}] 33/[\ion{Si}{2}] 34 \micron\ line ratios are 
very sensitive to the ionization parameters, 
I performed tests of both by modeling the \ion{H}{2} region line ratios with Cloudy 
{ (Ferland et al. 2017)}.
This is a well known procedure for estimating the effective temperatures 
of individual stars exciting small \ion{H}{2} regions and/or testing the SEDs produced 
by various stellar atmosphere codes (e.g., Simpson et al. 2004; Rubin et al. 2008). 
Here for the first time I estimate the ages of the four GC massive star clusters 
by employing the composite spectra suitable for the clusters 
from Starburst99 (Leitherer et al. 2014). 

There are two significant results: 
(1) { The ages estimated for the clusters, $4.6$ Myr for Sgr B1, $4.5$ Myr for the Arches Cluster, 
$3.0$ Myr for the Quintuplet Cluster, and $3.8$ Myr for Sgr C, 
do not correspond to the formation sequence predicted by recent models 
of the gas streams in the GC by Kruijssen et al. (2015).
In the orbiting gas stream model, the positions of Sgr B1 and Sgr C past pericenter approach to Sgr A*  
correspond to ages of  
$\sim 1.5$ and $\sim 3.6$ Myr, respectively (Barnes et al. 2017; Kruijssen et al. 2015);
the molecular clouds at the locations of the Arches and Quintuplet Clusters 
are intermediate (Kruijssen et al. 2015).
The two clusters 
must have formed a full orbit earlier than their current positions would suggest
(Kruijssen et al. 2015; Habibi et al. 2014).

Considering the overlap in models of different ages seen in Figures 10 -- 13 
and the spread in acceptable models in Table~13,
I estimate that there is an uncertainty of $\sim 0.5$ Myr in the ages estimated for the clusters 
through the fitting of these models (that is, one step in either direction in this fairly coarse model grid).
With these uncertainties, the only real outlier among the four sources is Sgr B1. 
Here the 4.6 Myr age is not in particular disagreement with its dispersed appearance,
but it is in disagreement with the velocity measurements that indicate that Sgr B1 is 
part of the same system with the known much-younger Sgr B2 (e.g., Lang et al. 2010; Mehringer et al. 1992) 
or if not, that it is on the back stream past Sgr B2 but still with a positional age of only 1.5 Myr
(Barnes et al. 2017).
%I suggest that the velocity agreement is accidental and that the Sgr B1 stars are, in fact, 
I suggest that the velocity agreement is accidental and that Sgr B1 is, in fact, 
of a larger age and lies on a continuation of the stream from Sgr C past Sgr A to higher Galactic longitudes
on the front side of the orbit with its lesser extinction.

Another problem concerns the respective ages for the Arches and Quintuplet Clusters, 
estimated by Schneider et al. (2014) to be $3.5 \pm 0.7$  and $4.8 \pm 1.1$ Myr, respectively,
in contrast to the ages estimated from the model fits.
Comparing the [\ion{Ne}{3}] 15/[\ion{S}{3}] 19 \micron\ ratios for the Arched Filament region 
and the Quintuplet Cluster region seen in Figures 10 and 11, 
the excitation is without doubt substantially higher in the gas excited by the Quintuplet Cluster compared 
to the gas excited by the Arches Cluster. 
If this excitation is caused by stellar photons, the only possible conclusion is that the Quintuplet Cluster
contains hotter stars than the Arches Cluster, and in models of the spectral evolution of the SEDs 
produced by massive clusters such as predicted by Starburst99 (e.g., Leitherer et al. 2014), 
hotter stars mean the cluster is younger.
However, if the Quintuplet Cluster really is older, 
as seems to be convincingly shown by Schneider et al. (2014),
its higher excitation would have to be produced by some means other than stellar photons.
Shocks from stellar winds or a possible supernova remnant are the most likely possibilities.
Thus I conclude that the current cluster models as applied to these \ion{H}{2} regions are inadequate.
Consequently, because our detailed studies of the Galaxy should be informing us 
on the conditions in the rest of the Universe, 
I suggest that such additional sources of excitation should be considered
by modelers of extragalactic \ion{H}{2} regions in starburst galaxies.
}

(2) It is not possible to predict {\it all} the observed line ratios from the SEDs 
from the 2014 version of Starburst99 --- 
additional photons with energies $\gtrsim 55$ eV are needed to produce 
the wide-spread detected [\ion{O}{4}] 26 \micron\ line.
The 55 -- 100 eV photons required for the models to fit the data 
are probably the low energy tails of the X-rays observed by {\it Chandra},  
{\it XMM-Newton},  {\it Suzaku}, etc.
These X-rays could have several possible origins: 
the low-mass X-ray sources of the old stars populating most of the mass of the GC, 
X-rays from the energetic shocks of supernova remnants, 
or the Starburst99 model SEDs inadequately representing the interactions of the stellar winds 
of the multiple massive stars found in the massive clusters of their models.
Probably there are contributions from all these sources of high energy photons in the GC.

Future observers may well want to use the observations discussed in this paper 
to plan additional observations of GC sources. 
In addition to that found at numerous ground-based observatories, 
there will be MIR spectroscopic capability out to 28.5 \micron\ with 
the Mid-Infrared Instrument (MIRI) on NASA's {\it James Webb Space Telescope} ({\it JWST}), 
to be launched in { 2020}. 
For longer wavelengths it will be necessary to use the instruments on 
the Stratospheric Observatory for Infrared Astronomy ({\it SOFIA}), 
such as for additional observations of the [\ion{S}{3}] 33 and [\ion{Si}{2}] 34 \micron\ lines, 
which are very bright at many locations. 
Above all, what is needed is a more complete identification 
of the OB and WR stars that are not in clusters but 
that contribute the diffuse ionizing photons that are ubiquitous in the GC.

\acknowledgments

This work is based on observations made with the {\it Spitzer
Space Telescope}, which is operated by the Jet Propulsion Laboratory,
California Institute of Technology, under a contract with NASA.
The IRS was a collaborative venture between Cornell University
and Ball Aerospace Corporation
funded by NASA through the Jet Propulsion Laboratory and Ames Research Center.
SMART was developed by the IRS Team at Cornell University and is available
through the Spitzer Science Center at Caltech.
This research has made use of the NASA/IPAC Infrared Science Archive, which is operated 
by the Jet Propulsion Laboratory, California Institute of Technology, 
under contract with the National Aeronautics and Space Administration.
I acknowledge {\it SOFIA} grant 04\_0113 for the payment of the page charges.  
I thank Angela Cotera for reading the manuscript 
{ and the referee for the thoughtful and detailed comments that improved the presentation in this paper}.

\vspace{5mm}
\facility{Spitzer(IRS)}

\software{Cloudy \citep{cloudy17},
CUBISM \citep{cubism},
PAHFIT \citep{pahfit}
}

\end{document}

%% file: tab1.tex
\setcounter{table}{0}
\begin{table*}
\centering
\begin{minipage}{130mm}
\caption{Line Parameters
}
\begin{tabular}{@{}lccccll@{}}
\hline
\hline
Line & Wavelength & $IP$\tablenotemark{a} & $E_{upper}$ & $\tau_\lambda/\tau_{9.6}$ & References & {Notes} \\
     & (\micron)  & (eV) & (cm$^{-1}$) &                        &&  \\
\hline
H$_2$ S(0) & 28.219  & 0 & 354.37 & 0.369 & 19 & 1 \\ %   Wolniewicz et al. (1998)
H$_2$ S(1) & 17.035  & 0 & 705.69 & 0.502 & 19 & 1  \\ %    ``
H$_2$ S(2) & 12.279  & 0 & 1168.80 & 0.344 & 19 & 1  \\ %   ``
%H$_2$ S(3) & 9.665  & 0 & 1740.36 & 0.997 & 19 & 1  \\ %   ``
%H$_2$ S(5) & 6.909  & 0 & 3187.64 & 0.273 & 19 & 1  \\ %   ``
H$_2$ S(7) & 5.512  & 0 & 5002.04 & 0.290 & 19 & 1  \\ %   ``
\mbox{H\,{\sc i}} 7-6    & 12.372  & 13.60 & 107440.45 & 0.332 & 17 & 2  \\ %   Storey \& Hummer 1995
\mbox{[O\,{\sc iv}]}  $^2$P$_{3/2}-^2$P$_{1/2}$   & 25.890 & 54.94 & 386.245  & 0.404 & 6, 12 & 3, 4, 7   \\ % F97 Liang et al.
\mbox{[Ne\,{\sc ii}]}  $^2$P$_{1/2}-^2$P$_{3/2}$  & 12.814 & 21.56 & 780.42 & 0.313 & 9, 11 & 2   \\ % KL95  Griffin et al.01
\mbox{[Ne\,{\sc iii}]} $^3$P$_1-^3$P$_2$       & 15.555 & 40.96 & 642.88 & 0.417 & 6, 13 & 2, 4, 7    \\ %  F97 McLau2011
\mbox{[Ne\,{\sc iii}]} $^3$P$_0-^3$P$_1$       & 36.014 & 40.96 & 920.55 & 0.304 & 6, 13 & 2, 4    \\ %  F97 McLau2011
\mbox{[Ne\,{\sc v}]} $^3$P$_1-^3$P$_0$         & 24.318 & 97.19 & 298.68 & 0.428 & 5, 6 & 4   \\ % F97 Dance2013
\mbox{[Ne\,{\sc v}]} $^3$P$_2-^3$P$_1$         & 14.322 & 97.19 & 833.06 & 0.309 & 5, 6 & 4   \\ % F97 Dance2013 
\mbox{[Si\,{\sc ii}]} $^2$P$_{3/2}-^2$P$_{1/2}$  & 34.815 & 8.15 & 287.23 & 0.306 & 1, 6 & 1, 2, 8   \\ %  F97 DK91
\mbox{[P\,{\sc iii}]}  $^2$P$_{3/2}-^2$P$_{1/2}$  & 17.885 & 19.77 & 559.13 & 0.539 & 16 & 2   \\ % SS99
\mbox{[S\,{\sc i}]} $^3$P$_1-^3$P$_2$          & 25.245 &     0 & 396.12 & 0.414 & 10 & 5   \\ % Haas et al.  
\mbox{[S\,{\sc iii}]} $^3$P$_1-^3$P$_0$         & 33.481 & 23.34 & 298.68 & 0.311 & 6, 8 & 2, 6, 7, 8   \\ %   F97 Grieve14
\mbox{[S\,{\sc iii}]} $^3$P$_2-^3$P$_1$         & 18.713 & 23.34 & 833.06 & 0.548 & 8, 11 & 2, 6, 7   \\ %  KL95 Grieve14
\mbox{[S\,{\sc iv}]}  $^2$P$_{3/2}-^2$P$_{1/2}$  & 10.511 & 34.86 & 951.43 & 0.777 & 16 & 2   \\ %   SS99
\mbox{[Cl\,{\sc ii}]} $^3$P$_1-^3$P$_2$       & 14.368 & 12.97 & 696.00 & 0.309 & 18 & 2   \\ % T04
\mbox{[Ar\,{\sc ii}]} $^2$P$_{1/2}-^2$P$_{3/2}$  & 6.985 & 15.76 & 1431.58 & 0.272 & 15 & 2   \\ % Pelan & Berrington 1995
\mbox{[Ar\,{\sc iii}]} $^3$P$_1-^3$P$_2$       & 8.991 & 27.63 & 1112.18 & 0.695 & 11, 14 & 2   \\ % F01 MB09
\mbox{[Ar\,{\sc iii}]} $^3$P$_0-^3$P$_1$       & 21.830 & 27.63 & 1570.26 & 0.467 & 7, 14 & 2   \\ % F01 MB09
\mbox{[Na\,{\sc iii}]}  $^2$P$_{1/2}-^2$P$_{3/2}$  & 7.318 & 47.29 & 1366.55 & 0.271 & 6 & 4   \\ % F97
\mbox{[Fe\,{\sc ii}]} $^6$D$_{7/2}-^6$D$_{9/2}$ & 25.988 & 7.90 & 384.79 & 0.403 & 4 & 1   \\ % Bautista2015
\mbox{[Fe\,{\sc ii}]} $^4$F$_{9/2}-^6$D$_{9/2}$ & 5.340 & 7.90 & 1872.57 & 0.295 & 4 & 1, 5   \\ % Bautista2015
\mbox{[Fe\,{\sc ii}]} $^4$F$_{7/2}-^4$F$_{9/2}$ & 17.936 & 7.90 & 2340.10 & 0.540 & 4, 11 & 1, 2, 5    \\ % KL95 Bautista2015
\mbox{[Fe\,{\sc iii}]} $^5$D$_3-^5$D$_4$       &  22.925 & 16.20 & 436.21 & 0.448  & 2, 3 & 2   \\ % Bautista2010
\hline
\end{tabular}
\tablenotetext{a}{Ionization Potential ($IP$) is the energy required to produce the ground state of the molecule or ion.}
\tablerefs{
(1) Aggarwal \& Keenan (2014), % [Si II] 
(2) Badnell \& Ballance (2014), % [Fe III]
(3) Bautista et al. (2010), % Fe III As and omegas
(4) Bautista et al. (2015), % Fe II As and omegas
%(1) Bautista \& Pradhan (1996),  % Fe II
(5) Dance et al. (2013), % [Ne V]
% (2) Dufton \& Kingston (1991), % Si II
(6) Feuchtgruber et al. (1997),
(7) Feuchtgruber et al. (2001),
% Galavis et al. (1995) % [Ar III] used by Cloudy
(8) Grieve et al. (2014), % [S III] 
(9) Griffin, Mitnik, \& Badnell (2001), % [Ne II]
(10) Haas et al. (1991) % [S I] 
(11) Kelly \& Lacy (1995),
(12) Liang et al. (2012), % [O IV]
(13) McLaughlin et al. (2011), % [Ne III]
(14) Munoz Burgos et al. (2009), % MB09 [Ar III] 
% (15) Nussbaumer \& Storey (1988), %[Fe II] As
(15) Pelan \& Berrington (1995), % [Ar II] 
% (10) Quinet (1996), % Fe III As
% (11) Quinet, Le Dorneuf, \& Zeippen (1996), % [Fe II] As
% (12) Ramsbottom et al. (2005),  % Fe II
(16) Saraph \& Storey (1999), % [S IV], [P III]
(17) Storey \& Hummer (1995), % H 7-6
(18) Tayal (2004), % [Cl II]
% (16) Tayal (2006), % O IV
(19) Wolniewicz, Simbotin, \& Dalgarno (1998), % H2
% (19) Zhang (1996), % Fe III
% Krueger \& Czyzak (1970) % [P III] used by Cloudy
}
\tablecomments{{ 
The following notes describe the various uses of each line that was measured in the Galactic Center:
(1) PDR and low-energy shock properties.
(2) Ionic abundances.
(3) High-energy excitation by X-rays or shocks.
(4) High-energy excitation by shocks or the central stars of planetary nebulae.
(5) Low-energy shocks.
(6) Electron densities and lower limits to the extinction.
(7) Indicators of the radiation-field spectrum used for estimates of stellar $T_{\rm eff}$ or age through comparison to models.
(8) Indicators of the radiation-field energy-density used for estimates of distances of the ionized gas from the exciting stars.
}}
\end{minipage}
\end{table*}

%\end{document}

%% file: tab2.tex
%Only a few positions near Sgr AE have the Fe or H2 S(7) lines. Omit from this table unless there is a detection from lines-isap. There are LOTs of spectra with Ar or Ne+H2 S(3) that are really bad at the shorter wavelengths due to saturation of the peak-up arrays and so even the Fe or S(7) errors are bad.

\setcounter{table}{1}
%\begin{longrotatetable}
%\begin{table*}
\begin{splitdeluxetable*}{lcccccccBcccc}
\tabletypesize{\scriptsize}
\tablecolumns{10}
\tablewidth{170mm}
%\centering
%\begin{minipage}{185mm}
\tablecaption{Intensities of Lines Observed with the Short-Low Module, Second Order (SL2)
}
%\begin{tabular}{@{}lccccccccccc@{}}
%\hline
%\hline
\tablehead{\colhead{Position} & \colhead{RA} & \colhead{Dec} & \colhead{$\tau_{9.6 \micron}$} & \colhead{AORKEY} & \colhead{Aperture} & \colhead{[Fe\,{\sc ii}] 5.34 \micron\ } & \colhead{Error} & \colhead{H$_2$ S(7) 5.51 \micron\ } & \colhead{Error} & \colhead{[Ar\,{\sc ii}] 6.98 \micron\tablenotemark{a}} & \colhead{Error} \\
 \colhead{}    & \colhead{(J2000)}  & \colhead{(J2000)} & \colhead{}  & \colhead{} & \colhead{(sr)} & \colhead{(W m$^{-2}$ sr$^{-1}$)} & \colhead{(W m$^{-2}$ sr$^{-1}$)} & \colhead{(W m$^{-2}$ sr$^{-1}$)} & \colhead{(W m$^{-2}$ sr$^{-1}$)}  & \colhead{(W m$^{-2}$ sr$^{-1}$)} & \colhead{(W m$^{-2}$ sr$^{-1}$)} }
%Position & RA & Dec & $\tau_{9.6 \micron}$ & AORKEY & Aperture & [Fe\,{\sc ii}] 5.34 \micron\ & Error & H$_2$ S(7) 5.51 \micron\ & Error & [Ar\,{\sc ii}] 6.98 \micron\tablenotemark{a} & Error \\
%     & (J2000)  & (J2000) & & & (sr) & (W m$^{-2}$ sr$^{-1}$) &  (W m$^{-2}$ sr$^{-1}$) &  (W m$^{-2}$ sr$^{-1}$) & (W m$^{-2}$ sr$^{-1}$) &  (W m$^{-2}$ sr$^{-1}$) &  (W m$^{-2}$ sr$^{-1}$)   \\
%\hline
\startdata
G359.1763$-$0.0247 & 17 43 44.46 & $-$29 39 04.4 &  2.82 & 23970304 &  4.831e$-$10 & ... & ... & ... & ... &   2.822e$-$07 & 2.695e$-$08 \\
G359.1772$-$0.0240 & 17 43 44.44 & $-$29 39 00.7 &  2.82 & 23970304 &  4.828e$-$10 & ...  & ... & ... & ... &   1.347e$-$07 & 3.165e$-$08 \\
G359.1788$-$0.0228 & 17 43 44.39 & $-$29 38 53.4 &  2.82 & 23970304 &  4.832e$-$10 & ... & ... & ... & ... &   1.541e$-$07 & 2.040e$-$08 \\
G359.1796$-$0.0222 & 17 43 44.36 & $-$29 38 49.7 &  2.82 & 23970304 &  4.815e$-$10 & ... & ... & ... & ... &   1.519e$-$07 & 1.274e$-$08 \\
G359.1813$-$0.0210 & 17 43 44.31 & $-$29 38 42.3 &  2.82 & 23970304 &  4.828e$-$10 & ... & ... & ... & ... &   2.359e$-$07 & 8.217e$-$08 \\
%\hline
%\end{tabular}
\enddata
\tablenotetext{a}{May be blended with H$_2$ S(5) 6.91 \micron.}
\tablecomments{(This table is available in its entirety in machine-readable form.)}
%\end{minipage}
%\end{table*}
\end{splitdeluxetable*}

%% file: tab3.tex
%Only a few positions near Sgr AE have the Fe or H2 S(7) lines. Omit from this table unless there is a detection from lines-isap. There are LOTs of spectra with Ar or Ne+H2 S(3) that are really bad at the shorter wavelengths due to saturation of the peak-up arrays and so even the Fe or S(7) errors are bad.

\setcounter{table}{2}
\begin{deluxetable*}{lccccccccc}
\tabletypesize{\scriptsize}
\tablecolumns{10}
\tablewidth{0pt}
%\begin{table*}
%\centering
%\begin{minipage}{185mm}
\tablecaption{Intensities of Lines Observed with the Short-Low Module, First Order (SL1)
}
%\begin{tabular}{@{}lccccccccc@{}}
%\hline
%\hline
%Position & RA & Dec & $\tau_{9.6 \micron}$ & AORKEY & Aperture & H$_2$ S(2) 12.28 \micron\ & Error & [Ne\,{\sc ii}] 12.81 \micron\ & Error \\
%     & (J2000)  & (J2000) & & & (sr) &  (W m$^{-2}$ sr$^{-1}$) &  (W m$^{-2}$ sr$^{-1}$)  &  (W m$^{-2}$ sr$^{-1}$) &  (W m$^{-2}$ sr$^{-1}$)   \\
\tablehead{\colhead{Position} & \colhead{RA} & \colhead{Dec} & \colhead{$\tau_{9.6 \micron}$} & \colhead{AORKEY} & \colhead{Aperture} & \colhead{H$_2$ S(2) 12.28 \micron\ } & \colhead{Error} & \colhead{[Ne\,{\sc ii}] 12.81 \micron} & \colhead{Error} \\
 \colhead{}    & \colhead{(J2000)}  & \colhead{(J2000)} & \colhead{}  & \colhead{} & \colhead{(sr)} & \colhead{(W m$^{-2}$ sr$^{-1}$)} & \colhead{(W m$^{-2}$ sr$^{-1}$)} & \colhead{(W m$^{-2}$ sr$^{-1}$)} & \colhead{(W m$^{-2}$ sr$^{-1}$)} }
%\hline
\startdata
G359.1576$-$0.0386 &  17 43 45.03 &  $-$29 40 28.0 &   2.47 & 23970304 &   4.818e$-$10 & ... & ...  & 3.798e$-$07 &  5.794e$-$08 \\
G359.1585$-$0.0380 &  17 43 45.00 &  $-$29 40 24.3 &   2.47 & 23970304 &   4.830e$-$10 &  ... & ...  &          4.180e$-$07 &  2.685e$-$08 \\
G359.1593$-$0.0373 &  17 43 44.98 &  $-$29 40 20.6 &   2.47 & 23970304 &   4.830e$-$10 & ... & ...  &         4.512e$-$07 &  2.768e$-$08 \\
G359.1601$-$0.0367 &  17 43 44.95 &  $-$29 40 17.0 &   2.47 & 23970304 &   4.830e$-$10 & ... & ...  &          3.904e$-$07 &  3.603e$-$08 \\
G359.1609$-$0.0361 &  17 43 44.93 &  $-$29 40 13.3 &   2.47 & 23970304 &   4.830e$-$10 &   5.603e$-$08 &  1.358e$-$08 &  3.787e$-$07 &  4.258e$-$08 \\
\enddata
%\hline
%\end{tabular}
\tablecomments{(This table is available in its entirety in machine-readable form.)}
%\end{minipage}
\end{deluxetable*}

%% file: tab4.tex
\setcounter{table}{3}
%\begin{table*}
%\centering
%\begin{minipage}{185mm}
\begin{splitdeluxetable*}{lccccccccBccccccccBcccccccc}
\tabletypesize{\scriptsize}
\tablecolumns{25}
\tablewidth{140mm}
\tablecaption{Intensities of Lines Observed with the Short-High Module (SH)
}
%\begin{tabular}{@{}lcccccccccccc@{}}
%\hline
%\hline
%Position & RA & Dec & $\tau_{9.6 \micron}$ & AORKEY & [S\,{\sc iv}] 10.5 \micron\ & H$_2$ S(2) 12.3 \micron\ & H 7-6 12.4 \micron\ & [Ne\,{\sc ii}] 12.8 \micron\ & [Cl II] 14.4 \micron\ & [Ne\,{\sc iii}] 15.6 \micron\ & H$_2$ S(1) 17.0 \micron\ & [P\,{\sc iii}] 17.9 \micron\  & [Fe\,{\sc ii}] 17.9 \micron\  & [S\,{\sc iii}] 18.7 \micron\ \\
%     & (J2000)  & (J2000) & & & (W m$^{-2}$ sr$^{-1}$) &  (W m$^{-2}$ sr$^{-1}$) & (W m$^{-2}$ sr$^{-1}$) & (W m$^{-2}$ sr$^{-1}$) &  (W m$^{-2}$ sr$^{-1}$) &  (W m$^{-2}$ sr$^{-1}$) & (W m$^{-2}$ sr$^{-1}$) & (W m$^{-2}$ sr$^{-1}$)   &  (W m$^{-2}$ sr$^{-1}$) & (W m$^{-2}$ sr$^{-1}$) & (W m$^{-2}$ sr$^{-1}$)   \\
\tablehead{\colhead{Position} & \colhead{RA} & \colhead{Dec} & \colhead{$\tau_{9.6 \micron}$} & \colhead{AORKEY} & \colhead{[S\,{\sc iv}] 10.5 \micron\ } & \colhead{Error} & \colhead{H$_2$ S(2) 12.3 \micron\ } & \colhead{Error} & \colhead{H 7--6 12.4 \micron} & \colhead{Error}  & \colhead{[Ne\,{\sc ii}] 12.8 \micron} & \colhead{Error}  & \colhead{[Cl\,{\sc ii}] 14.4 \micron} & \colhead{Error}  & \colhead{[Ne\,{\sc iii}] 15.6 \micron} & \colhead{Error}  & \colhead{H$_2$ S(1) 17.0 \micron} & \colhead{Error} & \colhead{[P\,{\sc iii}] 17.9 \micron} & \colhead{Error}   & \colhead{[Fe\,{\sc ii}] 17.9 \micron} & \colhead{Error} & \colhead{[S\,{\sc iii}] 18.7 \micron} & \colhead{Error} \\
 \colhead{}    & \colhead{(J2000)}  & \colhead{(J2000)} & \colhead{}  & \colhead{} & \colhead{(W m$^{-2}$ sr$^{-1}$)} & \colhead{(W m$^{-2}$ sr$^{-1}$)} & \colhead{(W m$^{-2}$ sr$^{-1}$)} & \colhead{(W m$^{-2}$ sr$^{-1}$)}  & \colhead{(W m$^{-2}$ sr$^{-1}$)} & \colhead{(W m$^{-2}$ sr$^{-1}$)} & \colhead{(W m$^{-2}$ sr$^{-1}$)} & \colhead{(W m$^{-2}$ sr$^{-1}$)} & \colhead{(W m$^{-2}$ sr$^{-1}$)} & \colhead{(W m$^{-2}$ sr$^{-1}$)}  & \colhead{(W m$^{-2}$ sr$^{-1}$)} & \colhead{(W m$^{-2}$ sr$^{-1}$)}  & \colhead{(W m$^{-2}$ sr$^{-1}$)} & \colhead{(W m$^{-2}$ sr$^{-1}$)}  & \colhead{(W m$^{-2}$ sr$^{-1}$)} & \colhead{(W m$^{-2}$ sr$^{-1}$)}  & \colhead{(W m$^{-2}$ sr$^{-1}$)} & \colhead{(W m$^{-2}$ sr$^{-1}$)}  & \colhead{(W m$^{-2}$ sr$^{-1}$)} & \colhead{(W m$^{-2}$ sr$^{-1}$)} }
%\hline
%\hline
\startdata
G359.0411+0.1265 & 17 42 49.31 & $-$29 41 12.8 &  1.33 & 3826432 & ... & ... & 1.381e$-$07 & 1.867e$-$09 & 3.390e$-$08 & 1.877e$-$09 & 9.738e$-$08 & 9.730e$-$10 & ... &  ... & 2.071e$-$08  & 1.110e$-$09 & 1.945e$-$07 & 9.375e$-$10 & ... & ... & ... & ... &  2.451e$-$08 & 9.342e$-$10 \\
G359.4536$-$0.0535 & 17 44 31.30 & $-$29 25 48.6 &  3.93 & 3826432  & ... & ... &  8.326e$-$08 & 1.988e$-$09 & 1.514e$-$08 & 2.001e$-$09 & 2.355e$-$07 & 1.031e$-$09 & ... &... & 1.457e$-$08 & 9.408e$-$10 & 8.163e$-$08 & 9.014e$-$10 & ... & ... & ... & ... &  4.974e$-$08 &  1.088e$-$09 \\
G359.4990$-$0.0304 & 17 44 32.41 & $-$29 22 45.6 &  3.50 & 3826432 &    4.830e$-$09 &  1.964e$-$09 &  4.883e$-$08 &  2.093e$-$09 &  1.311e$-$08 &  1.874e$-$09 &  2.418e$-$07 &  9.693e$-$10 & ... & ... & 1.592e$-$08 &  1.144e$-$09 &  6.734e$-$08 &  8.855e$-$10 & ... & ... & ... & ... &  5.727e$-$08 &  1.181e$-$09 \\
G359.5842+0.0100 & 17 44 35.20 & $-$29 17 08.4 &  1.28 & 3826432 & ... & ... &  8.018e$-$08 &  2.272e$-$09 &  1.372e$-$08 &  1.938e$-$09 &  2.938e$-$07 &  1.310e$-$09 &  ... & ... & 1.180e$-$08 &  8.105e$-$10 &  1.093e$-$07 &  1.178e$-$09 & ... & ... & ... & ...  &  9.443e$-$08 &  1.089e$-$09 \\
G359.6789$-$0.1901 & 17 45 35.79 & $-$29 18 33.4 &  3.00 & 3826432 & ... & ... &  1.686e$-$07 &  2.059e$-$09 & ... & ... &  4.776e$-$07 &  1.686e$-$09 & ... & ... &  2.415e$-$08 &  8.805e$-$10 &  2.016e$-$07 &  1.055e$-$09 & ... & ...  & ... & ... &  3.199e$-$07 &  1.073e$-$09 \\
%\end{tabular}\
\enddata
\tablecomments{(This table is available in its entirety in machine-readable form.)}
%\end{minipage}
%\end{table*}
\end{splitdeluxetable*}

%% file: tab5.tex
\setcounter{table}{4}
%\begin{table*}
%\centering
\begin{splitdeluxetable*}{lcccccccBcccc}
\tabletypesize{\scriptsize}
\tablecolumns{10}
\tablewidth{140mm}
%\begin{minipage}{185mm}
\tablecaption{Intensities of Lines Observed with the Long-Low Module, Second Order (LL2)
}
%\begin{tabular}{@{}lcccccccc@{}}
%\hline
%\hline
%Position & RA & Dec & $\tau_{9.6 \micron}$ & AORKEY & Aperture & [Ne\,{\sc iii}] 15.6 \micron\  & H$_2$ S(1) 17.0 \micron\ & [S\,{\sc iii}] 18.7 \micron\ \\
%     & (J2000)  & (J2000) & & & (sr) & (W m$^{-2}$ sr$^{-1}$) &  (W m$^{-2}$ sr$^{-1}$) &  (W m$^{-2}$ sr$^{-1}$)   \\
\tablehead{\colhead{Position} & \colhead{RA} & \colhead{Dec} & \colhead{$\tau_{9.6 \micron}$} & \colhead{AORKEY} & \colhead{Aperture} & \colhead{[Ne\,{\sc iii}] 15.6 \micron\ } & \colhead{Error} & \colhead{H$_2$ S(1) 17.0 \micron\ } & \colhead{Error} & \colhead{[S\,{\sc iii}] 18.7 \micron} & \colhead{Error} \\
 \colhead{}    & \colhead{(J2000)}  & \colhead{(J2000)} & \colhead{}  & \colhead{} & \colhead{(sr)} & \colhead{(W m$^{-2}$ sr$^{-1}$)} & \colhead{(W m$^{-2}$ sr$^{-1}$)} & \colhead{(W m$^{-2}$ sr$^{-1}$)} & \colhead{(W m$^{-2}$ sr$^{-1}$)}  & \colhead{(W m$^{-2}$ sr$^{-1}$)} & \colhead{(W m$^{-2}$ sr$^{-1}$)} }
%\hline
\startdata
G359.1383+0.0216 &  17 43 28.08 &  $-$29 39 33.8 &   2.20 &  23974144 &   3.639e$-$09 &   5.209e$-$08 &  3.487e$-$09 &  1.849e$-$07 &  6.827e$-$09 &  5.481e$-$08 &  3.579e$-$09 \\
G359.1397+0.0191 &  17 43 28.86 &  $-$29 39 34.0 &   2.20 &  23974144 &   3.641e$-$09 &   4.581e$-$08 &  2.956e$-$09 &  1.830e$-$07 &  3.529e$-$09 &  5.430e$-$08 &  2.925e$-$09 \\
G359.1411+0.0167 &  17 43 29.64 &  $-$29 39 34.3 &   2.20 &  23974144 &   3.639e$-$09 &   4.955e$-$08 &  3.664e$-$09 &  1.777e$-$07 &  9.341e$-$09 &  6.025e$-$08 &  4.304e$-$09 \\
G359.1426+0.0143 &  17 43 30.42 &  $-$29 39 34.5 &   2.20 &  23974144 &   3.641e$-$09 &   4.795e$-$08 &  2.491e$-$09 &  1.717e$-$07 &  7.897e$-$09 &  5.635e$-$08 &  2.763e$-$09 \\
G359.1440+0.0118 &  17 43 31.20 &  $-$29 39 34.7 &   2.20 &  23974144 &   3.639e$-$09 &   3.825e$-$08 &  3.311e$-$09 &  1.719e$-$07 &  4.204e$-$09 &  5.076e$-$08 &  4.218e$-$09 \\
\enddata
%\hline
%\end{tabular}\
\tablecomments{(This table is available in its entirety in machine-readable form.)}
%\end{minipage}
%\end{table*}
\end{splitdeluxetable*}

%% file: tab6.tex
\setcounter{table}{5}
%\begin{table*}
\begin{splitdeluxetable*}{lcccccccBcccccc}
\tabletypesize{\scriptsize}
\tablecolumns{10}
\tablewidth{140mm}
%\centering
%\begin{minipage}{185mm}
\tablecaption{Intensities of Lines Observed with the Long-Low Module, First Order (LL1)
}
%\begin{tabular}{@{}lccccccccc@{}}
%\hline
%\hline
%Position & RA & Dec & $\tau_{9.6 \micron}$ & AORKEY & Aperture & [O\,{\sc iv}] 25.9 \micron \tablenotemark{a}  & H$_2$ S(0) 28.3 \micron\ & [S\,{\sc iii}] 33.5 \micron\ & [Si\,{\sc ii}] 34.8 \micron\ \\
%     & (J2000)  & (J2000) & & & (sr) & (W m$^{-2}$ sr$^{-1}$) & (W m$^{-2}$ sr$^{-1}$) &  (W m$^{-2}$ sr$^{-1}$) &  (W m$^{-2}$ sr$^{-1}$)   \\
\tablehead{\colhead{Position} & \colhead{RA} & \colhead{Dec} & \colhead{$\tau_{9.6 \micron}$} & \colhead{AORKEY} & \colhead{Aperture} & \colhead{[O\,{\sc iv}] 25.9 \micron\tablenotemark{a}} & \colhead{Error} & \colhead{H$_2$ S(0) 28.3 \micron\ } & \colhead{Error} & \colhead{[S\,{\sc iii}] 33.5 \micron} & \colhead{Error}  & \colhead{[Si\,{\sc ii}] 34.8 \micron} & \colhead{Error} \\
 \colhead{}    & \colhead{(J2000)}  & \colhead{(J2000)} & \colhead{}  & \colhead{} & \colhead{(sr)} & \colhead{(W m$^{-2}$ sr$^{-1}$)} & \colhead{(W m$^{-2}$ sr$^{-1}$)} & \colhead{(W m$^{-2}$ sr$^{-1}$)} & \colhead{(W m$^{-2}$ sr$^{-1}$)}  & \colhead{(W m$^{-2}$ sr$^{-1}$)} & \colhead{(W m$^{-2}$ sr$^{-1}$)}   & \colhead{(W m$^{-2}$ sr$^{-1}$)} & \colhead{(W m$^{-2}$ sr$^{-1}$)} }
%\hline
\startdata
G359.1649$-$0.0240 &  17 43 42.65 &  $-$29 39 38.2 & 2.47 & 23974144 & 3.641e$-$09 & ... & ... &     5.409e$-$08 &  3.979e$-$09 &  4.126e$-$07 &  1.184e$-$08 &  5.645e$-$07 &  2.550e$-$08 \\
G359.1663$-$0.0264 &  17 43 43.43 &  $-$29 39 38.4 & 2.47 & 23974144 & 3.641e$-$09 & ... & ... &     5.524e$-$08 &  2.602e$-$09 &  5.324e$-$07 &  2.831e$-$08 &  5.222e$-$07 &  3.122e$-$08 \\
G359.1678$-$0.0289 &  17 43 44.21 &  $-$29 39 38.7 & 2.47 & 23974144 & 3.639e$-$09 & ... & ... &     7.988e$-$08 &  2.882e$-$09 &  6.332e$-$07 &  3.068e$-$08 &  7.107e$-$07 &  3.467e$-$08 \\
G359.1692$-$0.0313 &  17 43 44.99 &  $-$29 39 38.9 & 2.47 & 23974144 & 3.637e$-$09 & ... & ... &     1.138e$-$07 &  2.965e$-$09 &  6.073e$-$07 &  6.346e$-$08 &  9.768e$-$07 &  3.270e$-$08 \\
G359.1706$-$0.0337 &  17 43 45.77 &  $-$29 39 39.1 & 2.47 & 23974144 & 3.641e$-$09 & ... & ... &     9.482e$-$08 &  3.702e$-$09 &  6.204e$-$07 &  6.129e$-$08 &  1.052e$-$06 &  7.553e$-$08 \\
\enddata
%\hline
%\end{tabular}\
\tablenotetext{a}{The [O\,{\sc iv}] 25.9 \micron\ line is blended with the [Fe\,{\sc ii}] 26.0 \micron\ line and cannot be distinguished at the resolution of the Long-Low (LL) module. 
There is a weak emission feature in most LL1 spectra whose intensity cannot be accurately measured on account of its low line/continuum ratio. 
Consequently, the estimated intensities of these features are not in this table; neither were they used in the model comparisons. 
All emission features at this wavelength identified as the [O\,{\sc iv}] 25.9 \micron\ line in this table are substantially stronger than any of the weak, probably mostly [Fe\,{\sc ii}], lines in the other spectra, and the centroid wavelength is 25.9 \micron\ or shorter.}
\tablecomments{(This table is available in its entirety in machine-readable form.)}
%\end{minipage}
%\end{table*}
\end{splitdeluxetable*}

%% file: tab7.tex
\setcounter{table}{6}
%\begin{table*}
%\centering
%\begin{minipage}{185mm}
\begin{splitdeluxetable*}{lccccccccBcccccccc}
\tabletypesize{\scriptsize}
\tablecolumns{17}
\tablewidth{0pt}
\tablecaption{Intensities of Lines Observed with the Long-High Module (LH)
}
%\begin{tabular}{@{}lcccccccccc@{}}
%\hline
%\hline
%Position & RA & Dec & $\tau_{9.6\,\micron}$ & AORKEY & [Fe\,{\sc iii}] 22.9 \micron\ & [O\,{\sc iv}] 25.9 \micron\ & [Fe\,{\sc ii}] 26.0 \micron\ & H$_2$ S(0) 28.3 \micron\ & [S\,{\sc iii}] 33.5 \micron\ & [Si\,{\sc ii}] 34.8 \micron\ \\
%     & (J2000)  & (J2000) & & &  (W m$^{-2}$ sr$^{-1}$) & (W m$^{-2}$ sr$^{-1}$) & (W m$^{-2}$ sr$^{-1}$) & (W m$^{-2}$ sr$^{-1}$) &  (W m$^{-2}$ sr$^{-1}$) &  (W m$^{-2}$ sr$^{-1}$)   \\
\tablehead{\colhead{Position} & \colhead{RA} & \colhead{Dec} & \colhead{$\tau_{9.6 \micron}$} & \colhead{AORKEY} & \colhead{[Fe\,{\sc iii}] 22.9 \micron\ } & \colhead{Error}  & \colhead{[O\,{\sc iv}] 25.9 \micron} & \colhead{Error}  & \colhead{[Fe\,{\sc ii}] 26.0 \micron} & \colhead{Error}   & \colhead{H$_2$ S(0) 28.3 \micron} & \colhead{Error} & \colhead{[S\,{\sc iii}] 33.5 \micron} & \colhead{Error}  & \colhead{[Si\,{\sc ii}] 34.8 \micron} & \colhead{Error} \\
 \colhead{}    & \colhead{(J2000)}  & \colhead{(J2000)} & \colhead{}  & \colhead{} & \colhead{(W m$^{-2}$ sr$^{-1}$)} & \colhead{(W m$^{-2}$ sr$^{-1}$)} & \colhead{(W m$^{-2}$ sr$^{-1}$)} & \colhead{(W m$^{-2}$ sr$^{-1}$)}  & \colhead{(W m$^{-2}$ sr$^{-1}$)} & \colhead{(W m$^{-2}$ sr$^{-1}$)} & \colhead{(W m$^{-2}$ sr$^{-1}$)} & \colhead{(W m$^{-2}$ sr$^{-1}$)} & \colhead{(W m$^{-2}$ sr$^{-1}$)} & \colhead{(W m$^{-2}$ sr$^{-1}$)}  & \colhead{(W m$^{-2}$ sr$^{-1}$)} & \colhead{(W m$^{-2}$ sr$^{-1}$)} }
%\hline
\startdata
G359.0409+0.1274 &  17 42 49.07 &  $-$29 41 11.8 &   1.33 &  3823104 &    2.973e$-$09 &  4.040e$-$10 &  1.209e$-$09 &  3.659e$-$10 &  1.117e$-$08 &  4.517e$-$10 &  6.637e$-$08 &  5.823e$-$10 &  6.615e$-$08 &  1.660e$-$09 &  1.822e$-$07 &  2.228e$-$09 \\
G359.4535$-$0.0527 &  17 44 31.10 &  $-$29 25 47.5 &   3.93 &  3823104 &    4.399e$-$09 &  5.979e$-$10 &  3.173e$-$09 &  4.379e$-$10 &  1.329e$-$08 &  6.639e$-$10 &  6.044e$-$08 &  9.394e$-$10 &  2.487e$-$07 &  2.344e$-$09 &  6.007e$-$07 &  4.639e$-$09 \\
G359.4989$-$0.0296 &  17 44 32.22 &  $-$29 22 44.4 &   3.50 &  3823104 &    5.452e$-$09 &  3.911e$-$10 & ... & ... &                      1.444e$-$08 &  6.306e$-$10 &  5.476e$-$08 &  9.807e$-$10 &  2.585e$-$07 &  2.173e$-$09 &  4.821e$-$07 &  4.586e$-$09 \\
G359.5840+0.0106 &  17 44 35.03 &  $-$29 17 07.8 &   1.28 &  3823104 &    5.780e$-$09 &  5.229e$-$10 & ... & ... &                      1.974e$-$08 &  7.694e$-$10 &  5.153e$-$08 &  7.834e$-$10 &  2.519e$-$07 &  2.501e$-$09 &  5.938e$-$07 &  4.404e$-$09 \\
G359.6788$-$0.1895 &  17 45 35.60 & $-$29 18 32.6 &   3.00 &  3823104 &    3.340e$-$09 &  5.114e$-$10 & ... & ... &                          1.673e$-$08 &  8.464e$-$10 &  4.767e$-$08 &  8.699e$-$10 &  5.084e$-$07 &  2.512e$-$09 &  4.117e$-$07 &  4.144e$-$09 \\
%\hline
%\end{tabular}\
\enddata
\tablecomments{(This table is available in its entirety in machine-readable form.)}
%\end{minipage}
%\end{table*}
\end{splitdeluxetable*}

%% file: tab8.tex
\setcounter{table}{7}
\begin{deluxetable*}{lccccccccc}
\tabletypesize{\scriptsize}
%\centering
%\begin{minipage}{185mm}
\tablecaption{Positions with Measured High-Excitation [\ion{Ne}{5}]
}
\tablecolumns{10}
\tablewidth{140mm}
%\begin{tabular}{@{}lcccccccccc@{}}
%\hline
%\hline
\tablehead{\colhead{Position} & \colhead{RA} & \colhead{Dec} & \colhead{AORKEY} & \colhead{Module} & \colhead{[Ne\,{\sc v}] 14.32 \micron\ } & \colhead{Error} & \colhead{[Ne\,{\sc v}] 24.32 \micron} & \colhead{Error} & \colhead{Reference} \\
 \colhead{}    & \colhead{(J2000)}  & \colhead{(J2000)} & \colhead{} & \colhead{}  &  \colhead{(W m$^{-2}$ sr$^{-1}$)} &  \colhead{(W m$^{-2}$ sr$^{-1}$)}  &  \colhead{(W m$^{-2}$ sr$^{-1}$)} &  \colhead{(W m$^{-2}$ sr$^{-1}$)} & \colhead{} }
%\hline
\startdata
G359.5666$-$0.1561 & 17 45 11.7 & $-$29 23 15 & 23971584 & LL & ... & ... & 1.94e-08 &  1.31e-10 &  \\
\\
G359.6500$-$0.0786 & 17 45 05.5 & $-$29 16 33 & 23974144 & LL & ... & ... & 4.63e-07 &  1.29e-09 & 1,2 \\
\\
%G000.0968$-$0.0505 & 17 46  2.8  & $-$28 52 47  & 23970048 & LH & ... & ... & 3.93e-07 &  1.33e-09 & 1,3 \\
%G000.0978$-$0.0502 & 17 46  2.9  & $-$28 52 44  & 23970048 & SH &  5.36e-06 &  1.18e-07 & ... & ... &  \\
G000.0967$-$0.0511 & 17 46 02.9 & $-$28 52 49 & 23970048 & LH & ... & ... & 7.55e-07 &  6.24e-09 & 1,3 \\
G000.0980$-$0.0502 & 17 46 02.9 & $-$28 52 43 & 23970048 & SH &  6.88e-06 &  2.41e-08 & ... & ... &  \\
G000.0993$-$0.0520 & 17 46 03.5 & $-$28 52 42 & 23970048 & LL & ... & ... & 5.49e-07 &  3.99e-09 &  \\
G000.0987$-$0.0504 & 17 46 03.1 & $-$28 52 41 & 23972864 & LL &  4.97e-07 &  3.10e-08 & ... & ... &  \\
G000.0993$-$0.0520 & 17 46 03.5 & $-$28 52 43 & 23970048 & LL &  1.48e-06 &  9.06e-08 & ... & ... &  \\
\\
G000.1118$-$0.0948 & 17 46 15.3 & $-$28 53 24 & 28146944 & LH & ... & ... & 1.18e-08 &  1.45e-09 &  \\
G000.1120$-$0.0947 & 17 46 15.3 & $-$28 53 23 & 28146944 & SH &  5.86e-09 &  1.03e-09 & ... & ... &  \\
G000.1142$-$0.0970 & 17 46 16.2 & $-$28 53 21 & 23972864 & LL &  1.24e-07 &  8.62e-09 & ... & ... &  1 \\
%G000.1118$-$0.0944 & 17 46 15.2  & $-$28 53 23  & 28146944 & LH & ... & ... & 5.70e-09 &  6.61e-10 &  \\
%G000.1118$-$0.0944 & 17 46 15.2  & $-$28 53 23  & 28146944 & SH &  5.69e-09 &  1.53e-09 & ... & ... &  \\
\\
G000.2004$-$0.0768 & 17 46 23.8 & $-$28 48 17 & 23970560 & LL &  3.29e-07 &  2.27e-08 & ... & ... &  \\
\\
G000.2443+0.0344 & 17 46 04.0 & $-$28 42 35 & 23968512 & LL & ... & ... & 4.18e-07 &  1.69e-09 & 4  \\
\\
%G000.3938+0.2137 & 17 45 43.4  & $-$28 29 19 & 23969792 & LH & ... & ... & 3.59e-08 &  8.10e-10 &  \\
G000.3938+0.2131 & 17 45 43.5 & $-$28 29 21 & 23969792 & LH & ... & ... & 8.28e-08 &  5.48e-09 &  \\
G000.3952+0.2143 & 17 45 43.4 & $-$28 29 14 & 23969792 & SH &  3.21e-08 &  7.54e-09 & ... & ... &  \\
G000.4014+0.2012 & 17 45 47.4 & $-$28 29 20 & 28146688 & SH &  1.32e-08 &  3.58e-09 & ... & ... &  \\
G000.4017+0.2013 & 17 45 47.4 & $-$28 29 18 & 28146688 & LH & ... & ... & 1.39e-08 &  2.46e-09 &  \\
%G000.4012+0.2011 & 17 45 47.4  & $-$28 29 20 & 28146688 & LH & ... & ... & 6.96e-09 &  6.63e-10 &  \\
%G000.4012+0.2010 & 17 45 47.4  & $-$28 29 20 & 28146688 & SH &  1.63e-08 &  2.91e-09 & ... & ... &  \\
G000.3986+0.2082 & 17 45 45.4 & $-$28 29 15 & 23969792 & LL & ... & ... & 1.50e-07 &  2.35e-10 &  \\
G000.3764+0.2109 & 17 45 41.6 & $-$28 30 18 & 23969792 & LL &  2.83e-08 &  2.66e-10 & ... & ... &  \\
G000.4076+0.2246 & 17 45 42.8 & $-$28 28 16 & 23969792 & LL &  1.75e-08 &  2.35e-10 & ... & ... &  \\
G000.4082+0.1905 & 17 45 50.9 & $-$28 29 19 & 23969792 & LL &  1.27e-08 &  9.72e-11 & ... & ... &  \\
G000.3989+0.2079 & 17 45 45.5 & $-$28 29 15 & 23969792 & LL &  1.73e-07 &  7.05e-09 & ... & ... &  \\
\\
%G000.4805$-$0.0295 & 17 46 52.4  & $-$28 32 28  & 23968768 & LH & ... & ... & 1.79e-08 &  6.49e-10 &  \\
%G000.4805$-$0.0295 & 17 46 52.4  & $-$28 32 28  & 23968768 & SH &  3.23e-08 &  1.11e-09 & ... & ... &  \\
G000.4805$-$0.0301 & 17 46 52.6 & $-$28 32 29 & 23968768 & LH & ... & ... & 3.68e-08 &  1.28e-09 &  \\
G000.4808$-$0.0295 & 17 46 52.5 & $-$28 32 27 & 23968768 & SH &  3.01e-08 &  1.15e-09 & ... & ... &  \\
G000.4817$-$0.0321 & 17 46 53.2 & $-$28 32 29 & 23969024 & LL &  1.77e-07 &  6.47e-09 & ... & ... &  \\
G000.4826$-$0.0312 & 17 46 53.1 & $-$28 32 24 & 23976448 & LL &  1.91e-07 &  5.20e-09 & ... & ... &  \\
G000.4832$-$0.0316 & 17 46 53.3 & $-$28 32 23 & 23968512 & LL &  1.57e-07 &  1.29e-08 & ... & ... &  \\
%\hline
\enddata
%\end{tabular}\
%\tablecomments{Table 8 is published in its entirety in the machine-readable format. A portion is shown here for guidance regarding its form and content.}
\tablerefs{
(1) Source in P$\alpha$ map (Dong et al. 2011).
(2){Source in radio map (Lang et al. 2010).}
(3) Source in radio map (Yusef-Zadeh et al. 2004).
(4) Source in radio map (Immer et al. 2012).
}
%\end{minipage}
\end{deluxetable*}

%% file: tab9.tex
\setcounter{table}{8}
\startlongtable
\begin{deluxetable*}{lccccc}
%\centering
%\begin{minipage}{120mm}
%\caption{Positions with Substantial High-Excitation [O IV] but No Measured [Ne V]
\tablecaption{Positions with Substantial High-Excitation [\ion{O}{4}] but No Measured [\ion{Ne}{5}]
}
\tablecolumns{6}
\tablewidth{130mm}
%\begin{tabular}{@{}lccccc@{}}
%\hline
%\hline
%Position & RA & Dec  & AORKEY & Module & Reference \\
%     & (J2000)  & (J2000) & &  & \\
\tablehead{\colhead{Position} & \colhead{RA} & \colhead{Dec} & \colhead{AORKEY} & \colhead{Module} & \colhead{Reference} \\
\colhead{} & \colhead{(J2000)} & \colhead{(J2000)} & \colhead{} & \colhead{} &  \colhead{}}
%\hline
\startdata
G359.4036+0.0010  & 17 44 11.2  & -29 26 38  & 23964416  & LH  & \\
   \\
G359.8334+0.0618  & 17 44 58.9  & -29 02 46  & 10464256  & LL1  & 1 \\
   \\
G359.9627$-$0.1202  & 17 46 00.0  & -29 01 50  & &  & 2 \\
G359.9631$-$0.1198  & 17 46 00.0  &  -29 01 48   & 23969536  & SH\tablenotemark{a}  & \\
G359.9621$-$0.1201 &  17 45 59.9  &  -29 01 52  &  23969536  &  LH\tablenotemark{a}  & \\
G359.9625$-$0.1200  &  17 45 59.9  & -29 01 51  & 10458624 &  LL2  & \\
G359.9636$-$0.1199  &  17 46 00.0  & -29 01 47  & 23969536  & LL2   & \\
G359.9627$-$0.1204  & 17 46 00.0  & -29 01 51   & 10458624  & LL1  & \\
G359.9637$-$0.1199  & 17 46 00.1  & -29 01 47  & 23969536  & LL1  & \\
G359.9633$-$0.1205  & 17 46 00.1  & -29 01 49  & 23969536  & SL2\tablenotemark{b}  & \\
G359.9633$-$0.1205  & 17 46 00.1  & -29 01 49  & 23969536  & SL1\tablenotemark{c}  & \\
  \\
G000.0200+0.0460  &  17 45 29.2  & -28 53 43  &  &  &  1 \\ 
G000.0209+0.0438  &  17 45 29.9   & -28 53 44  &  23975680   & SH   & \\
G000.0209+0.0438  &  17 45 29.9   & -28 53 44  &  23975680  &  LH   & \\
G000.0178+0.0456   &  17 45 29.0  & -28 53 50  &  23967232  &   LL2  & \\
G000.0189+0.0460  &  17 45 29.1  & -28 53 46   & 23971328   &   LL2  & \\
G000.0198+0.0459  &  17 45 29.2  & -28 53 43   & 23967488  &    LL2  & \\
G000.0177+0.0459  &  17 45 28.9  & -28 53 50   & 23967232   &   LL1  & \\
G000.0177+0.0457  &  17 45 29.0  & -28 53 50   & 23967232   &   LL1  & \\
G000.0190+0.0460   &  17 45 29.1  & -28 53 46   & 23971328   &  LL1  & \\
G000.0204+0.0463  &  17 45 29.2  & -28 53 41   & 23967488    &   LL1  & \\
G000.0204+0.0472  &   17 45 29.1  & -28 53 39   &   23969280 &  LL1  & \\
   \\
G000.0860+0.1749  & 17 45 08.6  & -28 46 18  &  &  &   \\
G000.0870+0.1745   &  17 45 08.8  &  -28 46 15    & 23974656  & SH     & \\
G000.0928+0.1638   &  17 45 12.1   & -28 46 18    & 23975424  & LH     & \\
   \\
G000.396$-$0.067  & \\
G000.3960$-$0.0675  &  17 46 49.3   & -28 37 59   &  28146688  & SH  & \\
G000.3975$-$0.0686   &  17 46 49.8  & -28 37 56   &  23976448  & LL2   & \\
G000.3964$-$0.0668  &   17 46 49.2  & -28 37 56   &  23976448  & LL1    & \\
G000.3983$-$0.0697  &   17 46 50.2  & -28 37 56   &  23969792  & SL1    & \\
 \tablebreak
G000.455$-$0.005 &  &  &  &  &   \\
G000.4544$-$0.0049  &   17 46 43.0  & -28 33 02  &  23964416   &   LL1     & \\
G000.4551$-$0.0063  &   17 46 43.4  & -28 33 03  &   23964416  &  LL2      & \\
G000.4570$-$0.0051  &   17 46 43.4  & -28 32 55  &  23976448   & LL2     & \\
G000.4564$-$0.0041  &   17 46 43.1  & -28 32 55   & 23976448  &  LL2   & \\
% \tablebreak
\\
G000.5709$-$0.0922  &  17 47 19.9  & -28 29 47  &   23968512   & LL2\tablenotemark{d} & \\
   \\
G000.6177+0.1823  &   17 46 22.5  & -28 18 50  &  28146688  &  SH  & \\
G000.6231+0.1686   &  17 46 26.5  & -28 18 59 &   23969792  & SH  & \\
G000.6385+0.1753  &  17 46 27.0  & -28 17 59  &   28146688  & SH  & \\
G000.6081+0.1627   &   17 46 25.7  & -28 19 56  &  28146688   &  LH  & \\
G000.6178+0.1823   &  17 46 22.5  & -28 18 50   &  28146688  &   LH  & \\
G000.6221+0.1683   &   17 46 26.4  & -28 19 02  & 23969792   &  LH  & \\
G000.6267+0.1530   &   17 46 30.6  & -28 19 17   &  28146688   &  LH  & \\
G000.6385+0.1754   &   17 46 27.0  & -28 17 59  &  28146688   &   LH  & \\
%\hline
\enddata
%\end{tabular}\
\tablenotetext{a}{The [\ion{Ne}{3}] 15.6 \micron\ line in the SH spectrum is saturated. The intensity of the [\ion{Ne}{3}] 36.015 \micron\ line in the LH spectrum is $3.10  \pm 0.38 \times 10^{-7}$ W m$^{-2}$ sr$^{-1}$.}
\tablenotetext{b}{The intensity of the [\ion{Na}{3}] 7.32 \micron\ line is $4.88 \pm 0.26 \times 10^{-7 }$ W m$^{-2}$ sr$^{-1}$.}
\tablenotetext{c}{The intensity of the [\ion{Ar}{3}] 8.99 \micron\ line is $4.57 \pm 0.18 \times 10^{-7}$ W m$^{-2}$ sr$^{-1}$.}
\tablenotetext{d}{Observed in LL2 only, but the [\ion{Ne}{3}] 15.6 \micron\ line is brighter than the [\ion{S}{3}] 18.7 \micron\ line.}
\tablerefs{
(1) Source in P$\alpha$ map (Dong et al. 2011).
(2) Source in radio map (Yusef-Zadeh et al. 2004).
}
%\end{minipage}
\end{deluxetable*}

%% file: tab10.tex
\setcounter{table}{9}
\begin{table*}
\centering
\begin{minipage}{100mm}
\caption{Positions with Measured [S I]
}
\begin{tabular}{@{}lccccc@{}}
\hline
\hline
Position & RA & Dec & AORKEY &  [S\,{\sc i}] 25.25 \micron\ & Error \\
%   & & & & $\times 10^{-18}$  \\
     & (J2000)  & (J2000)  &  &  (W m$^{-2}$ sr$^{-1}$)  &  (W m$^{-2}$ sr$^{-1}$) \\
\hline
G359.0409+0.1274   & 17 42 49.1 & -29 41 12 & 3823104  &  1.51e-09 &  3.54e-10 \\
G359.4553$-$0.1165 & 17 44 46.4 & -29 27 42 & 23965952 &  1.78e-09 &  5.51e-10 \\
G359.4603$-$0.1215 & 17 44 48.2 & -29 27 36 & 23966720 &  4.55e-09 &  6.26e-10 \\
G359.5682$-$0.2527 & 17 45 34.6 & -29 26 11 & 28146688 &  1.23e-08 &  1.51e-09 \\
G359.5918$-$0.2410 & 17 45 35.2 & -29 24 37 & 28146688 &  1.17e-08 &  1.78e-09 \\
G359.7792$-$0.0922 & 17 45 27.2 & -29 10 22 & 23966720 &  2.24e-09 &  7.83e-10 \\
G359.9474$-$0.0725 & 17 45 46.6 & -29 01 08 & 23966208 &  1.49e-08 &  4.64e-09 \\
G359.9683$-$0.0743 & 17 45 50.0 & -29 00 07 & 28146688 &  1.70e-08 &  3.04e-09 \\
G359.9887$-$0.0796 & 17 45 54.2 & -28 59 14 & 28146688 &  1.81e-08 &  4.39e-09 \\
G359.9916$-$0.0782 & 17 45 54.3 & -28 59 03 & 23966208 &  1.14e-08 &  4.33e-09 \\
G000.1304$-$0.0972 & 17 46 18.5 & -28 52 31 & 28146944 &  4.91e-09 &  1.22e-09 \\
G000.3361+0.0998   & 17 46 01.7 & -28 35 50 & 23974400 &  2.94e-09 &  1.15e-09 \\
G000.4562$-$0.0685 & 17 46 58.1 & -28 34 55 & 23975936 &  3.15e-09 &  8.11e-10 \\
G000.4607$-$0.0356 & 17 46 51.1 & -28 33 40 & 23973632 &  1.99e-09 &  7.91e-10 \\
G000.6063$-$0.0214 & 17 47 08.4 & -28 25 45 & 23976192 &  1.77e-09 &  6.84e-10 \\
G000.6486$-$0.0309 & 17 47 16.6 & -28 23 53 & 23976192 &  1.65e-09 &  5.69e-10 \\
G000.6684$-$0.0908 & 17 47 33.4 & -28 24 44 & 23965696 &  5.15e-09 &  1.43e-09 \\
G000.6689$-$0.0576 & 17 47 25.7 & -28 23 40 & 23976192 &  2.59e-09 &  7.29e-10 \\
G000.6728$-$0.1006 & 17 47 36.3 & -28 24 48 & 23976192 &  2.57e-09 &  5.88e-10 \\
G000.6783$-$0.0829 & 17 47 32.9 & -28 23 58 & 23976192 &  3.93e-09 &  1.26e-09 \\
G000.4840$-$0.0272 & 17 46 39.7 & -28 30 31  & 3822848  &  1.84e-09 &  4.96e-10 \\
G000.6791$-$0.1999 & 17 48 00.3 & -28 27 34  & 3822592  &  1.56e-09 &  4.82e-10 \\
G000.7626$-$0.0479 & 17 47 36.6 & -28 18 34  & 3822592  &  6.95e-09 &  5.97e-10 \\
G000.8269$-$0.0995 & 17 47 57.7 & -28 16 51  & 3822592  &  2.59e-09 &  3.39e-10 \\
\hline
\end{tabular}\
\end{minipage}
\end{table*}

%% file: tab11.tex
\setcounter{table}{10}
\begin{table*}
\centering
\begin{minipage}{110mm}
\caption{Sources with Ice Absorption Features at 6.0 and 6.8 \micron\ 
}
\begin{tabular}{@{}lccc@{}}
\hline
\hline
Position & RA & Dec & AORKEY  \\
     & (J2000)  & (J2000)  &   \\
\hline
 G359.4357$-$0.1019 & 17 44 40.09 &  -29 28 14.4 &  23970816   \\
 G359.4372$-$0.1008 & 17 44 40.05 &  -29 28 07.8 &  23970816   \\
 G359.6128+0.0217 & 17 44 36.57 &  -29 15 18.6 &  23965440  \\
 G359.6878$-$0.0238 & 17 44 58.03 &  -29 12 53.8 &  23973120  \\
 G359.8690$-$0.0179 & 17 45 22.62 &  -29 03 26.3 &  10455808   \\
 G359.9323$-$0.0632 & 17 45 42.29 &  -29 01 36.8 &  23970304  \\
 G000.0023$-$0.0686 & 17 45 53.56 &  -28 58 11.7 &  23969792   \\
 G000.2566$-$0.0161 & 17 46 17.50 &  -28 43 31.7 &  23965184 \\
 G000.3277$-$0.0065 & 17 46 25.37 &  -28 39 35.0 &  23966976 \\
 G000.4470$-$0.0061 & 17 46 42.22 &  -28 33 27.2 &  23974656 \\
 G000.4529$-$0.0020 & 17 46 42.11 &  -28 33 01.4 &  23974656  \\
 G000.5293$-$0.0064 & 17 46 53.95 &  -28 29 14.6 &  23965696   \\
 G000.5530$-$0.0634 & 17 47 10.62 &  -28 29 47.9 &  23965184  \\
 G000.5683$-$0.0529 & 17 47 10.34 &  -28 28 41.4 &  23965184  \\
 G000.6003$-$0.0370 & 17 47 11.17 &  -28 26 32.9 &  23968256  \\
% G000.6234+0.1680 & 17 46 26.62 &  -28 18 58.8 &  23969792 \\
 G000.6284$-$0.0261 & 17 47 12.58 &  -28 24 46.1 &  23965696   \\
 G000.6304$-$0.0096 & 17 47 09.01 &  -28 24 09.1 &  23968256  \\
 G000.6465$-$0.0878 & 17 47 29.54 &  -28 25 45.5 &  23968256 \\
 G000.6532$-$0.0702 & 17 47 26.37 &  -28 24 52.0 &  23964672  \\
 G000.6545$-$0.0722 & 17 47 27.03 &  -28 24 51.6 &  23971584  \\
 G000.6548$-$0.0996 & 17 47 33.47 &  -28 25 41.9 &  23965696  \\
% G000.6549$-$0.0690 & 17 47 26.34 &  -28 24 44.6 &  23964672   \\
 G000.6579$-$0.0419 & 17 47 20.44 &  -28 23 44.7 &  23965696  \\
% G000.6661$-$0.0505 & 17 47 23.62 &  -28 23 35.5 &  23964672   \\
 G000.6679$-$0.0377 & 17 47 20.87 &  -28 23 06.3 &  23965184  \\
 \hline
\end{tabular}\
\end{minipage}
\end{table*}

%% file: tab12.tex
\setcounter{table}{11}
\begin{table*}
\centering
\begin{minipage}{120mm}
\caption{Average Electron Densities and Ionic Abundance Ratios
}
\begin{tabular}{@{}lcccc@{}}
\hline
\hline
H II Region & Electron Density & Ne/H\tablenotemark{a} & S/H\tablenotemark{b} & Si$^+$/H$^+$ \\
     & (cm$^{-3}$)  &  $\times 10^{-4}$ & $\times 10^{-5}$ &  $\times 10^{-5}$ \\
\hline
Arched Filaments\tablenotemark{c}  & 600 & $1.70 \pm 0.22$ & $2.09 \pm 0.29$ & $1.17 \pm 0.01$ \\
%Densities and abundances from paper version 9/26/2017 (initial ApJ submission)
%GC Filaments\tablenotemark{d}      & 420 & $1.27 \pm 0.19$ & $1.31 \pm 0.20$  & $1.07 \pm 0.04$ \\
%Quintuplet Region\tablenotemark{e} & 390 & $1.29 \pm 0.23$ & $1.12 \pm 0.20$ & $1.05 \pm 0.05$ \\
%Sgr B1\tablenotemark{f}            & 90 &  $1.50 \pm 0.25$ & $1.25 \pm 0.21$ & $0.87 \pm 0.03$ \\
%Sgr C\tablenotemark{g}             & 200 & $1.22 \pm 0.20$ & $0.57 \pm 0.09$ & $0.96 \pm 0.04$ \\
GC Filaments\tablenotemark{d}      & 270 & $1.69 \pm 0.35$ & $1.75 \pm 0.36$  & $1.41 \pm 0.08$  \\
Quintuplet Region\tablenotemark{e} & 310 & $1.80 \pm 0.45$ & $1.50 \pm 0.37$ & $1.39 \pm 0.08$ \\
Sgr B1\tablenotemark{f}            & 290 & $1.96 \pm 0.45$ & $1.87 \pm 0.41$ & $1.35 \pm 0.07$ \\
Sgr C\tablenotemark{g}             & 300 & $1.54 \pm 0.34$ & $0.86 \pm 0.18$ & $1.47 \pm 0.08$ \\
\hline
\end{tabular}
\tablenotetext{a}{(Ne$^+$ + Ne$^{++}$)/H$^+$} 
\tablenotetext{b}{(S$^{++}$ + S$^{3+}$)/H$^+$}
\tablenotetext{c}{Positions 29 -- 34 in Simpson et al. (2007)}
{ 
\tablenotetext{d}{Thirty positions within $0.05 < l < 0.25$ and $-0.01 < b < +0.012$}
\tablenotetext{e}{Ten positions within $0.11 < l < 0.25$ and $-0.07 < b < -0.01$}
\tablenotetext{f}{Twenty-nine positions within $0.45 < l < 0.60$ and $-0.11 < b < -0.01$}
\tablenotetext{g}{Twenty-eight positions within $-0.62 < l < -0.42$ and $-0.20 < b < 0.00$}
}
\end{minipage}
\end{table*}

%\end{document}

%% file: tab13.tex
\setcounter{table}{12}
\begin{table*}
\movetabledown=2.0in
\begin{rotatetable*}
\begin{center}
%\begin{minipage}{220mm}
\caption{Galactic Center H II Region Models}
\begin{tabular}{@{}lccccccccccccc@{}}
\hline
Source &  $N_{\rm LyC}$ & Starburst0 & $R_{\rm inner}$ & Filling Factor & Log age & Log $T_{\rm BB}$ & Log $L_{\rm BB}$ & Log $U$ & $Chisq$ & $\frac{{\rm [Ne \ III]} 15 \micron\ }{{\rm [S \ III]} 19 \micron\ }$ & $\frac{{\rm [S \ III]} 33 \micron\ }{{\rm [Si \ II]} 34 \micron\ }$ & $\frac{{\rm [O \ IV]} 26 \micron\ }{{\rm [S \ III]} 33 \micron\ }$  \\ 
       & photons s$^{-1}$ & Model & (pc) & & (Myr) & (K) & (erg s$^{-1}$) & &  & & &   \\
\hline
\hline

Arched Filaments & $4 \times 10^{50}$ \\
&& 67\_65192 & 10.0 & 0.316 & 6.7 & 6.5 & 38.5 & -2.72 & 0.049 &  0.046 & 0.883 & 0.0042  \\
&& 665\_65102 & 10.0 & 0.1 & 6.65 & 6.5 & 38.5  & -2.90 & 0.063 & 0.046 & 0.912 & 0.0059 \\
&& 665\_65192 & 10.0 & 0.316 & 6.65 & 6.5 & 38.5  & -2.72 & 0.071 & 0.051 & 0.955 & 0.0064 \\
&& 67\_6041  & 3.163 & 0.1 & 6.7 & 6.0 & 37.5 & -2.78 & 0.016 &  0.042 &  1.377 & 0.0042 \\
&& 665\_6010 & 1.0 & 0.1 & 6.65 & 6.0 & 37.0 & -2.75 & 0.047 &  0.036 & 1.466 & 0.0036 \\
&& 665\_6041 & 3.163 & 0.1 & 6.65 & 6.0 & 37.5 & -2.77 & 0.086 &  0.057 &  1.462 & 0.0071  \\
\
Quintuplet & $5 \times 10^{50}$ \\
&& 64\_65192 & 10.0 & 0.316 & 6.4 & 6.5 & 38.5 & -2.71 & 0.065 & 0.261 & 1.303 & 0.0090  \\
&& 64\_65102  & 10.0 & 0.1 & 6.4 & 6.5 & 38.5  & -2.90 & 0.066 & 0.239 & 1.291 & 0.0073 \\
&& 65\_65192 & 10.0 & 0.316 & 6.5 & 6.5 & 38.5 & -2.71 & 0.072 & 0.157 & 1.186 & 0.0089 \\
&& 65\_6041 & 3.163 & 0.1 & 6.5 & 6.0 & 37.5 & -2.76 & 0.134 &  0.228 & 1.730  & 0.0088 \\
&& 66\_6041 &  3.163 & 0.1 & 6.6 & 6.0 & 37.5 & -2.75 & 0.152 & 0.142 & 1.560 & 0.0095 \\
&& 65\_6040  & 1.0 & 0.1 & 6.5 & 6.0 & 37.5 & -2.75  & 0.197 & 0.244 & 1.750 & 0.0124 \\

Sgr B1 & $3 \times 10^{50}$ \\
&& 665\_6570  & 1.0 & 0.1 & 6.65 & 6.5 & 38.0 & -2.76 & 0.022 & 0.049 & 1.524 & 0.0116 \\
&& 67\_65101  & 3.1623 & 0.1 & 6.7 & 6.5 & 38.5 & -2.79  & 0.027 & 0.065 & 1.239 & 0.0176 \\
&& 665\_65161 & 3.1623 & 0.0316 & 6.65 & 6.5 & 38.5 & -3.13 & 0.100 & 0.060 & 0.834 & 0.0150 \\
&& 67\_6070  & 1.0 & 0.1 & 6.7 & 6.0 & 38.0 & -2.78 & 0.095 & 0.114 & 1.355 & 0.0348 \\
&& 665\_6070  & 1.0 & 0.1 & 6.65 & 6.0 & 38.0 & -2.77 & 0.168 & 0.137 &  1.432 &  0.0401 \\
&& 665\_6071  & 3.1623 & 0.1 & 6.65 & 6.0 & 38.0 & -2.78  & 0.182 & 0.133 &  1.409 &  0.0251 \\

Sgr C & $1 \times 10^{50}$ \\
&& 66\_65202 &  10.0 & 0.1 & 6.65 & 6.5 & 39.0 & -2.92 & 0.130 & 0.180 &  0.673 &  0.0243 \\
&& 65\_65221 & 3.1623 & 0.01 & 6.5 & 6.5 & 39.0  & -3.49 & 0.130 & 0.178 & 0.670 &  0.0248 \\   
&& 665\_65202  & 10.0 & 0.1 & 6.65 & 6.5 & 39.0 & -2.93 & 0.136 &  0.143 &  0.601 &  0.0179 \\
&& 66\_6080 & 1.0 & 0.01 & 6.6 & 6.0 & 38.0 & -3.48 & 0.146 &  0.135 &  0.578 &  0.0176 \\
&& 65\_6080 & 1.0 & 0.01 & 6.5 & 6.0 & 38.0 & -3.48 & 0.148 & 0.171 & 0.671 &   0.0151 \\
&& 66\_6081 & 3.1623 & 0.01 & 6.6 & 6.0 & 38.0 & -3.48 & 0.178 & 0.135 &  0.577 &   0.0134 \\

\hline
\end{tabular}
\end{center}
%\end{minipage}
\end{rotatetable*}
\end{table*}